\documentclass[preprints,review,accept,oneauthor,pdftex]{Definitions/mdpi} 
\newcommand{\com}{{\cal C}}
\usepackage{float}

\DeclareMathOperator{\sinc}{sinc}
\DeclareMathOperator\erf{erf}
\usepackage{amsthm}
\usepackage{amssymb}
\newcommand{\be}{\begin{equation}}
\newcommand{\ee}{\end{equation}}
\newcommand{\bea}{\begin{eqnarray}}
\newcommand{\eea}{\end{eqnarray}}

\newcommand{\RomanNumeralCaps}[1]{\MakeUppercase{\romannumeral #1}}

\firstpage{1} 
\pubvolume{1}
\issuenum{1}
\articlenumber{0}
\doinum{}
\pubyear{2021}
\copyrightyear{2021}
\history{}
\usepackage{url}
\usepackage{hyperref}
\Title{PBH formation from spherically symmetric hydrodynamical perturbations: a review}

\Author{Albert Escrivà $^{1,2,3}$\orcidA{}}

\address{%
$^{1}$ \quad Service de Physique Th\'eorique, Universit\'e Libre de Bruxelles (ULB), Boulevard du Triomphe, CP225, 1050 Brussels, Belgium. \textbf{albert.escriva@ulb.be}\\ 

$^{2}$ \quad Institut d'Estudis Espacials de Catalunya (IEEC), Edifici Nexus I, C/ Gran Capit\`a, 2-4, desp. 201, 08034 Barcelona, Spain. \\

$^{3}$ \quad Institut de Ciències del Cosmos, Universitat de Barcelona, Martí i Franquès 1, 08028 Barcelona, Spain.}

\corres{Correspondence: albert.escriva@ulb.be}

\abstract{Primordial black holes, which could have been formed in the very early Universe due to the collapse of large curvature fluctuations, are nowadays one of the most attractive and fascinating research areas in cosmology for their possible theoretical and observational implications. This review article presents the current results and developments on the conditions for primordial black hole formation from the collapse of curvature fluctuations in spherical symmetry on a Friedman-Lemaître-Robertson-Walker background and its numerical simulation. We review the appropriate formalism for the conditions of primordial black hole formation, and we detail a numerical implementation. We then focus on different results regarding the threshold and the black hole mass using different sets of curvature fluctuations. Finally, we present the current state of analytical estimations for the primordial black hole formation threshold, contrasted with numerical simulations.
}


\begin{document}
%%%%%%%%%%%%%%%%%%%%%%%%%%%%%%%%%%%%%%%%%%
%\setcounter{section}{-1} %% Remove this when starting to work on the template.

\section{Introduction}\label{sec:intro}

One of the great mysteries in science is the composition of dark matter, which accounts for $27\%$ of the present Universe {\cite{10.1093/ptep/ptaa104}}. Although there are different theories and candidates that try to explain it {\cite{Bertone:2016nfn}}, still, the answer remains elusive. One of the most promising possibilities is Primordial Black Holes (PBHs), i.e., Black Holes (BH) generated at earlier than star formation times and therefore not of stellar origin. For the current observational status and constraints of PBHs in the form of dark matter, we refer the reader to \cite{Carr:2020gox,Carr:2021bzv,Carr:2020mqm,Carr:2020erq,Carr:2009jm,Keith:2020jww}.

PBHs were first considered in \cite{hawking1,hawking2,acreation1}. They could have formed in the very early Universe due to the gravitational collapse of cosmological perturbations. Under this scenario, PBHs could have been generated as a consequence of high non-linear peaks in the primordial distribution of density perturbations, leaving open the possibility that PBHs could constitute a significant fraction of Dark Matter (DM) \cite{Carr1,darkmatter1,darkmatter2,Sasaki:2018dmp,darkmatter4,darkmatter5,darkmatter6,Bird,darkmatter8,Clesse:2016vqa,Bernal:2017vvn,Clesse:2015wea,Clesse:2018ogk,Tada:2019amh,Atal:2020yic,Atal:2020igj,DeLuca:2020qqa,Bartolo:2018rku,Clesse:2017bsw,Ezquiaga:2019ftu,Takhistov:2020ssb,Jedamzik:2020ypm,Vallejo-Pena:2019lfo,Kashlinsky:2016sdv,Kashlinsky:2018mnu,Kashlinsky:2020ial,Pi:2017gih}.

PBHs with a size smaller than $M_{\rm PBH} <10^{-20}$ \(\textup{M}_\odot\) would have already evaporated due to Hawking radiation \cite{Carr:2020xqk}. Therefore, those with higher masses can account for dark matter. In addition, they may be responsible for seeding supermassive black holes at the centre of galaxies \cite{PhysRevD.66.063505,Bernal:2017nec}, generating large-scale structures through Poisson statistics \cite{Meszaros:1975ef}, or changing the thermal history of the Universe \cite{10.1093/mnras/194.3.639}.

Although PBHs were theorised a long time ago, it has not been until now that their popularity has increased exponentially. The current big interest in PBHs (and their golden age) is due to the first detection of gravitational waves from a black hole merger by the LIGO/Virgo collaborations \cite{LIGO}. Soon after this remarkable discovery, it was suggested that the constituents of the black hole merger could have had a primordial origin \cite{Bird,Sasaki:2016jop}. Later on, several analyses of the data signals suggested that the population of black holes detected could have a primordial origin \cite{Atal:2020igj,Franciolini:2021tla,Sasaki:2021iuc,LIGOScientific:2021job,Chen:2021nxo,DeLuca:2020sae} (see also the references therein). On the other hand, recent results from the NANOGrav experiment \cite{NANOGrav:2020bcs} have also been connected with PBHs \cite{Atal:2020yic,Vaskonen:2020lbd,DeLuca:2020agl,Kohri:2020qqd,Bian:2020urb,Sugiyama:2020roc,Domenech:2020ers,Bhattacharya:2020lhc,Inomata:2020xad}.

If formed by the collapse of inflationary perturbations, the abundance of PBHs is exponentially sensitive to the threshold of the gravitational collapse $\delta_{c}$ \cite{Germani:2018jgr} (where $\delta_{c}$ is the minimum amplitude of the gravitational potential peak related to the perturbation undergoing gravitational collapse and leading to a BH). In order to obtain the necessary precision on $\delta_{c}$, a numerical analysis of PBH formation is an obvious way to go.

Numerical simulations of PBH formation started some time ago, the first one being performed in \cite{nadezhin}, where the non-linear behaviour of the gravitational collapse was studied. Later on, in~\cite{Niemeyer1,Niemeyer2}, $\delta_{c}$ was computed and a scaling behaviour for the PBH mass was found whenever the amplitude of the perturbation was close to $\delta_{c}$~\cite{Niemeyer1,Niemeyer2,hawke2002,musco2009}. The value of the scaling exponent found was consistent with the one given in the literature and computed from a perturbative treatment \cite{renormalizationcriticalcollapse,Maison:1995cc} or from numerical simulations \cite{gamma1} in asymptotically flat spacetime. While the constant of proportionality appearing in the scaling law depends on the specific shape of the perturbation considered, the scaling exponent is a universal quantity only dependent on the type of fluid. Earlier simulations of PBHs were mainly based on the application of a Lagrangian hydrodynamic code based on finite differences and developed from an earlier work~\cite{maywhite}, solving Misner--Sharp equations~\cite{misnersharp}, which describe the motion of a relativistic fluid in a curved spacetime. An inconvenience of this procedure is the appearance of a coordinate singularity after forming the apparent horizon, which breaks down the simulation. To resolve that issue, Hernandez--Misner equations \cite{misnerhernandez} (which are the Misner--Sharp equations in null coordinates) are used to evade the formation of an apparent horizon and follow the subsequent evolution to determine the value of the black hole mass. The method is based on \cite{baumgarte}. As was shown in \cite{musco2005}, the formation of PBHs from the collapse of initial fluctuations at super-horizon scales does not form shocks (a discontinuous solution) in comparison with the earlier result of \cite{hawke2002}, where shock waves were found, but using initial conditions on sub-horizon scales.

On the other hand, even if a numerical code allows computing the threshold, the realisation of a numerical simulation can be quite expensive to be actually useful for statistical applications such as the calculations of PBH abundances. Therefore, an analytical expression would drastically simplify the problem. There are some analytical estimations of $\delta_c$ (e.g., \cite{carr75,harada}) that were based on analytical models under certain assumptions. However, numerical simulations have shown that, even for a fixed equation of state, $\delta_c$ is not universal \cite{musco2005,hawke2002,refrencia-extra-jaume,Niemeyer2,Shibata:1999zs,Nakama:2014fra,musco2018,escriva_solo}. The main argument is that $\delta_c$ is dependent on the specific detail of the initial curvature fluctuation \cite{musco2018}, i.e., on the scale dependence or ``shape'' of the perturbation, something that these analytical estimations do not take into account.

Some works have tried to characterise the threshold for black hole formation in terms of the initial density perturbation profile. In \cite{Hidalgo:2008mv}, the amplitude of the curvature fluctuation and its second radial derivate at the centre were used to characterise the initial configuration profiles, showing that the probability of PBH formation is sensitive to them. Later on, in \cite{Nakama_2014}, numerical simulations were used to argue the formation of a PBH from an over-density peak in an FLRW Universe assuming the perturbation to be initially at super-cosmological scales, and it only depends on two master parameters: the integral of the initial curvature perturbation and the edge of the over-density distribution. The paper \cite{musco2018} recently refined the arguments of \cite{Nakama_2014} by showing that these parameters may be more conveniently given in terms of the amplitude of the ``gravitational potential'' at its maximum ($r=r_m$), as already noticed in \cite{Shibata:1999zs}, and that it mainly depends on the functional form (shape) of the gravitational potential. More precisely, in \cite{musco2018}, the threshold $\delta_{c}$ was identified with the peak of the ``compaction function'' \cite{Shibata:1999zs} at super-horizon scales. The compaction function closely resembles the Schwarzschild gravitational potential and is defined as twice the local excess-mass over the co-moving areal radius.

Following the aim to have a more accurate analytical estimation for the threshold of PBH formation, recently, in \cite{universal1}, it was argued that the threshold for primordial black hole formation should be quite insensitive to the physics beyond $r_m$: the threshold is the amplitude for which a ``black hole'' with zero mass would be formed, taking an infinite time. This is an asymptotically and ideal limit. Therefore, all the over-density beyond $r_m$ will be diffused away while that just in the proximity of $r_m$ will block the collapse.

With those arguments, it was shown that during a radiation-dominated epoch (equation of state $p=w\rho$ with $w=1/3$), with a very good approximation, there exists a universal threshold value for the volume-averaged compaction function, which is actually shape-independent. Taking into account that the volume average is dominated by scales near the maximum of the compaction function, in \cite{universal1}, it was shown that it is enough to parametrise the profile dependence of $\delta_c$ by the curvature shape around the compaction function at its maximum. This result was used to make an analytical approximation for $\delta_c$ taking into account the shape dependence on the curvature fluctuation, which matches that found in simulations to within a few percent \cite{universal1}.

From the results in \cite{universal1}, there was the question of whether or not this universality on the averaged critical compaction function is generic and not only an accident of radiation. Thus, in \cite{Escriva:2020tak}, the universality question beyond the case of a radiation fluid was addressed. In \cite{Escriva:2020tak}, the arguments of \cite{universal1} were generalised to provide a new analytical estimation for the threshold in terms of the equation of state and the shape profile. This analytical formula was tested in front of numerical simulations and was found with an accuracy better than a $6\%$.

The other half-part of the problem is the determination of the final PBH mass. When a perturbation collapses, it will firstly form an apparent horizon with an associated mass $M_{PBH,i}$. After that, due to the surrounding fluid of the FLRW background, there is an accretion process up to a final mass $M_{PBH,f}$ when the PBH is formed. As we have indicated, for fluctuations whose amplitude is near the threshold (the critical regime), a scaling law was found for the PBH mass in terms of the amplitude of the perturbation \cite{Niemeyer1,Niemeyer2,hawke2002,musco2009}. Although the importance of this result was pointed out in the context of critical collapse~\cite{Gundlach:1999cu}, currently, it is substantially relevant because, as indicated in the current literature~\cite{Germani:2018jgr}, the PBHs with a higher probability of formation are those that are formed in the critical regime. The probability distribution function is exponentially suppressed beyond the threshold, and therefore, the scaling law can be directly used for estimations of PBH~abundances.

On the other hand, some studies have also considered the maximum size of PBHs at the formation time and the accretion effect from the surrounding background in the final mass of the PBH $M_{PBH,f}$. It was already noticed in \cite{size1,carr75} that the accretion effect for small PBHs should be small. Specifically, in \cite{size1}, some analytical expressions were derived for the upper bound of a PBH formed by the collapse of hydrodynamical perturbations. In \cite{Escriva:2021pmf}, the $M_{PBH,f}$ was computed numerically for different curvature profiles and the accretion effect was quantified, which could be substantially large for PBHs formed beyond the critical regime $O(10\%)$.

In conclusion, considering the recent developments in the field and future perspectives, we present this review paper to give the current state-of-the-art regarding the analytical, numerical, and theoretical results about PBH formation from the collapse of hydrodynamical perturbations on an FLRW background under spherical symmetry, currently the most common and popular scenario in the literature \cite{Carr:2020xqk}. We hope that this review paper allows the scientific community to be introduced to the topic of PBH formation and understand the main insights.

We want to tell the reader that there are fascinating and valuable review papers on the PBH topic with a different perspective than the one considered here. For an interesting historical review of the dark matter candidates, we refer the reader to \cite{Bertone:2016nfn}. Regarding reviews that focus on gravitational waves applied to PBHs and the different constraints to account for dark matter, we have \cite{Sasaki:2018dmp,darkmatter2,Carr:2020gox,Carr:2020xqk,Green:2020jor,Yuan:2021qgz,Villanueva-Domingo:2021spv,Domenech:2021ztg}. Another interesting review with potential applications to PBHs is \cite{Allahverdi:2020bys}, which is focused on scenarios with deviations from radiation domination in the early Universe.

This review paper is organised as follows: In Section \ref{sec:pbh_basics}, we present the basics of PBH formation. In Section \ref{abundances_sec}, we give some basic details about the statistical estimation of PBH abundances. In Section \ref{sec:set_up}, we give the necessary ingredients to set up the formalism of PBH formation, in particular: the differential equations that describe the collapse of the perturbations, the initial conditions, the definition of the compaction function, and the threshold criteria. In Section \ref{sec:numerics}, we discuss and comment on the different numerical techniques employed in the literature, and we focus on a specific method that is publicly available. In Section \ref{sec:results}, we give numerical results regarding the threshold and PBH mass in terms of the specific profiles using numerical simulations. In Section \ref{sec:analytics_on_the_thresholds}, we discuss in detail the different developments for the analytical estimation of the threshold. Finally, in Section \ref{sec:other_scenarios}, we enumerate other scenarios of PBH formation not explicitly considered in this review.

\section{Basics of PBH Formation}\label{sec:pbh_basics}

As we pointed out in the previous Section \ref{sec:intro}, PBHs could have been formed in the very early Universe during radiation domination due to the gravitational collapse of large curvature perturbations generated during inflation \cite{hawking1,hawking2}.

\begin{figure}[H]                    
\begin{center}                    
\includegraphics[width=0.6\columnwidth]{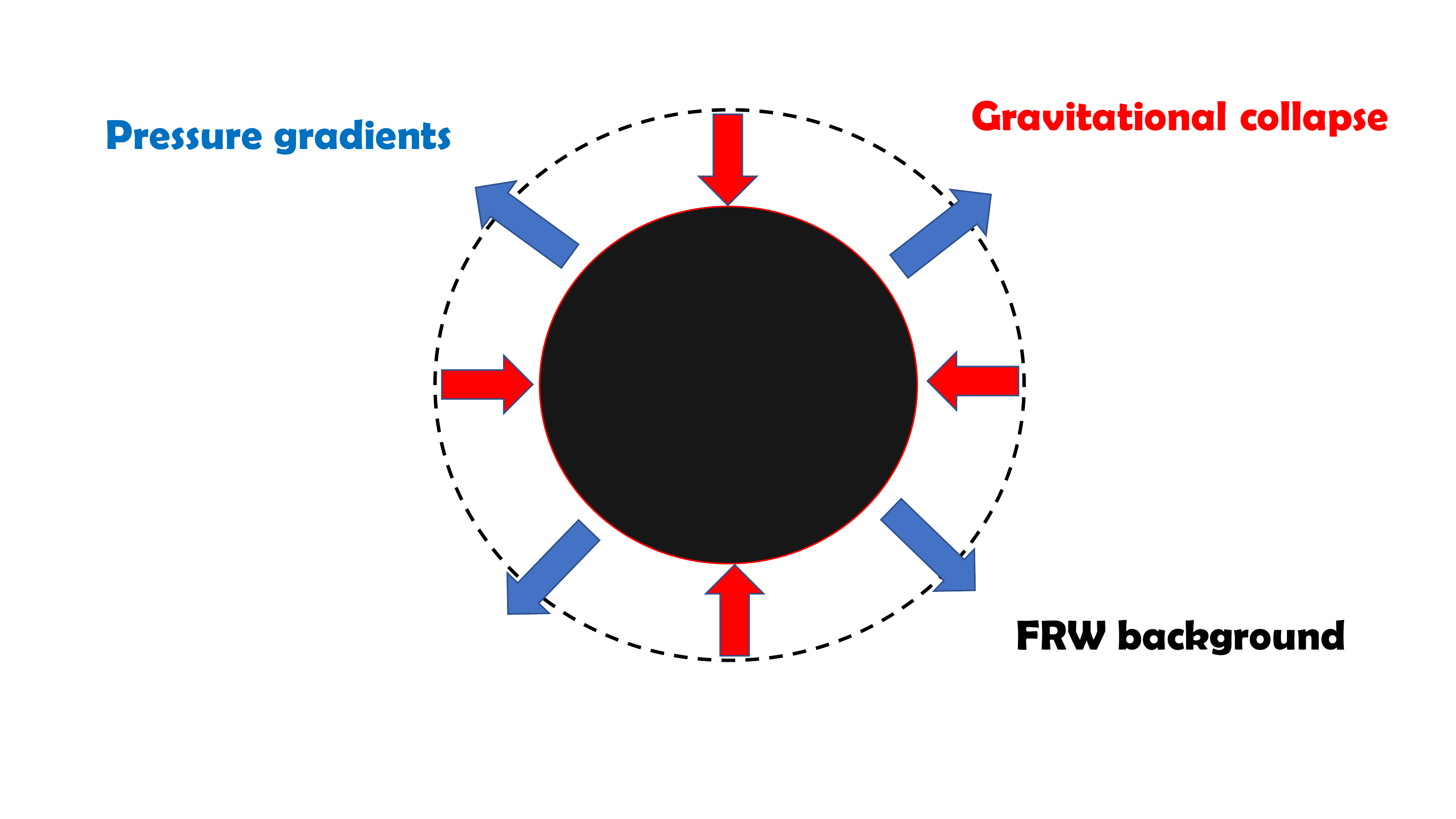}
\caption{Schematic picture of the collapse/dispersion of a perturbation on an FLRW background.} 
\label{fig:intro_colapse}                   
\end{center}                     
\end{figure}

Under this scenario, the collapse or dispersion of those hydrodynamic perturbations depends on the perturbation's strength: If its greater than a given threshold, the perturbation will collapse and form a BH after the perturbation {re-enters} the cosmological horizon. Otherwise, if it is lower, it will disperse because of pressure gradients preventing the collapse (a schematic picture can be found in Figure~\ref{fig:intro_colapse}). Both things could happen, i.e., the perturbation could undergo gravitational collapse and subsequently bounce. In this case, the fluid is dispersed and collapsed continuously, making rarefaction waves (we will see this in more detail in Section \ref{sec:results}). This behaviour is particularly evident when the initial strength of the perturbation is very close to its threshold value. Therefore, pressure gradients play a crucial role in the collapse of the perturbation. If pressure gradients are strong enough, they will prevent gravitational collapse, which implies having a large threshold compared to the situation when the pressure gradients are small.

A big effort has been made during the past few decades to find the correct PBH formation criteria. The first estimation of a threshold was given by \cite{hawking1}, using a Jeans length argument and Newtonian gravity. The criterion for the formation of a BH used was that the size of an over-density at the maximum expansion should be larger than the Jeans length, but also smaller than the particle horizon. This translates to the requirement that the peak value of the density contrast at scales smaller than the cosmological horizon must be at least $w$ (where $w$ is the equation of state of the perfect fluid that fills the FLRW Universe) in order to collapse.

A simple, intuitive picture of PBH formation and its threshold connected with the Jeans length argument can be found in \cite{Sasaki:2018dmp}, and we summarise it here. Consider a cosmological perturbation at super-horizon scales. The spacetime metric can be written as a closed FLRW Universe (we will see this in more detail in Section \ref{sec:set_up}) with a radial dependence on $K(r)$,

\begin{equation}
ds^2 = -dt^2 + a^2(t) \left[\frac{dr^2}{1-K(r) r^2}+r^2 d\Omega^2 \right].
\label{frw_example}
\end{equation}
where $a(t)$ is the scale factor. If we ignore the radial derivative of $K(r)$, the time--time component of the Einstein equations is given by,
\begin{equation}
H^{2}+\frac{K(r)}{a^{2}} = \frac{8 \pi}{3}\rho,
\end{equation}
which basically corresponds to the FLRW equation, but with a small non-homogeneity term given by $K(r)$. The previous equation allows defining a density contrast as,
\begin{equation}
\frac{\delta \rho}{\rho_{b}}= \frac{\rho-\rho_{b}}{\rho_{b}} = \frac{3K}{8\pi \rho_{b}a^{2}}=\frac{K}{(aH)^{2}},
\end{equation}
{where $\rho_b$ is the energy density of the background}. A region will collapse instead of expanding when $K>0$ if we ignore the radial dependence on $K(r)$. This precisely happens when $3K/a^{2}= 8 \pi \rho$, which corresponds roughly to when the Hubble scale becomes equal to the length scale of the curved region. In this case, the assumption made with Equation~\eqref{frw_example} is no longer valid, but still can be assumed to obtain some threshold estimation.

When $\delta \rho (t_c)/ \rho_{b}=1$ is the time when a homogeneous and isotropic Universe would stop expanding, we therefore assume that this epoch is the time of black hole formation $t=t_{c}$, in particular,
\begin{equation}
\frac{\delta \rho}{\rho_{b}}(t_c) = \frac{K}{k^{2}c^{2}_{s}}=1.
\end{equation}
The criterion is that perturbations cannot collapse if they have smaller scales than the Jeans length, which immediately implies that $k^{2}=a^{2}H^{2}/c^{2}_{s}$, where $c_s= \sqrt{w}$ is the sound speed of the fluid (perfect fluid in our case) and $k$ is the wavelength mode associated with the perturbation. Therefore, the condition that should satisfy the density contrast at the time when the perturbation re-enters the horizon ($H^{2}(t_m) a^{2}(t_m)=k^{2}$) is given by:
\begin{equation}
\left(\frac{\delta \rho}{\rho_{b}} (t_m)\right)_{c} = \frac{K}{H^{2}(t_m) a^{2}(t_m)} = c^{2}_{s} \frac{k^{2}}{H^{2}(t_m) a^{2}(t_m)} = c^{2}_{s} = w,
\end{equation}
which is basically the threshold value estimation obtained in \cite{hawking1}.

The second analytical estimation came years later in \cite{harada}. The authors considered a compensated ``three-zone'' model, i.e., a central over-dense region followed by an under-dense layer (which compensates the over-density), and finally, the FLRW background, to estimate the threshold for BH formation. Following an argument about the sound waves at the maximum expansion of the perturbation, it was found that the threshold {of the over-density (at $r=0$)} when it renters the horizon must be $ (3(1+w)/(5+3w)) \cdot \sin^{2}(\pi \sqrt{w}/(1+3w))$.

Later on, however, it became clear that the threshold depends on the shape of the curvature perturbation \cite{Niemeyer2,musco2009,Nakama_2014}. {Moreover, the threshold also depends on the specific equation of state of the fluid since pressure gradients play a role in its determination: the larger is $w$, the stronger are the pressure gradients, and therefore, the higher is the threshold.}

In \cite{Shibata:1999zs}, a new criterion for PBH formation was introduced. Importantly, it was found that the peak of the compaction function (the average mass excess on a given volume) was a good criterion for PBH formation, and it currently has become the standard definition of the threshold, which we call $\delta_{c}$. Prior to this criterion, some other measures for the threshold ``definition'' were not empty of some issues. See \cite{musco2018} for a historical perspective about that.

In \cite{musco2018}, simulations were performed for a radiation fluid with a set of curvatures profiles leading to a range of threshold values given by $0.41 \lesssim \delta_{c} \leq 2/3$ using the criterion of the compaction function's peak as the threshold definition. Later on, in \cite{universal1}, the minimum threshold value was refined, obtaining $0.4$ rather than $0.41$ in the case of a radiation fluid. This difference was due to the choice {of unphysical profiles in some extreme limit in \cite{musco2018}}.

Although $\delta_c$ is a profile-dependent quantity, in \cite{Escriva:2020tak,universal1}, it was shown that there exists an (approximately) universal value given by the averaged critical compaction function, which allows building a sufficiently accurate analytical expression for $\delta_{c}$, which only depends on the type of fluid and the curvature around the peak of the compaction function.

On the other hand, if a perturbation is sufficiently strong, it will collapse and form an apparent horizon. Later on, it will follow a process of accretion of energy density from the surrounded FLRW background until reaching a stationary state, when the final mass of the PBH is achieved. As was shown in \cite{Niemeyer2,Niemeyer1}, the final PBH mass follows a critical scaling law when the amplitude of the perturbation is close to its threshold value (the critical regime),
\begin{equation}
M_{BH} = \mathcal{K} M_{H} (\delta-\delta_{c})^{\gamma},
\end{equation}
with values of $\gamma$ consistent with the previous numerical computation \cite{Gundlach:1999cu,Neilsen:1998qc} $\gamma \approx 0.356$. The scaling exponent is universal and only depends on the type of fluid, not on the initial condition. $M_{H}$ is the mass of the cosmological horizon at the time of horizon crossing (when the length scale of the perturbations equals the cosmological horizon). The constant $\mathcal{K}$ depends on each initial condition used. As in the case of the threshold, the values of $\mathcal{K}$ are different depending on what the initial conditions are, but with values $O(1)$ \cite{musco2013}.

The simulations in \cite{Niemeyer2} were performed for $\delta-\delta_{c} \gtrsim 10^{-5}$, which was not sufficient to test the critical regime up to very small values, i.e., up to machine precision $\delta-\delta_{c} \approx 10^{-15}$. In~\cite{musco2005}, the scaling law up to machine precision was finally verified, and this was tested with an explicit example of a Gaussian perturbation.

\section{The Abundance of PBHs}\label{abundances_sec}

This section gives a brief review of the two approaches commonly used in the literature to perform PBH statistics (formed in a radiation-dominated Universe) and estimate their abundances. In this review, we do not focus on the statistical estimation of PBH abundances, for which there are extensive works in this direction, with different approaches and methods. If the reader is interested, we suggest checking these references and the ones therein \cite{nonlinear,DeLuca:2019qsy,newnicolla,Erfani:2021rmw,Wu:2020ilx,DeLuca:2020ioi,Young:2020xmk,Yoo:2020dkz,yoo,Gow:2020bzo,Young:2020xmk,Young:2019osy,Young:2019yug,Young:2013oia,Young:2014ana,Germani:2018jgr,Garriga,Suyama:2019npc,Ando:2018qdb,Zaballa:2006kh,Yokoyama:1998xd,Tada:2021zzj}. However, to show this topic at the basic level, we basically describe the Press--Schechter \cite{Press:1973iz} and the peak theory \cite{peak_theory} procedures. This section aims to clearly point out the relation between the threshold and PBH mass on the estimation of PBH abundances, especially why the threshold should be known with enough accuracy to be useful for statistics.

In both cases, we consider the amplitude of the peak of density contrast in the comoving slicing {at super-horizon scales} (usually denoted in the literature as $\Delta \equiv \delta \rho$ \mbox{$(r=0,t)$}$ /\rho_{b}(t)$~\cite{Green:2004wb,Germani:2018jgr}), as a statistically distributed variable. The power spectrum associated with the density contrast is defined as,

\begin{equation}
 P_{\Delta}(k,t)\delta_{D}(k,k') =\frac{1}{(2 \pi)^{3}} <\Delta(k,t) \Delta(k',t)>,
\end{equation}
where $\delta_{D}$ is the Dirac delta and the moments $j$ of $P_{\Delta}$ are given by,
\begin{equation}
\sigma_{j}^{2}(t) = \int \frac{k^{2} dk}{2 \pi^{2}} P_{\Delta}(k,t) k^{2j}.
\end{equation}

In the Press--Schechter formalism, we make two assumptions: (i) the density contrast field $\Delta$ is a Gaussian variable; (ii) perturbations with $\Delta> \Delta_{c}$ will collapse and form a PBH. Therefore, basically, we integrate the probability distribution $P(\Delta)$ over the range $\Delta_{c} \leq \Delta < \Delta_{\rm max}$, where $\Delta_{\rm max}$ is the maximum allowed value. In practice, we integrate up to $\Delta_{\rm max} \rightarrow \infty$ since the probability distribution is a rapidly decreasing function above $\Delta_{c}$, and therefore does not change the result and allows simplifying the computation. It is important to notice that $\Delta_{c} \neq \delta_{c}$, since we are comparing the critical density contrast with the {peak of the critical compaction function (as we will see later, it is related to the averaged density contrast).}

Finally, consider a Gaussian distribution probability distribution,

\begin{equation}
P(\Delta) = \frac{1}{\sqrt{2 \pi} \sigma_{0}}e^{-\frac{\Delta^{2}}{2 \sigma_{0}^{2}}};
\end{equation}
the abundance of PBHs can be computed as,
\begin{equation}
\begin{split}
\beta & = \frac{\rho_{PBH}}{\rho_{\rm tot}} = 2 \int_{\Delta_{c}}^{\infty} \frac{M_{PBH}}{M_{H}} P(\Delta) d \Delta = 2 \int_{\Delta_{c}}^{\infty} \mathcal{K} (\Delta-\Delta_{c})^{\gamma} P(\Delta) d \Delta \\
& = \frac{1}{\sqrt{\pi}}2^{-\frac{1+\gamma}{2}} \mathcal{K} \Delta_{c} \sigma^{-1+\gamma}_{0} \Gamma(1+\gamma) \Gamma_{U} \left(1+\frac{\gamma}{2},\frac{3}{2},\frac{\Delta^{2}_{c}}{2 \sigma^{2}_{0}}\right) e^{-\frac{\Delta^{2}_{c}}{2 \sigma_{0}^{2}}},
\end{split}
\label{eq:abundance}
\end{equation}
where $\Gamma_{U}(a,b,c)$ is the confluent hypergeometric function, \footnote{The confluent hypergeometric function is defined as $\Gamma_{U}(a,b,z) = \frac{1}{\Gamma(a)}\int_{0}^{\infty}e^{-zt}t^{a-1}(1+t)^{b-a-1}dt$.} {and the factor $2$ at the beginning of the integral put by hand is introduced to avoid the under-counting that happens in Press--Schechter theory.} It has been assumed that the scaling law is accurate always even when $\Delta \gg \Delta_{c}$, something not true, as we will show in Section \ref{sec:results}. However, since the probability distribution is exponentially smaller for large $\Delta$, Equation~\eqref{eq:abundance} is accurate. In the case of considering a monochromatic mass spectrum, i.e., all the PBHs formed to have the same mass (in particular, the horizon mass $M_{H}$), the abundance is given by \footnote{The $\erf$ function is defined as $\erf(x) = \frac{2}{ \sqrt{\pi}}\int_{0}^{z} e^{-t^{2}}dt$.},
\begin{equation}
 \beta = \bar{\mathcal{K}} \erf\left(\frac{\Delta_{c}}{\sqrt{2}\sigma_{0}}\right).
 \label{eq:psekeres}
\end{equation}
{This} %Is the noindent format necessary?
 can be obtained by simply taking the limit $\gamma \rightarrow 0$ and $\mathcal{K}=\bar{\mathcal{K}}$ in Equation~\eqref{eq:abundance}.

The procedure in the peak theory approach is a bit different. It makes statistics on counting the numbers of over-threshold peaks on the over-density; another approach used the curvature fluctuation $\zeta$ peaks instead \cite{Yoo:2018kvb,Yoo:2020dkz}.

We summarise the procedure performed in \cite{Germani:2018jgr} to estimate the abundances through\linebreak peak~theory.

First of all, we define the variable $\nu = \Delta/\sigma_{0}$, for which in the rare peak assumption, we have that $\nu \gg 1$. The number of rare peaks is given by \cite{peak_theory},
\begin{equation}
\mathcal{N}(\nu) = \frac{1}{a(t_{f})^{3}}\frac{1}{4 \pi^{2}} \left(\frac{\sigma_{1}}{3 \sqrt{\sigma_{0}}}\right)^{3} \nu^{3} e^{-\nu^{2}/2} \theta(\nu-\nu_{c}),
\end{equation}
where $t_{f}$ is the time when the PBHs are formed. It is clear that we consider only that peaks higher than the threshold value $\nu_{c}$ will contribute to the formation of PBHs. Therefore, the abundance of PBHs measured at formation with respect to the background energy-density can be computed as,
\begin{equation}
\beta = \int_{\nu_{c}}^{\infty} \frac{\rho_{PBH}(\nu)}{\rho_{b}(t_{f})}d\nu = \int_{\nu_{c}}^{\infty} \frac{M_{PBH}(\nu) \mathcal{N}(\nu)}{\rho_{b}(t_{f})}d\nu.
\end{equation}
{The} %Is the noindent format necessary?
 scaling law mass, in terms of $\nu$, is given by,
\begin{equation}
M_{PBH} = \mathcal{K} M_{H} \left(\frac{\tilde{\sigma}_{0}}{a_{m}^{2}H_{m}^{2}}\right)^{\gamma}(\nu-\nu_{c})^{\gamma},
\end{equation}
where $\tilde{\sigma}_{0}= \sigma_{0}a^{2}H^{2}$ and $t_m$ is the time of horizon crossing. {Taking into account that numerical simulations have shown $a_{f} \approx a_{m} \approx 3$ \cite{musco2018}, $a_{f}$ is weakly dependent on $\nu$, as pointed out in \cite{Germani:2018jgr}, and using the saddle point approximation \cite{Carr} $\nu_{s} \approx \nu_{c}+\gamma/\nu_{c}$, finally, we obtain},
\begin{equation}
\beta \approx \sqrt{\frac{2}{\pi}} \mathcal{K} \left(\frac{k_{*}}{a_{m} H_{m}}\right)^{3} \left(\frac{\tilde{\sigma}_{0}}{ a^{2}_{m} H^{2}_{m} }\right)^{\gamma} \nu_{c}^{1-\gamma} \gamma^{\gamma+1/2} e^{-\nu_{c}^{2}/2},
\label{11_abundance}
\end{equation}
where $k_{*}=\sigma_{1}/(\sqrt{3} \sigma_{0})$. From Equations \eqref{eq:abundance}, \eqref{eq:psekeres}, and \eqref{11_abundance}, we clearly see the exponential dependence on the threshold for PBH formation and the linear dependence on the constant $\mathcal{K}$ associated with the scaling law. This is the reason why an accurate determination of the threshold is essential, since the abundances depend exponentially on it. These two methods are however only an approximation of the true statistics, which consider the fact that each statistical realisation of the profiles has a different threshold. This was developed in \cite{nonlinear}, so we will not discuss it here as it is beyond the scope of this review paper. However, as can be seen in \cite{nonlinear}, one finds again the generic behaviour that $\beta$ is exponentially sensitive to the threshold. {We also mention that, in principle, a window function has to be used to estimate PBH abundances correctly. It is necessary to relate the power spectrum in Fourier space to the probability distribution function in real space by coarse-graining the cosmological perturbations. The necessity of considering a window function is remarkably important in the case of having a broad power spectrum (where several wavelength scales $k$ are involved). Despite that, the choice of the window function is not unique, and the result depends on the choice used. We refer the reader to \cite{Young:2019osy,Ando:2018qdb,Tokeshi:2020tjq} to check the details.}

\section{Cosmological Setup for PBH Formation}\label{sec:set_up}

This section reviews the needed ingredients to simulate the formation of PBHs in an FLRW Universe under spherical symmetry and filled by a perfect fluid. We will see what differential equations describe the gravitational collapse, the consistent initial conditions for PBH formation we need to use, and the boundary conditions to supply. Later on, we will study the formation of the apparent horizon and characterise it in terms of the expansion of the congruences. Finally, we will describe the process of the accretion of the PBH mass from the FLRW background {and the estimation of the final mass.}

\subsection{Misner--Sharp Equations}\label{sec:2_sharp}
The Misner--Sharp equations \cite{misnersharp} describe the motion of a relativistic fluid with spherical symmetry. This corresponds to the Einstein field equations written in the comoving gauge (the gauge we use in this review paper). To obtain them, first of all, we need to consider a perfect fluid with the energy-momentum tensor,
\begin{equation}
T^{\mu \nu} = (p+\rho)u^{\mu}u^{\nu}+pg^{\mu\nu},
 \label{eq:tensor_energy}
\end{equation}
and with the following spacetime metric in spherical symmetry:
\begin{equation}
\label{2_metricsharp}
ds^2 = -A(r,t)^2 dt^2+B(r,t)^2 dr^2 + R(r,t)^2 d\Omega^2,
\end{equation}
where $R(r,t)$ is the areal radius, $A(r,t)$ is the lapse function, and $d\Omega^{2} = d\theta^2+\sin^2(\theta) d\phi^2$ is the line element of a two-sphere. The components of the four-velocity $u^{\mu}$ are given by $u^{t}=1/A$ and $u^{i}=0$ for $i=r,\theta,\phi$, since we are considering comoving coordinates (comoving gauge). Throughout the review paper, we use units $G_{N}=c=1$.

Solving the Einstein field equations for this problem, the following quantities appear:
\begin{align}
\frac{1}{A(r,t)}\frac{\partial R(r,t)}{\partial t} &\equiv D_{t}R \equiv U(r,t),\nonumber \\
\frac{1}{B(r,t)}\frac{\partial R(r,t)}{\partial r} &\equiv D_{r}R \equiv \Gamma(r,t),
\label{2_covariantR}
\end{align}
where $D_{t}$ and $D_{r}$ are the proper time and distance derivatives, respectively. $U$ is the radial component of the four-velocity associated with an Eulerian frame (so not comoving), which measures the radial velocity of the fluid with respect to the origin of the coordinates. The Misner--Sharp mass $M(r,t)$ is introduced as:
\begin{equation}
M(R) \equiv \int_{0}^{R} 4\pi \tilde{R}^{2} \rho \, d\tilde{R}\, ,
\end{equation}
which is related with $\Gamma$, $U$ and $R$ though the constraint:
\begin{equation}
\Gamma = \sqrt{1+U^2-\frac{2 M}{R}},
\end{equation}
where $\Gamma$ is called the generalised Lorentz factor, {which includes the gravitational potential energy and kinetic energy per unit mass. This also can be seen in the Newtonian limit,}
\begin{equation}
\Gamma \sim 1+\frac{1}{2}\frac{v^{2}}{c^{2}}-\frac{GM}{c^{2}R}.
\end{equation} 
{Finally}%Is the noindent format necessary? please check all in this manuscript.
, the differential equations governing the evolution of a spherically symmetric collapse of a perfect fluid in general relativity are:
\begin{align}
D_{t}U &= -\left[\frac{\Gamma}{(\rho+p)}D_{r}p+\frac{M}{R^{2}}+4\pi R p \right], \\
D_{t} R &= U, \\
D_{t}\rho &= -\frac{(\rho+p)}{\Gamma R^{2}}D_{r} (U R^{2}), \\
D_{t}M &= -4 \pi R^{2} U p, \\
\label{2_eqconstraint}
D_{r}M &= 4\pi \Gamma \rho R^{2}, \\
\label{2_lapse}
D_{r} A &= \frac{-A}{\rho+p}D_{r}p\, .
\end{align}
We refer the reader to \cite{misnersharp} to see the details of the derivation. The boundary conditions in the Misner--Sharp equation are $R(r=0,t)=0$, leading to $U(r=0,t)=0$ and $M(r=0,t)=0$. Then, by spherical symmetry, we have $D_{r} p (r=0,t)=0$, and also, we consider a perfect fluid $p=w \rho$.

At $r \rightarrow \infty$, we should recover the solution of the FLRW background. However, in a numerical simulation, we have to handle the discretisation of time and space. Therefore, the condition $D_{r} p (r=r_f,t) = 0$ is used (where $r_{f}$ is the outer point of the grid) to match with the FLRW solution and avoid reflections from density waves. On the other hand, Equation~\eqref{2_eqconstraint} is called the Hamiltonian constraint, which is commonly used to check the correct solution of the numerical equations. For the case $p=w \rho$, the lapse equation in Equation~\eqref{2_lapse} can be solved considering $A(r_f,t) = 1$ to match with the FLRW spacetime,
\begin{equation}
A(r,t) = \left(\frac{\rho_{b}(t)}{\rho(r,t)}\right)^{\frac{w}{w+1}},
\end{equation}
where $\rho_{b}(t) = \rho_{0}(t_{0}/t)^{2}$ is the energy density of the FLRW background and $\rho_{0}=3 H_{0}^{2}/8\pi$.

Actually, the Misner--Sharp equations can be written in a more advantageous way to perform numerical simulations using the definitions of Equation~\eqref{2_covariantR},
\begin{align}
\label{eq:u_simply}
\dot{U} &= -A\left[\frac{w}{1+w}\frac{\Gamma^2}{\rho}\frac{\rho'}{R'} + \frac{M}{R^{2}}+4\pi R w \rho \right], \\
\label{eq:r_simply}
\dot{R} &= A U, \\
\label{eq:rho_simply}
\dot{\rho} &= -A \rho (1+w) \left(2\frac{U}{R}+\frac{U'}{R'}\right), \\
\label{eq:m_simply}
\dot{M} &= -4\pi A w \rho U R^{2}, \\
M' &= 4 \pi \rho R^{2} R',
\end{align}
where the dot $\dot{U}$ represents the time derivate $\partial U /\partial t$ and $\rho'$ the radial derivative $\partial \rho /\partial r$.

\subsection{The Gradient Expansion Approximation, Initial Conditions, and Compaction Function}

The gradient expansion method \cite{PhysRevD.42.3936} (also called long-wavelength approximation) was used in \cite{musco2007} to set up consistent initial conditions for PBH formation on an FLRW background at super-horizon scales. First of all, let us consider a cosmological perturbation at super-horizon scales, i.e., with a length scale $R_{m}$ much larger than the Hubble horizon. We can define a parameter $\epsilon$ to relate the two scales: the Hubble horizon $R_{H}$ and the length scale of the perturbation $R_{m}$,
\begin{equation}
\epsilon = \frac{R_{H}(t)}{R_{m}(t)},
\end{equation}
where $R_{H}=1/H$. It is clear that at super-horizon scales, we will have $\epsilon \ll 1$. The gradient expansion method allows expanding these inhomogeneities in the spatial gradient in terms of $\epsilon$. In the limit $\epsilon \rightarrow 0$, the spacetime locally corresponds to the FLRW metric, when the perturbation is smoothed out at sufficiently large scales $R_{m}$.

Consider a general spacetime metric in the $3+1$ Arnowitt--Deser--Misner (ADM) formalism \cite{Arnowitt:1962hi,PhysRev.116.1322},
\begin{equation}
ds^{2} = -\alpha^{2} dt^{2}+\gamma_{ij}(dx^{i}+\beta^{i} dt)(dx^{j}+\beta^{j} dt);
\label{ADM_metric}
\end{equation}
in general, the spatial part of the metric can be decomposed into:
\begin{equation}
\gamma_{ij} = a^{2}(t) e^{2 \zeta(t,x^{i})} \tilde{\gamma}_{ij}
\end{equation}
where $\gamma_{ij}$, $\beta_{i}$, and $\alpha$ are the spatial metric, shift-vector, and lapse function, respectively. $\zeta$ is usually called the curvature fluctuation. Under the gradient expansion method, it was shown in \cite{Tanaka:2007gh,Shibata:1999zs,Lyth:2004gb,Sugiyama:2012tj} that $\beta = O(\epsilon)$, $\alpha = 1+O(\epsilon^{2})$, and $\tilde{\gamma}_{ij} = \delta_{ij}+O(\epsilon^{2})$. Therefore, in spherical symmetry and in the limit $\epsilon \rightarrow 0$, the metric of Equation~\eqref{ADM_metric} can be written as,
\begin{equation}
ds^{2} =-dt^{2}+a^{2}(t) e^{2 \zeta(\tilde{r})}(d\tilde{r}^{2}+\tilde{r}^{2} d \Omega^{2}).
\label{metrica_222}
\end{equation}
\textls[-10]{Notice the time independence of $\zeta(\tilde{r})$ since at super-horizon scales, it is shown that \mbox{$\dot{\zeta}=O(\epsilon^{2})$}. It is important to point out that although, we consider the comoving gauge, other gauges are possible. In particular, in \cite{Shibata:1999zs}, the constant mean curvature gauge was used instead. However, as was shown in \cite{Lyth:2004gb,refrencia-extra-jaume}, the differences between the two gauges in $\zeta$ are of $O(\epsilon^{2})$.}

Finally, the metric Equation~\eqref{metrica_222} can be also written in other coordinates as an FLRW metric with a non-constant curvature $K(r)$,
\begin{equation}
\label{2_FLRWmetric5}
ds^2 = -dt^2 + a^2(t) \left[\frac{dr^2}{1-K(r) r^2}+r^2 d\Omega^2 \right].
\end{equation}
The change of coordinates $r, \tilde{r}$ between the two metrics and the conversion between $K(r), \zeta(\tilde{r})$ are given by the following relations, which were obtained in several works previously (not necessarily in the context of PBH formation for all of them) \cite{musco2018,enea-sasaki-alexei,Hidalgo:2008mv,Garriga}:
\begin{align}
\label{transforms_tilde_K}
r &= \tilde{r} e^{\zeta(\tilde{r})}, \\ \nonumber
\frac{dr}{d \tilde{r}} &= e^{\zeta(\tilde{r})} \left[1+\tilde{r}\zeta'(\tilde{r}) \right], \\ \nonumber
K(r) r^{2} &= -\tilde{r} \zeta'(\tilde{r}) \left[2+\tilde{r}\zeta'(\tilde{r}) \right], \\ \nonumber
 \zeta(\tilde{r}) &= \int_{\infty}^{r} \left(1-\frac{1}{\sqrt{1-K(\hat{r}) \hat{r}^{2}}}\right) \frac{d\hat{r}}{\hat{r}}, \\ \nonumber
 \tilde{r} &= r \, exp\left[\int_{\infty}^{r} \left(\frac{1}{\sqrt{1-K(\hat{r})\hat{r}^{2}}}-1\right) \frac{d\hat{r}}{\hat{r}}\right]
\end{align}
Therefore, the cosmological perturbation will be characterised by the curvature perturbation $K(r), \zeta(\tilde{r})$. The Misner--Sharp equations can be solved at leading order in $\epsilon \ll 1$ using the long wavelength approximation as was performed in \cite{musco2007},
\begin{align}
\label{2_expansion}
A(r,t) &= 1+\epsilon^2(t) \tilde{A},\nonumber\\
R(r,t) &= a(t)r(1+\epsilon^2(t) \tilde{R}),\nonumber\\ 
U(r,t) &= H(t) R(r,t) (1+\epsilon^2(t) \tilde{U} ),\\ 
\rho(r,t) &= \rho_{b}(t)(1+\epsilon^2(t)\tilde{\rho}),\nonumber\\ 
M(r,t) &= \frac{4\pi}{3}\rho_{b}(t) R(r,t)^3 (1+\epsilon^2(t) \tilde{M} ),\nonumber 
\end{align}
where for $\epsilon \rightarrow 0$, we recover the (FLRW) solution. The perturbations of the tilde variables at first order in gradient expansion were computed in \cite{musco2007}, which we summarise here:
\begin{align}
\label{eq:2_perturbations}
\tilde{\rho}&= \frac{3(1+w)}{5+3\omega}\left[K(r)+\frac{r}{3}K'(r)\right] r^2_{m},\nonumber \\
\tilde{U} &= -\frac{1}{5+3\omega}K(r) r^2_{m},\nonumber\\
\tilde{A} &= -\frac{w}{1+w} \tilde{\rho},\\
\tilde{M} &= -3(1+w) \tilde{U},\nonumber\\
\tilde{R} &= -\frac{w}{(1+3\omega)(1+w)}\tilde{\rho}+\frac{1}{1+3\omega}\tilde{U}.\nonumber 
\end{align}
The perturbations in the $\tilde{r}$ coordinate with $\zeta(\tilde{r})$ are \cite{musco2018},
\begin{align}
\label{eq:tilde_perturb}
\tilde{\rho} &= -\frac{2(1+w)}{5+3w}\frac{\exp{ \left[2 \zeta(\tilde{r}_{m})\right]}}{\exp{\left[2 \zeta(\tilde{r})\right]}}\left[\zeta''(\tilde{r})+\zeta'(\tilde{r})\left(\frac{2}{\tilde{r}}+\frac{\zeta'(\tilde{r})}{2}\right)\right]\tilde{r}^{2}_{m}, \\ \nonumber
\tilde{U} &=\frac{1}{5+3w}\frac{\exp{\left[2 \zeta(\tilde{r}_{m})\right]}}{\exp{\left[2 \zeta(\tilde{r})\right]}}\zeta'(\tilde{r})\left[\frac{2}{\tilde{r}}+\zeta'(\tilde{r})\right] \tilde{r}^{2}_{m} \nonumber.
\end{align}

From Equations~\eqref{eq:2_perturbations} and \eqref{eq:tilde_perturb}, an expression for the density contrast at super-horizon scales in terms of the curvature fluctuations can be obtained \cite{musco2018}, given by:
\begin{align}
\label{eq:tilde_perturb_contrast}
\frac{\delta \rho}{\rho_{b}}(r) &= \frac{3(1+w)}{5+3w} \left(\frac{1}{aH}\right)^{2}\left[K(r)+\frac{r}{3}K'(r)\right], \\ \nonumber
\frac{\delta \rho}{\rho_{b}}(\tilde{r}) &= -\frac{3(1+w)}{5+3w} \left(\frac{1}{aH}\right)^{2}\exp{\left[-2 \zeta(\tilde{r})\right]} \left[\zeta''(\tilde{r})+\zeta'(\tilde{r})\left(\frac{2}{\tilde{r}}+\frac{\zeta'(\tilde{r})}{2}\right)\right]. \nonumber 
\end{align}

The areal radius of the length scale $R_m$ is given by $R_m(t)=a(t) r_m$ or also $R_m(t)=a(t)\tilde{r}_{m}e^{\zeta(\tilde{r}_{m})}$ with the $\tilde{r}$ coordinate taking into account Equation~\eqref{transforms_tilde_K}. As was shown in \cite{Polnarev:2012bi}, where the gradient expansion method beyond first order was applied using an iterative scheme (see \cite{Polnarev:2012bi} for more terms in the $\epsilon$ expansion in comparison with Equation~\eqref{2_expansion}), the long-wavelength approximation at first order is accurate enough if a sufficiently small $\epsilon$ parameter is taken.

When we take $\epsilon=0$, we recover the background solution equations: $H(t)=H_{0} t_{0}/t$, $a(t)= a_{0}(t/t_{0})^{\alpha}$, and $R_{H}(t) = R_{H}(t_{0})(t/t_{0})$ where $a_{0} = a(t_{0})$, $H_{0}=H(t_{0}) = \alpha /t_{0}$, and $R_{H}(t_{0}) = 1/H_{0}$. Moreover, we define $\alpha = 2/3(1+w)$. We also consider as the initial condition that $r_m = \Lambda R_{H}(t_{0})$, where $\Lambda$ is some number (usually taken as $O(10)$ in simulations). A time scale given when $\epsilon(t_{m}) =1$ is commonly defined in the literature, which precisely corresponds to the time when the Hubble horizon equals the length scale of the perturbation (the time of horizon crossing) $t_m = t_{0}(a_{0} r_m/R_H(t_0))^{1/(1-\alpha)}$. {Notice that $t_{0}$ is the initial time, and it is assumed to be a time where the perturbation is at super-horizon scales.}

On the other hand, the amplitude of a cosmological perturbation can be measured as the averaged density contrast within a spherical region:
\begin{equation}
\label{2_massexcess}
\bar{\delta}(R) = \frac{1}{V}\int_{0}^{R} 4\pi \frac{\delta \rho}{\rho_{b}} \hat{R}^{2}\, d\hat{R},
\end{equation}
where $\delta \rho = \rho-\rho_{b}$ and $V=4\pi R^{3}/3$.

Related to that, in \cite{Shibata:1999zs}, the compaction function $\com(r)$ was defined, which gives a measure of the mass excess on a given volume. In particular,
\begin{equation}
\com(r,t) = \frac{2 \left[M(r,t)-M_{b}(r,t)\right]}{R(r,t)},
\label{2_compactionfunction}
\end{equation}
where $M_{b}(r,t)$ is the mass of the FLRW background in the volume $V$. The compaction function can also be written in terms of $\bar{\delta}$, as was shown in \cite{refrencia-extra-jaume},
\begin{equation}
\com(R) = \bar{\delta} (RH)^{2}. 
\end{equation}
At leading order in $\epsilon$,
\begin{equation}
\bar{\delta}(r,t)=\left(\frac{1}{a H r_{m}}\right)^{2} \delta(r),
\end{equation}
where:
\begin{align}
\label{eq:2_delta_perturbations}
\delta(r) &= f(w)K(r)r^{2}_{m}, \\
\label{eq:2_fw}
f(w) &= 3(1+w)/(5+3w).
\end{align}
From the above definitions, we finally define the compaction function at super-horizon scales at leading order of gradient expansion as:
\begin{equation}
\com(r,t) \simeq \com(r)=f(w) K(r) r^{2} = \frac{r^2 }{r^2_{m}}\delta(r),
\label{C_r}
\end{equation}
which yields $\com(r_{m}) = \delta(r_{m})=\delta_{m}$, i.e., $\delta_{m}=f(w) K(r_m)r^{2}_m$. {The previous equation can be obtained introducing the initial condition of Equation~\eqref{2_expansion} into Equation~\eqref{2_compactionfunction} and making an expansion in $\epsilon$. Notice in Equation~\eqref{C_r} the time-independence of the compaction function at super-horizon scales at leading order in gradient expansion, since the curvature fluctuations remain frozen at super-horizon scales, as we have mentioned before}. The compaction function at super-horizon scales is an essential magnitude, since it allows {defining the threshold}. In particular, we now define the location of the maximum of $\com(r)$ as $r_{m}$, and its value $\com_{\rm max}=\com(r_{\rm m})$ is used as a criterion for PBH formation, as was proposed in \cite{Shibata:1999zs}, while $r_m$ ($\tilde{r}_m$ in the case of using the $\tilde{r}$ coordinate) is considered the length scale of the perturbation. The threshold for PBH formation will correspond to the peak value of the critical compaction function, i.e., $\com_{c}(r_{m})= \delta_{c}$. Therefore, we define the ``amplitude'' of the perturbation as $\delta_{m} = \com(r_m)$. Perturbations with $\delta_{m}>\delta_{c}$ will collapse and form a PBH. In the opposite case, perturbations with $\delta_{m} < \delta_{c}$ will disperse on the FLRW background with no black hole formation. Notice that from Equation~\eqref{2_FLRWmetric5}, we have a bound on the amplitude value $\delta_m$ at super-horizon scales given by $\delta_{\rm m,max}=f(w)$, since $K_{\rm max}(r_m)r^{2}_m=1$.

Because of the previous definitions, the value of $r_{m}$ is given by the solution of $\com'(r)=0$:
\begin{equation}
\label{2_cmax}
K(r_{m})+\frac{r_{m}}{2}K'(r_{m}) = 0.
\end{equation}

The definition of the compaction function at super-horizon scales using the $\zeta(\tilde{r})$ perturbation instead of $K(r)$ leads to a slightly different definition,

\begin{equation}
\com(\tilde{r}) = f(w) \left[1-\left(1+\tilde{r} \zeta'(\tilde{r})\right)^{2} \right],
\label{eq:c_zeta}
\end{equation}
which yields the following condition for the peak $\tilde{r}_m$:
\begin{equation}
\label{eq:tilde_max}
\zeta'(\tilde{r}_m)+\tilde{r}_m \zeta''(\tilde{r}_{m})=0,
\end{equation}
and therefore, $\delta_m=-f(w) \left[2+\tilde{r}_{m}\zeta'(\tilde{r}_m)\right]\tilde{r}_m \zeta'(\tilde{r}_m)$. Interestingly, in \cite{musco2018}, $\bar{\delta}(r_m,t)= 3 \delta \rho(r_m,t)/\rho_{b}$ was found at super-horizon scales, which relates the averaged mass excess with the local density contrast at $r_m$.

In this review paper, we focus on type I PBHs, which are those that fulfil that the areal radius $R$ is a monotonic function, i.e., $R'>0$. Instead, type II PBHs satisfy $R'<0$ \cite{Kopp:2010sh}. Type II PBHs are still unexplored in the literature from a numerical point of view or even with a general analytical treatment.

\subsection{Horizon Formation}\label{sec:horizons}

If an initial perturbation at super-horizon scales has an amplitude $\delta_m$ bigger than its threshold $\delta_{c}$, the perturbation will continue growing and, at some point, a trapped surface will be formed. This indicates the onset of gravitational collapse. 

To identify when trapped surfaces are formed, we have to consider the expansion $\Theta^{\pm}$ of null geodesics' congruences $k^{\pm}$, orthogonal to a spherical surface $\Sigma$. The expansion $\Theta^{\pm}$ is defined as $\Theta^{\pm} \equiv h^{\mu \nu} \nabla_{\mu}k_{\nu}^{\pm}$, where $h^{\mu \nu}$ is the spacetime metric induced on $\Sigma$. There are two congruences: we call them inwards $k_{\mu}^{+}$ and outwards $k_{\mu}^{-}$, whose components are $k_{\mu}^{\pm}=(A,\pm B, 0,0)$ with $k^{+} \cdot k^{-} =-2$.

In the case of flat spacetime, $\Theta^{-}<0$ and $\Theta^{+}>0$, and these surfaces $\Sigma$ are called normal surfaces. On the other hand, if $\Theta^{\pm}<0$, the surface is called trapped, while if both are positive $\Theta^{\pm}>0$, the surface is anti-trapped. In our case, we have that,
\begin{equation}
\Theta^{\pm} = \frac{2}{R}(U \pm \Gamma).
\end{equation}

In spherical symmetry, we can consider that any point $(r,t)$ is a closed surface $\Sigma$ with a proper radius $R$. These points can be classified as normal, trapped, and anti-trapped. Specifically, the transition from a normal to a trapped surface should satisfy $\Theta^{-}<0$ and $\Theta^{+}=0$, which corresponds to a marginally trapped surface, usually called the ``apparent horizon''. Taking into account that $\Theta^{+}\Theta^{-} = \frac{4}{R^{2}}(U^{2}-\Gamma^{2})$, the condition for the apparent horizon is given by $U^{2}=\Gamma^{2} \Rightarrow 2M=R$. 

\begin{figure}[H]
\centering
\includegraphics[width=0.6\linewidth]{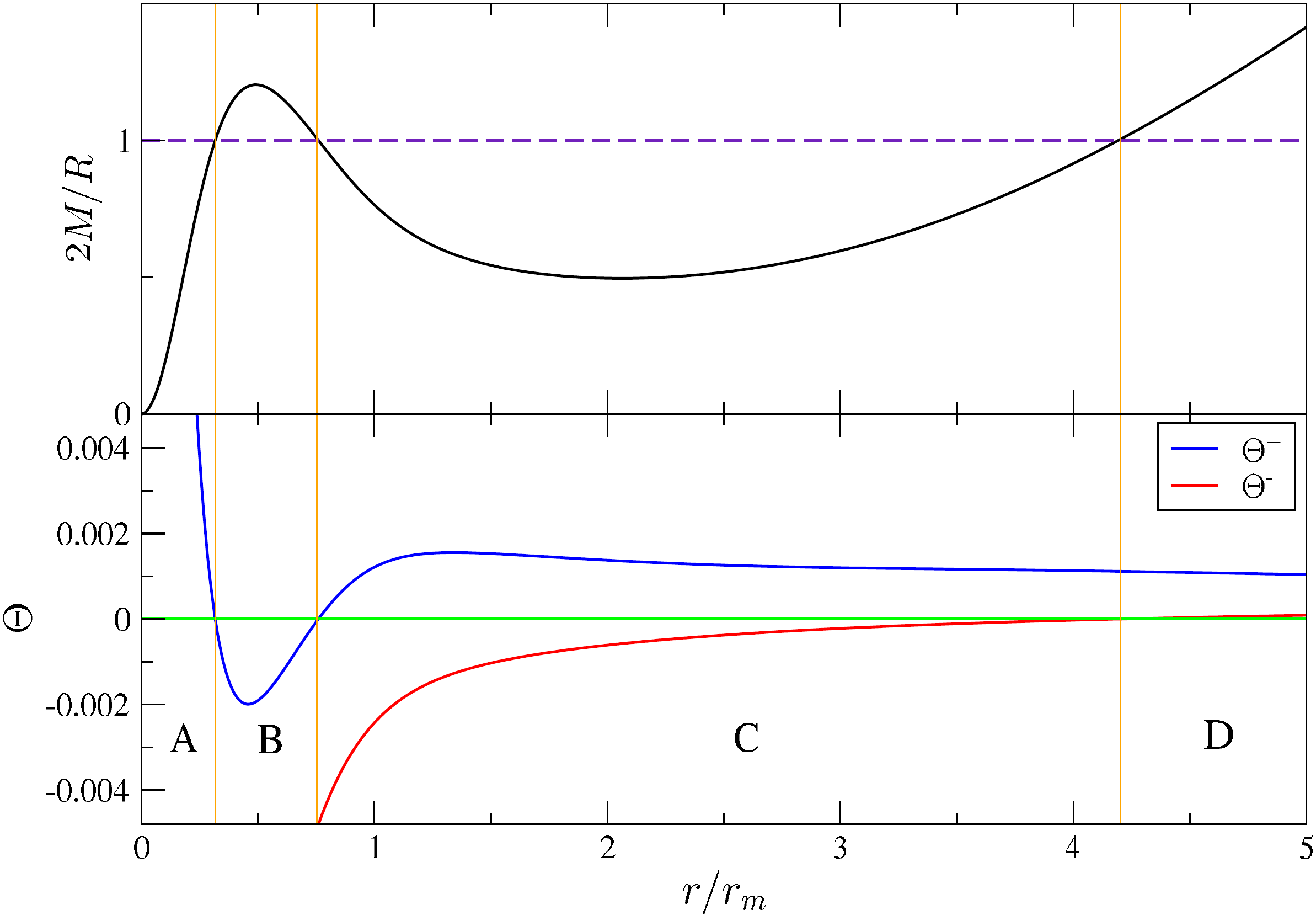} 
\caption{\textbf{{Top-panel}}: Profile of $2M/R$ once the trapped surface has been formed. The dashed line corresponds to $2M/R=1$. \textbf{Bottom-panel}: Profiles of the two expansions $\Theta^{\pm}$ at the same time $t$ as the top panel. The vertical orange lines correspond to points where $\Theta^{\pm}=0$. The initial fluctuation considered corresponds to a Gaussian profile.}
\label{fig:horizons}
\end{figure}

In Figure~\ref{fig:horizons}, {$2M/R$ and the congruences $\Theta^{\pm}$} for a supercritical (with an amplitude $\delta_m>\delta_{c}$) Gaussian curvature fluctuation can be seen, after the first horizon has been formed. The same qualitative behaviour can be applied for any perturbation of type I PBHs (with $ w \neq 0$).

Looking at Figure~\ref{fig:horizons}, we can see three points where $2M=R$, and therefore three different surfaces that we can characterise:

\begin{itemize}
\item Surface (A--B): In this region, we have a transition between $\Theta^{+}>0$ and $\Theta^{-}<0$ (normal region) to $\Theta^{+}<0$ and $\Theta^{-}<0$. This horizon, moves inwards of the computational domain, which eventually will encounter a singularity at $r=0$;

\item Surface (B--C): In this region, we have a transition between $\Theta^{+}<0$ and $\Theta^{-}<0$ to $\Theta^{+}>0$ and $\Theta^{-}<0$ (normal region). This horizon, typically called ``the apparent horizon'', moves outwards of the computational domain. Actually, this horizon and the previous one emerge from a single marginally trapped surface. During the evolution, the motion of the horizon can be followed to obtain the PBH mass $M(r_{*},t)$, where $r_{*}(t)$ is the location of the outer horizon in time. At the final stage of the gravitational collapse for very late times, this horizon will become static, i.e., $r_{*}=\text{const}$;

\item Surface (C--D): In the last region, we have a transition between $\Theta^{+}>0$ and $\Theta^{-}<0$ (normal region) to $\Theta^{+}>0$ and $\Theta^{-}>0$. This case corresponds to an anti-trapped surface, which corresponds to the cosmological horizon, moving outwards. We should mention that this horizon does not correspond to the cosmological horizon $R_H$ from the FLRW background, since the perturbed medium (deviating from the FLRW solution) affects the evolution of the cosmological horizon.
\end{itemize}

For a more formal discussion about horizons, we suggest the reader check \cite{Mello:2016irl,Helou:2016xyu,Dafermos2005,Williams:2007tp,Booth:2005ng,Ashtekar:2004cn,galaxies1030114,PhysRevD.49.6467,PhysRevD.85.084031,Yoo:2014boa}.

\subsection{PBH Mass and Accretion}

Once the apparent horizon has formed, the initial PBH mass, i.e., the mass of the PBH $M_{\rm PBH,i}$ at the moment of formation of the first apparent horizon $t_{AH}$, will start to grow to arrive at a stationary value $M_{\rm PBH,f}$. This situation is different from the case of dust collapse $w=0$, where the mass would increase forever since no pressure gradients would avoid the accretion of energy--matter. 
 
The process of accretion of energy--matter from the FLRW background has been studied in the past \cite{scalar_harada,acreation1,size1,hawking1,Custodio:1998pv} in different scenarios. To estimate the final mass of the PBH, one could follow the motion of the apparent horizon until very late times when the horizon remains static. However, since it could be computationally very expensive, we can use the Zeldovich--Novikov formula Equation~\eqref{eq:2_ZN_formula} \cite{acreation1}, which considers Bondi accretion \cite{acreation1,acreation2,acreation3}, and in our case, assume that the energy density right outside the apparent horizon decreases as in an FLRW Universe. It is essential to indicate that this is not valid at the time $t_{AH}$, since it neglects the cosmological expansion of the spacetime \cite{size3}, but we can apply from sufficiently late time for $t \gg t_{AH}$ considering an effective constant accretion efficiency rate $F$ \cite{acreation2,acreation3}, {which depends on the non-linear process of the accretion flow and the equation of state}. This approximation was already used numerically in the context of PBH formation from domain walls in \cite{Deng:2016vzb} and from curvature fluctuations in \cite{escriva_solo} with great success.

In particular, for very late times $t \gg t_{AH}$, the mass accretion follows,
\begin{equation}
\label{eq:2_ZN_formula}
\frac{dM}{dt} = 4 \pi F R^2_{\rm PBH} \rho_{b}(t)\ ,
\end{equation}
where $F$ is commonly numerically found to be of order $O(1)$. In the previous equation, the cosmological expansion is neglected, and also, it assumes a quasi-stationary flow onto a PBH. The inaccuracies of these assumptions are absorbed in $F$. We should also mention that this formula assumes that PBHs are at rest relative to the FLRW background or have negligible peculiar velocities, which is a good approximation in our case. We refer the reader to \cite{Custodio:1998pv} for a discussion about the scenario with relative velocities. 

Taking into account the condition of the apparent horizon in spherical symmetry $R_{\rm PBH}=2M_{\rm PBH}$, the previous equation is solved analytically as:
\begin{equation}
\label{eq:2_acretationformula}
M_{ PBH}(t) = \frac{1}{\frac{1}{M_{a}}+\frac{3}{2}F\left(\frac{1}{t}-\frac{1}{t_{a} }\right)}\ ,
\end{equation}
where $M_a$ is the initial mass when the asymptotic approximation is used at the time $t_a$. The value of $F$ has been found to weakly depend on the specific shape of the perturbation and its amplitude, but strongly on the equation of state $w$. Specifically, $F$ depends on the high non-linear process of accretion and the counterpart of pressure gradients. $F$ can be obtained numerically by fitting the evolution of $M_{PBH}(t)$ (at sufficiently late times after the formation of the apparent horizon, when it is valid). Using that, the PBH mass is estimated as the asymptotic mass value at $t\rightarrow \infty$, i.e.,

\begin{equation}
\label{eq:2_massfinalfinal}
M_{\rm PBH,f} = M_{ PBH}(t \rightarrow \infty) = \left(\frac{1}{M_{a}}-\frac{3 F}{2 t_{a}}\right)^{-1} \ .
\end{equation}

\begin{figure}[H]
\centering
\includegraphics[width=0.5\linewidth]{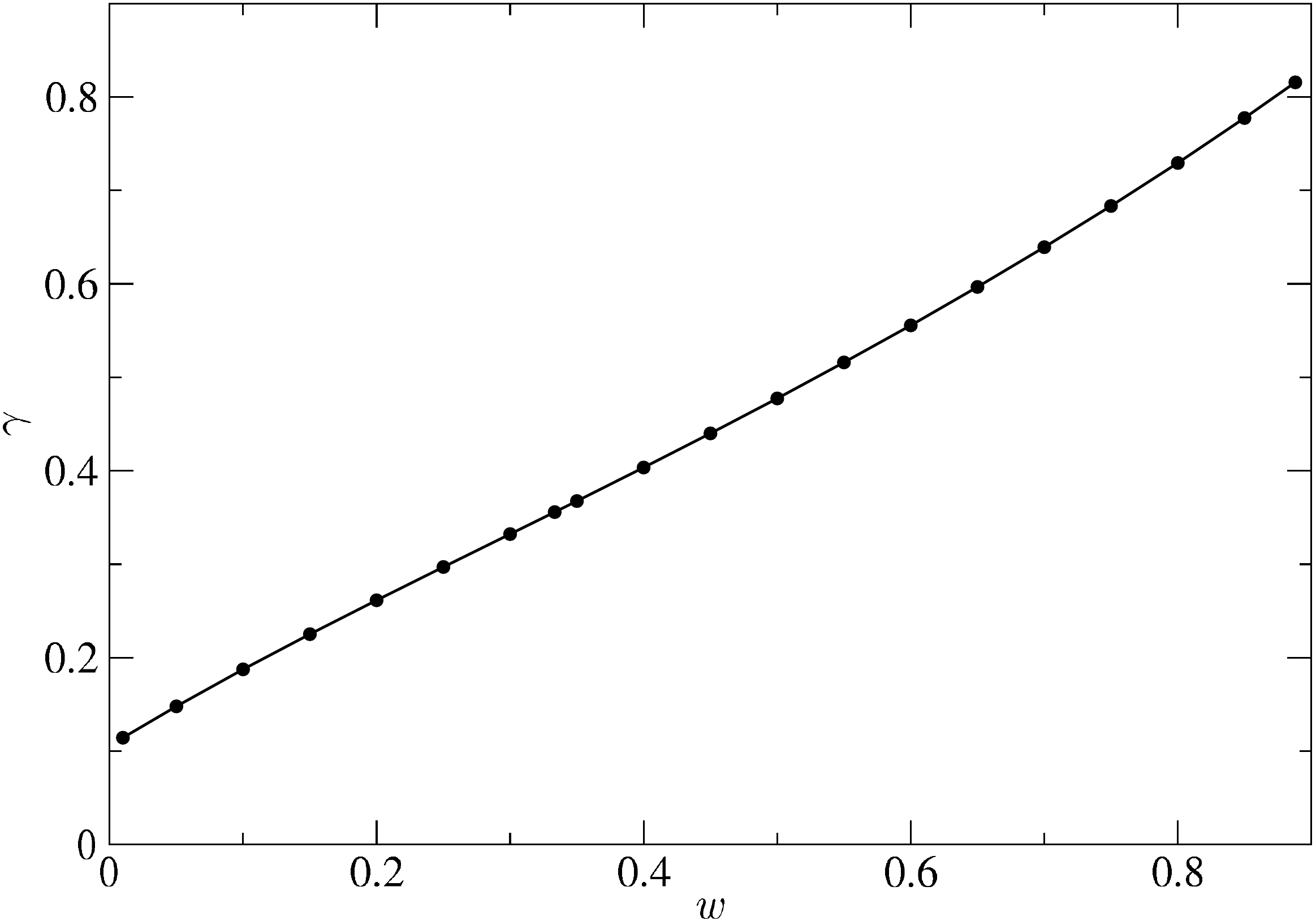} 
\caption{Values of $\gamma(w)$ plotted from the data of Table 1 in \cite{Maison:1995cc}.}
\label{fig:gammas}
\end{figure}

The final PBH mass will be dependent on the specific profile of the fluctuation, its initial amplitude $\delta_m$, and $w$. As was already shown in the past, in the critical regime where $\delta_m$ is very close to the critical value $\delta_{c}$, the black hole mass follows a scaling law~\cite{musco2009,Niemeyer2,hawke2002},
\begin{equation}
M_{\rm PBH,f} = {\cal K} M_{H} (\delta_m-\delta_{c})^{\gamma},
\label{eq:2_scaling}
\end{equation} 
where $\gamma \approx 0.356$ in radiation $w=1/3$. The values of $\gamma$ in terms of the equation of state $w$ were found semi-analytically in \cite{Maison:1995cc} and are plotted in Figure~\ref{fig:gammas}. These results were numerically confirmed in \cite{musco2013}.

In Equation~\eqref{eq:2_scaling}, the constant ${\cal K}$ depends on the specific shape of the profile considered and $M_H\equiv 1/2 H(t_m)$ is the Hubble mass, which is calculated at the time $t_{m}$. Specifically, $M_H$ in terms of the initial values of the perturbation can be written as:

\begin{equation}
M_H = \frac{3(1+w)}{4} t_0\left(a_0 \Lambda \right)^{\frac{3(1+w)}{1+3w}},
\label{eq:horizon_mass}
\end{equation} 
which will depend on the initial length scale $r_m$ and the equation of state $w$ (we can consider gauge values $a_0=1$ and $t_0=1$). In the case of using the $\tilde{r}$ coordinate, we just need to replace $\Lambda \rightarrow \Lambda e^{\zeta(\Lambda R_{H}(t_0))}$ in Equation~\eqref{eq:horizon_mass}. The scaling law starts to deviate at $(\delta_m-\delta_{c}) \gtrsim 2 \times 10^{-2}$ (for $w=1/3$), as was indicated in \cite{musco2013}. It was shown explicitly in \cite{escriva_solo} that for large $\delta_m$ beyond the critical case, the scaling law seems to deviate by order $O(10\%)$ (at least for $w=1/3$), but it can be smaller or larger depending on the profile considered. The deviation from the critical regime is expected to be also found for other $w$'s, but the quantification still is missing in the literature. As pointed out in \cite{Kuhnel:2015vtw}, it is important to consider the scaling law behaviour Equation~\eqref{eq:2_scaling} for the predictions of the PBH mass spectra.

\section{Numerical Techniques}\label{sec:numerics}

Most of the times, Partial Differential Equations (PDEs) cannot be solved analytically, and numerical methods are needed. This is especially common in general relativity and is the case we have here. Currently, several methods and tools can be employed in numerical relativity \cite{bookNR,Lehner:2001wq,random33,Gourgoulhon:2007ue,novak,Font:2000pp,Clough:2015sqa,Loffler:2011ay,Ruchlin:2017com}. In the case of PBH formation, there is the extra difficulty that the gravitational collapse happens on an FLRW background, so not on asymptotically flat~spacetime.

Different authors have employed different numerical procedures to simulate PBH formation from the collapse of curvature fluctuations under spherical symmetry, as we mentioned in Section \ref{sec:intro}. From a historical perspective, several works have used {the standard and very well-known} finite differences approach to solve Hernandez--Misner--Sharp equations in the comoving gauge \cite{Niemeyer1,Niemeyer2,musco2018,musco2007,musco2013,musco2009,Nakama_2014,Bloomfield:2015ila,Nakama:2014fra}. Alternatively, simulations were also performed at earlier times using a different gauge, in particular the mean curvature gauge \cite{Shibata:1999zs}. The numerical results of both procedures with different gauges were verified to be consistent with each other in \cite{refrencia-extra-jaume} {at leading order in gradient expansion}. A new numerical method came up recently using pseudospectral methods \cite{escriva_solo} to solve Misner--Sharp equations together with the use of an excision technique to obtain the final mass of the black hole.

In this review, we describe the pseudospectral collocation technique used in \cite{escriva_solo}. The basics of the numerical code are available here \cite{web} (currently, the only one in the literature publicly available), and therefore, the reader can reproduce or extend some of the results presented in this review.

\subsection{Pseudospectral Methods}\label{sec:2_spectral}

%Most of the times, PDEs can not be solved analytically and numerical method are needed. Actually, a numerical solution could be even better than an analytical solution, since could be very tedious to obtain an analytical solution.

Pseudospectral methods have been extensively used in computer science for solving complex physics phenomena \cite{doi:10.1142/3662,doi:10.1146/annurev.fl.19.010187.002011,Hesthaven2007SPECTRALMF}. The application to the field of numerical relativity has been also very popular to study different problems, from astrophysics to more oriented toward high energy physics \cite{PhysRevLett.116.041101,PhysRevD.82.044023,Santos-Olivan:2018smi,Santos-Olivan:2016djn,Meringolo:2020jsu,BONAZZOLA1999433,sengupta2021analysis,Moxon:2021gbv,Alcoforado:2020wnp,Alcoforado:2020wwy,Oltean:2018szc,Frauendiener:2016zwb,Santos-Olivan:2016djn,Schwabe:2020eac,Edwards:2018ccc,Musoke:2019ima}.

In what follows, we outline how to use the basics of the pseudospectral technique applied to our problem, which allows solving partially algebraically the differential equations describing the gravitational collapse. See also \cite{spectrallloyd,spectralmatlab} for much more details and mathematical foundations. In particular, we use the Chebyshev collocation method (but there are also other formulations).

Let us consider a function $f(x)$ and fit it with $N_{\rm cheb}$ Chebyshev polynomials (any other orthonormal function could be used as well). The function $f(x)$ can be composed by a sum of Chebyshev polynomials as,
\begin{equation}
f_{N_{\rm cheb}}(x) = \sum_{k=0}^{N_{\rm cheb}} c_{k} T_{k}(x),
\label{eq:cheb1}
\end{equation}
where $T_{k}(x)$ are the Chebyshev polynomials of order $k$. The coefficients $c_{k}$ with $k=0,1,\ldots,N_{\rm cheb}$ are then obtained by solving $f_{N_{\rm cheb}}(x_k)=f(x_k)$ where $x_{k}=\cos(k \pi/N_{\rm cheb})$. The points $x_{k}$, which satisfy $T'_{k}(x_k)=0$, are called Chebyshev collocation points.

Therefore,
\begin{align}
f_{N_{\rm cheb}}(x) &= \sum_{k=0}^{N_{\rm cheb}} L_{k}(x) f(x_{k}),\\
L_{k}(x) &= \frac{(-1)^{k+1}(1-x^2)T'_{N_{\rm cheb}}(x)}{\bar{c}_{k}N_{\rm cheb}^2(x-x_{k})} ,
\end{align}
where $\bar{c}_{k} =2$ if $k=0,N$ and $\bar{c}_{k}=1$ in the other cases. The functions $L_k$ are the Lagrange interpolation polynomials.

The derivate of $p$-order of the function will be given by, 
\begin{equation}
f^{(p)}_{N_{\rm cheb}}(x_{i}) = \sum_{k=0}^{N_{\rm cheb}} L^{(p)}_{k}(x_{i}) f_{N_{\rm cheb}}(x_{k}).
\end{equation}
We define the Chebyshev differentiation matrix as $D^{(p)}=\{L^{(p)}_{k}(x_{i})\}$. The components of the matrix are given by:
\begin{align}
D^{(1)}_{i,j} &= \frac{\bar{c}_{i}}{\bar{c}_{j}}\frac{(-1)^{i+j}}{(x_{i}-x_{j})} , (i \neq j), i,j = 1,\ldots,N_{\rm cheb}-1 ,\\
D^{(1)}_{i,i} &= -\frac{x_{i}}{2(1-x_{i}^2)} , i=1,\ldots,N_{\rm cheb}-1,\\ 
D^{(1)}_{0,0} &=-D^{(1)}_{N_{\rm cheb},N_{\rm cheb}} = \frac{2N_{\rm cheb}^2+1}{6}\,.
\end{align}
To improve round-off errors in the numerical computations \cite{spectralmatlab}, the following identity of the diagonal components of the matrix $D$ is used:
\begin{equation}
D^{(1)}_{i,i} = -\sum_{j=0,j \neq i}^{N_{\rm cheb}} D_{i,j}^{(1)}.
\end{equation}

%\bigskip

In spectral methods, the error decays exponentially with $N_{\rm cheb}$. This represents a significant improvement in comparison with finite differences, where the error decays algebraically as $1/N^{v}$, where $v>0$ and $N$ is the number of points of the grid. The exponential convergence in spectral methods is obtained thanks to the fact that the derivate at a given point is computed taking into account all the other points of the grid, instead of finite differences, which only consider the neighbours.

The radial domain $\Omega$ in our case corresponds to $\Omega = [r_{\rm min},r_{\rm max}]$ where $r_{\rm max} = N_{H} R_{H}(t_{0})$ and $r_{\rm min}=0$. $N_{H}$ is the number of initial cosmological horizons where we put the final point of the grid, for which it usually is enough to take $N_{H} \sim 10^{2}$. The domain of the Chebyshev polynomials is $[-1,1]$; therefore, we need to make a mapping between the physical domain and the spectral one. Different options are possible, but a linear~mapping is commonly used:
\begin{equation}
\tilde{x}_{k} = \frac{r_{max}+r_{min}}{2}+\frac{r_{max}-r_{min}}{2}x_{k}.
\end{equation}
$\tilde{x}_{k}$ are the new Chebyshev points rescaled to our physical domain $\Omega$. Furthermore, the Chebyshev matrix can be rescaled using the chain rule:
\begin{equation}
\tilde{D} = \frac{2}{r_{\rm max}-r_{\rm min}} D.
\end{equation}
The imposition of boundary conditions in spectral methods is straightforward. In the case of the Dirichlet boundary condition at $x_{k}$, such that $f(x=x_{k})=u_{D,bc}$, we just need to satisfy $f_{N_{\rm cheb}}(x=x_{k})=u_{D,bc}$. In the case of the Neumann boundary condition such that $f^{(1)}(x=x_{k})=u_{N,bc}$, we should satisfy that $(D \cdot f_{N_{\rm cheb}})(x=x_{k})=u_{N,bc}$. The numerical stability will depend on the values chosen for $dt$ and $N_{\rm cheb}$. We do not have complete ``freedom'' to choose those values. If the number of points of the grid is increased, it will require a reduction of the time step $dt$, in order to avoid instabilities during the evolution. The Courant--Friedrichs--Lax (CFL) condition for hyperbolic PDEs already indicates that the maximum allowed time step should fulfil $dt \propto 1/N^{2}$.

There are situations where pressure gradients are substantially large during the numerical evolution. This, for instance, can happen when the curvature fluctuation has very sharp profiles or when $w$ becomes large. In this situation, it is necessary to increase the numerical accuracy of the simulation. One possibility is to increase the number of points of the grid $N_{\rm cheb}$, but of course, this would imply reducing the time step $dt$ to ensure stability. The other one (which is actually better and more clever) is to use a composite Chebyshev grid: split the full domain into several Chebyshev grids depending on the density of points needed in specific regions of the domain. In particular, the full domain $\Omega$ is split into $M$ subdomains as $\Omega_{l} =[r_{l},r_{l+1}]$ with $l=0,1\ldots,M$. A mapping between the spectral and physical domain for each is required. We choose a linear mapping again as,
\begin{equation}
 \tilde{x}_{k,l} = \frac{r_{l+1}+r_{l}}{2}+\frac{r_{l+1}-r_{l}}{2}x_{k,l} ,
\end{equation}
where $\tilde{x}_{k,l}$ are the new Chebyshev points re-scaled to the subdomain $\Omega_{l}$. The Chebyshev differentiation matrix is re-scaled again using the chain rule:
\begin{equation}
 \tilde{D_{l}} = \frac{2}{r_{l+1}-r_{l}} D_{l}.
\end{equation}
The numerical evolution is performed independently in each subdomain, i.e., the spatial derivative is computed using the Chebyshev differentiation matrix $\tilde{D_{l}}$ associated with each subdomain, and the time integration is performed applying the Runge--Kutta 4 method on each~subdomain.

\begin{figure}[H]
\centering
\includegraphics[width=0.6\linewidth]{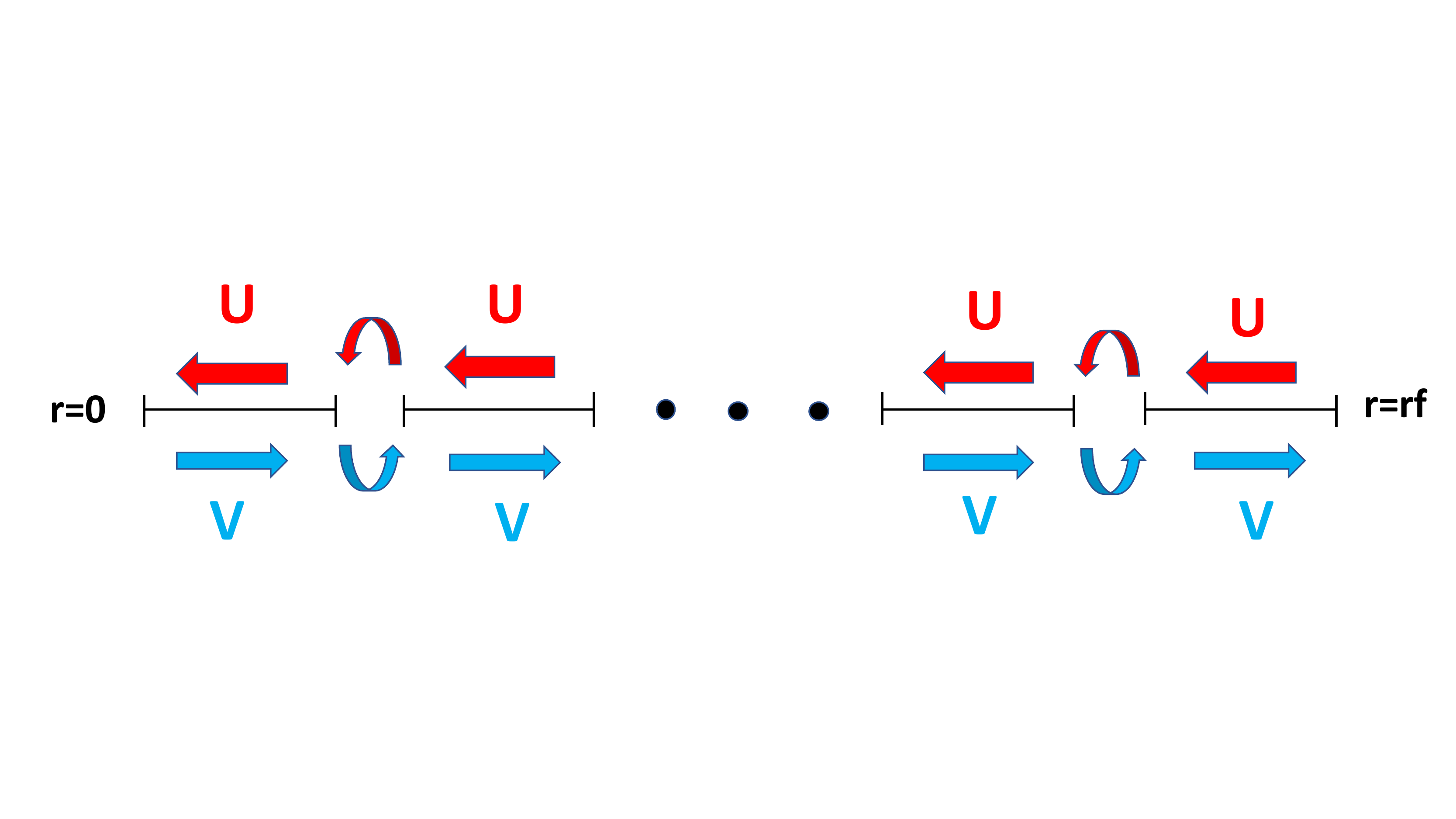} 
\caption{Sketch of the application of the boundary conditions between the different subdomain grids. Through an analysis of the characteristics, the flow of the field variables $u$ and $v$ can be identified {(in our case $u$ and $v$ would correspond to $\rho,M,R,U$)}, and we apply the prescription of Equation~\eqref{2_boundari_domains}.}
\label{fig:subdomains}
\end{figure}

To evolve across the different subdomains $\Omega_{l}$, it is necessary to impose boundary conditions. The approach followed is the one of \cite{Teukolsky}, which is based on performing an analysis of the characteristics of the field variables. The prescription is that the incoming fields' derivative is replaced by the time derivative of the outgoing fields from the neighbouring domain at the boundaries. It was checked in \cite{Escriva:2020tak} that all the fields of the Misner--Sharp equations are incoming except for the density field. Therefore, the boundary conditions that must be applied between each subdomain are:
\begin{align}
\label{2_boundari_domains}
\dot{M}(t,r_{l+1,i}) &= \dot{M}(t,r_{l,f}) ,\\
\dot{U}(t,r_{l+1,i}) &= \dot{U}(t,r_{l,f})\nonumber,\\ 
\dot{R}(t,r_{l+1,i}) &= \dot{R}(t,r_{l,f})\nonumber,\\ 
\dot{\rho}(t,r_{l,f}) &= \dot{\rho}(t,r_{l+1,i})\nonumber \,.
\end{align}
where $r_{l,f}$ and $r_{l+1,i}$ mean the last grid point in the subdomain $\Omega_{l}$ and the first grid point in the subdomain $\Omega_{l+1}$, respectively. A diagram of the imposition of the boundary conditions between the subdomains can be found in Figure~\ref{fig:subdomains}.

The use of multigrids in spectral methods has a similar idea to the use of an Adaptive Mesh Refinement (AMR) in finite differences, to increase the resolution of the grid in some regions. Actually, the use of AMR was essential in \cite{musco2013} to verify the scaling law behaviour for the PBH mass in the critical regime. {The procedure of the AMR is based on splitting the cell grids on a particular desired region and making subsequent subdivisions of the cells. A summarised and basic procedure is the following: (1) During the evolution, individual grid cells are tagged for refinement using some criteria (high-density regions have to be highly resolved, for instance). (2) All tagged cells are then refined, which means that a finer grid (inside each initial cell) is overlaid on the big one. (3) Finally, after refinement has been performed, individual grid patches on a single fixed level of refinement are evolved in time using the time-discretisation scheme, implemented from the differential equations of the problem. If the level of refinement implemented in a cell is greater than what is needed, the refinement of the cell can be reduced. The disadvantage as in spectral methods is that the stability of the method is modified, and a smaller time step will be required. Therefore, the AMR (and in general, any non-uniform grid procedure) allows resolving problems that are intractable due to the high amount of resolution needed for some regions of the domain.}

A known weak point of spectral methods is when they have to handle discontinuous solutions, in particular with the presence of shocks, since the derivative at each point is computed globally, and therefore, the discontinuity is quickly propagated to the other parts of the grids. In this situation, it is better to use a finite differences approach, although usually, a modification of the numerical implementation is still needed to be able to capture shocks, such as the introduction of artificial viscosity. 

However, as mentioned in Section \ref{sec:intro}, in our case, shocks are not formed in the simulations, since we should consider initial conditions at super-horizon scales \cite{musco2013}.

\subsection{Numerical Procedure}\label{numerical_procedure}

In all the simulations, we fixed the value of $w$, $t_{0}=1$ and $a_{0}=1$. From that, we automatically obtained $H_{0} = \alpha $, $R_{H}(t_{0})=1/H_{0}$, and recall that $\alpha = 2/3(1+w)$. For the initial length scale of the perpetuation, $r_{m}=10 R_{H}(t_{0})$ ($\Lambda=10$) is usually considered, which is sufficient to ensure the long wavelength approximation, as was indicated in \cite{musco2018}. 

First, we focus on describing the procedure to obtain the threshold $\delta_{c}$, and we leave the method to obtain the final mass $M_{\rm PBH,f}$ for the following subsection. To obtain the threshold, a bisection method was used, which compares different regimes of $\delta_m$ until finding the range in which the formation of the apparent horizon appears. The threshold $\delta_{c}$ corresponds to the midpoint of this range. 

An example of a successful procedure is the following:

\begin{itemize}
\item Create the different grid points $\tilde{x}_{k,l}$ for the different subdomains $\Omega_{l}$ with a given number of points $N_{\rm cheb,l}$, and compute the different Chebyshev differentiation matrices $D_{l}$. Introduce also the initial time step $dt_{0}$, which usually $dt_{0} \leq 10^{-3}$ is enough to ensure stability;

\item Specify a lower and an upper limit in $\delta_m$ of the bisection. The maximum value $\delta_{\rm m, max}=f(w)$ can be chosen for the upper limit, and for the lower bound, $\delta_{\rm m,min} \approx w$ (which corresponds to the threshold estimation using a Jeans length criteria). To reduce the computational time, a closer domain to $\delta_{c}$ can be chosen if such a domain is known;

\item A value $\delta_m$ is taken once the bisection starts. From that, the curvature profiles $K(r)$ or $\zeta(\tilde{r})$ can be built, and the location of the compaction function peak $r_m$ can be found, $\tilde{r}_m$, using Equations \eqref{2_cmax} and \eqref{eq:tilde_max} respectively. Subsequently, compute the perturbations of the hydrodynamical variables following Equations~\eqref{2_expansion} and \eqref{eq:2_perturbations};

\item Integrate Equations~\eqref{eq:u_simply} and \eqref{eq:m_simply} at each time step $dt$ using the fourth-order Runge--Kutta method, imposing at each internal time step the boundary conditions at \mbox{$r=0$} and $r=r_{f}$. Include the internal boundary conditions between the different Chebyshev grids Equation~\eqref{2_boundari_domains};

\item To distinguish numerically the formation of an apparent horizon or not, we used the peak value of the compaction function in time. When the peak approaches $\com_{\rm max} \approx 1$, this implies the formation of the apparent horizon \footnote{Technically, the condition of the formation of the apparent horizon is given when $2M(r,t)/R(r,t)=1$. However, for computationally efficiency, ot is easier to check whatever $\com \approx 1$, since $\com = 2M(r,t)/R(r,t)-2M_{b}(r,t)/R(r,t)$, which avoids finding the cosmological~horizon.}. In this case, the value of 
$\delta_m$ chosen would imply the formation of a black hole, i.e., 
$\delta_{\rm m,yes}$, and therefore, a lower value $\delta_m$ closer to $\delta_c$ must be found modifying the bound of the bisection in such way that $\delta_{c} \in [\delta_{c, \rm min} , \delta_{m, \rm yes}]$. In the opposite case, when $\com_{\rm max}$ decreases continuously for some time, this will imply the dispersion of the perturbation, and the bisection bound should be modified as 
$\delta_{c} \in [\delta_{m ,\rm no} , \delta_{ c,\rm max}]$;

\item The previous steps are iterated until the difference between $\delta_{m,\rm yes}$ and $\delta_{m,\rm no}$ is lower than the desired resolution, i.e., $\delta_{m,\rm yes}-\delta_{m,\rm no} \lesssim \delta(\delta_{c})$. From that, the threshold value is defined as $\delta_{c} = (\delta_{m,\rm yes}+\delta_{m,\rm no})/2 \pm \delta(\delta_{c})$. When $\delta_m$ approaches the critical value, in some cases, the grid resolution initially set up is not enough to follow the evolution; if so, then $\delta_m$ is shifted according to $\delta(\delta_{c})$ to avoid being close to $\delta_c$. Another alternative would be the dynamical inclusion of more subgrids in the desired region where gradients start to be large.
\end{itemize}

During the time evolution, a conformal time step $dt = dt_{0}(t/t_{0})^{\alpha}$ is used, which substantially improves the performance of the simulation. The two-norm of the Hamiltonian constraint Equation~\eqref{2_eqconstraint} is computed at each time step. It is expected that the constraint should remain constant from $t=t_{0}$ if the Einstein equations are correctly solved. In~particular,
\begin{align}
\mathcal{H} &= D_{r}M - 4\pi \Gamma \rho R^{2}, \\
\mid\mid\mathcal{H}\mid\mid_{2} &\equiv \sum_{l} \frac{1}{N_{\rm cheb,l}}\sqrt{\sum_{k} \Big| \frac{M_{k,l}'/R_{k,l}'}{4 \pi \rho_{k,l} R_{k,l}^{2}}-1 \Big|^2},
\label{eq:2_constraint}
\end{align}
where the label $l$ refers to the sum of the different Chebyshev grids and $k$ to the different points of the grid.

\subsection{Numerical Estimation of the Final PBH Mass}\label{sec:2_excision}

To obtain the final mass of the PBH $M_{\rm PBH,f }$, we used an excision technique, and we describe it with detail in this section. Saying that, other alternatives in the literature have used null coordinates with the Hernandez--Misner equations \cite{misnerhernandez,musco2005,Niemeyer2} to obtain the final mass. We refer the reader to these references, but the primary approach followed can be summarised as follows: (i) The simulation was set up with some initial data using the Misner--Sharp equations. (ii) Starting the evolution, an outgoing null radial ray is sent from the centre to the outer point of the grid, and the values of the field variables are stored once the ray crosses each grid point. (iii) The Misner--Hernandez equations are evolved using the initial conditions obtained in the previous step, which are basically the values stored. 

Let us focus now on excision. The main idea of an excision technique is based on the fact that nothing inside the event horizon can affect the physics outside. Therefore, the aim is to dynamically remove the segment of the computational domain inside the apparent horizon, to avoid large gradients appearing and breaking down the simulation.

There are different procedures to implement excision, but the one we used here is the following: We define two parameters, $\Delta r$ and $dr$ (we always consider $\Delta r>dr$), where $\Delta r$ is the distance separation between the apparent horizon and the excision surface that we establish after each redefinition of the excision surface, and $dr$ is the maximum allowed displacement of the apparent horizon before we create a new excision boundary.
At each time step, we locate the position of the apparent horizon, taking into account that $2M(r_{*},t)/R(r_{*},t)=1$, where $r_{*}(t)$ is the location of the apparent horizon in time. Since we were handling a discretised grid, we used a cubic spline interpolation to find the correct value of $r_{*}(t)$ (the difference of taking a quadratic spline interpolation is $O(0.01\%)$ for the PBH mass values).

In particular, the method used is the following:

For a supercritical perturbation, when the peak of the compaction function $\com_{\rm max} \approx 1.2$ (the final result is not affected by the exact value that we take, as long as we consider $\com_{\rm max} \approx O(1)$), part of the computation domain is removed, creating an excision surface close to the location of the apparent horizon, whose separation from the excision boundary is basically given by $\Delta r$. Once having done this, we created a dense, but small Chebyshev grid starting from the excised boundary. Another single Chebyshev grid can cover the other domain since large gradients are only developed near the excision surface.

We evolved in time the Chebyshev grids in the usual form independently, and we tracked the apparent horizon motion at each $dt$. Once the apparent horizon had been moved a distance greater than $dr$, we created a new excision surface separated a distance $\Delta r$ from the apparent horizon. During the evolution, we needed to continuously reduce the values $\Delta r$,$d r$, since the velocity of the apparent horizon decreases in time. This is especially important for very small $\delta_m-\delta_c$. We decided to reduce $\Delta r$,$d r$ when the Hamiltonian constraint starts to be continually violated. In particular, it was found that considering $\Delta r \approx 2 dr \approx O(10^{-2})$ worked nicely for our purposes.

There is no physical boundary condition to apply at the excision surface. Still, it was found that freezing {(leave it constant)} the gradient of $\rho$ {(i.e., $\rho'$)} at the excision surface, after each redefinition of the excision boundary, allowed increasing the stability of the method without changing the numerical results.

\section{Numerical Results}\label{sec:results}

In this section, we review some numerical results using the numerical procedure pointed out in the previous section. First of all, we introduce a set of curvature profiles that allows us to span all different possible thresholds for PBH formation. Later on, we will see the dynamical evolution of the collapse for three different cases in terms of the initial amplitude $\delta_m$ and with a specific profile chosen. Finally, we show the results regarding the final PBH mass in terms of the different profiles, as well as the effect of the accretion from the FLRW background.

The numerical results were obtained using the procedure of Section \ref{sec:numerics} with a computer, i-7 core, $16$ GB of RAM, and without parallelisation procedures. As an example of the efficiency, in the case of $w=1/3$ and with a Gaussian profile, to obtain the threshold $\delta_{c}$ with a resolution $\delta_m-\delta_{c} \approx 10^{-3}$, the running time was $ \sim$3--5 min using the bisection method in Section~\ref{numerical_procedure}. To estimate the final mass $M_{\rm PBH,f}$, we needed to run the excision method in Section~\ref{sec:2_excision} for $\sim$3 h for the largest PBHs. These values can differ when changing the profiles or the equation of state $w$. The typical size of the grid chosen was two or three grids with $\sim$100 points in each grid. However, more points could be needed for sharp initial profiles. With these configurations, the maximum resolution obtained for the threshold in the case of a Gaussian profile was $\delta_m-\delta_{c} \approx 10^{-5}$ (in the case of $w=1/3$). With sharper profiles or larger $w$, this resolution slightly reduced.

\subsection{Curvature Fluctuations}

The seed of the generation of the curvature fluctuation is the power spectrum, namely $P(k)$, where $k$ is the wavelength. We briefly review the procedure assuming the Gaussian statistics of \cite{peak_theory} applied to our purposes. We considered that the amplitudes of the curvature perturbations are statistically Gaussian distributed accordingly to the specific inflationary model considered, {and therefore, under this assumption, the rare over-densities leading to PBH formation are, with a very good approximation, spherically symmetric \cite{peak_theory}.}

From the power spectrum $P_{\zeta}(k)$, we can relate it to the curvature fluctuation $\zeta$ Equation~\eqref{metrica_222} as,
\begin{equation}
<\zeta_{k} \zeta_{k'}> = (2 \pi)^{3} \frac{2 \pi^{3} P_{\zeta}(k)}{k^{3}}\delta_{D}^{3}(k+k'),
\end{equation}
where $\zeta_{k}$ is the Fourier mode $k$ of the field $\zeta$,
\begin{equation}
\zeta(\vec{x}) = \int \frac{d \vec{k}}{(2 \pi)^{3}} e^{i \vec{k}\cdot \vec{x}}\zeta(\vec{k}).
\end{equation}
Notice that in linear theory and at super-horizon scales, the density contrast $\delta \rho / \rho_{b}(\tilde{r})$ is related to $\zeta(\tilde{r})$ as,
\begin{equation}
\frac{\delta \rho}{\rho_{b}}(\tilde{r}) \approx -\frac{2(1+w)}{5+3w}\frac{1}{a^{2}H^{2}}\nabla^{2} \zeta(\tilde{r}),
\end{equation}
and therefore:
\begin{equation}
\left(\frac{\delta \rho}{\rho_{b}}\right)_{k} \approx -\frac{2(1+w)}{5+3w}\frac{k^{2}}{a^{2}H^{2}}\zeta_{k}.
\end{equation}
From that, we can build the normalised two-point correlation function by,
\begin{equation}
\psi(\tilde{r}) = \frac{1}{\sigma_{0}^{2}} \int W(k) P_{\zeta}(k) \sinc(k \tilde{r}) \frac{dk}{k},
\end{equation}
{where $W(k)$ is the window function,} and the variance of the field $\zeta$ is given by,
\begin{equation}
<\zeta^{2}> = \sigma_{0}^{2} = \int \frac{dk}{k} W(k) P_{\zeta}(k).
\end{equation}
Therefore, the curvature fluctuation $\zeta(r)$ assuming peak theory can be built as,
\begin{equation}
\zeta(\tilde{r}) = \mu \psi(\tilde{r}) = \frac{\mu}{\sigma_{0}} \int W(k) P_{\zeta}(k) \sinc(k \tilde{r}) \frac{dk}{k},
\label{zeta_yo_power} 
\end{equation}
where $\mu$ is the peak value of $\zeta$.

In conclusion, using Equation~\eqref{zeta_yo_power}, we can relate the power spectrum to our curvature fluctuation $\zeta(\tilde{r})$ and, from that, set up the initial conditions following \mbox{Equations~\eqref{eq:2_perturbations} and \eqref{eq:tilde_perturb}}, or alternatively also build $K(r)$ using the transformations of Equation~\eqref{transforms_tilde_K}. {Although it goes beyond the scope of the review, we should mention that one could also consider the effect of the inclusion of non-Gaussianities in Equation~\eqref{zeta_yo_power}, which can have important consequences for the PBH abundance estimation}. There are vast and extensive works that consider its impact on the thresholds and PBH abundances \cite{Atal:2018neu,vicente-garriga,Hayato2,Cai:2017bxr,Bullock:1996at,Pattison_2017,PinaAvelino:2005rm,Riccardi:2021rlf,Young:2013oia,Young:2014ana,Young:2014oea,Young:2015cyn,yoo,Hidalgo:2007vk,Atal:2021jyo,Kitajima:2021fpq,Davies:2021loj,Taoso:2021uvl,Cai:2021zsp}, including some of them that have used numerical simulations following the basics shown in this \mbox{review \cite{rioto,garrigavicentejudithescriva}}.

Saying that, it is much more practical to already span a set of curvature profiles instead of considering different power spectrums when we are interested in studying the physical process of PBH formation. The reason is that setting up directly the curvature fluctuation allows us to modulate different shapes and therefore obtain all possible thresholds and estimate their effects on the PBH mass. In any case, some typical power spectrum templates can be found in \cite{Atal:2018neu,Germani:2018jgr}.

In the literature, exponential families of profiles have been commonly used to modulate $K(r)$ \cite{musco2018,Nakama_2014,Hidalgo:2008mv,Shibata:1999zs}. Another possibility is a polynomial family \cite{Escriva:2020tak}. Both can be written in a convenient way as:
\begin{align}
 K_{\rm pol}(r) &= \frac{\delta_m}{f(w)r_m^2}\frac{1 + 1/q}{1+\frac{1}{q}\left(\frac{r}{r_{m}}\right)^{2(q+1)}}, \label{4_basis_pol} \\
 K_{\rm exp}(r) &= \frac{\delta_m}{f(w) r_m^2}\,
\left(\frac{r}{r_{m}}\right)^{2\lambda}\,
e^{\frac{(1+\lambda)^{2}}{q}\left(1 - \left(\frac{r}{r_{m}}\right)^{\frac{2q}{1+\lambda}}\right)}. \ \label{eq:4_lamda}
\end{align}
The profiles of Equations~\eqref{4_basis_pol} and~\eqref{eq:4_lamda} correspond to polynomial and exponential (for $\lambda=0$) non-centrally peaked (for $\lambda \neq 0$) profiles, respectively. Notice that the profiles depend on the ratio $r/r_m$. The profiles are written in a convenient way with the amplitude $\delta_m$, in such away that when computing $\com(r)$ using Equation~\eqref{C_r}, this automatically gives $\com(r_m)=\delta_m$ when evaluating $r=r_m$.

\begin{figure}[H]
\centering
%\begin{adjustwidth}{-\extralength}{0cm}
\includegraphics[width=0.48\linewidth]{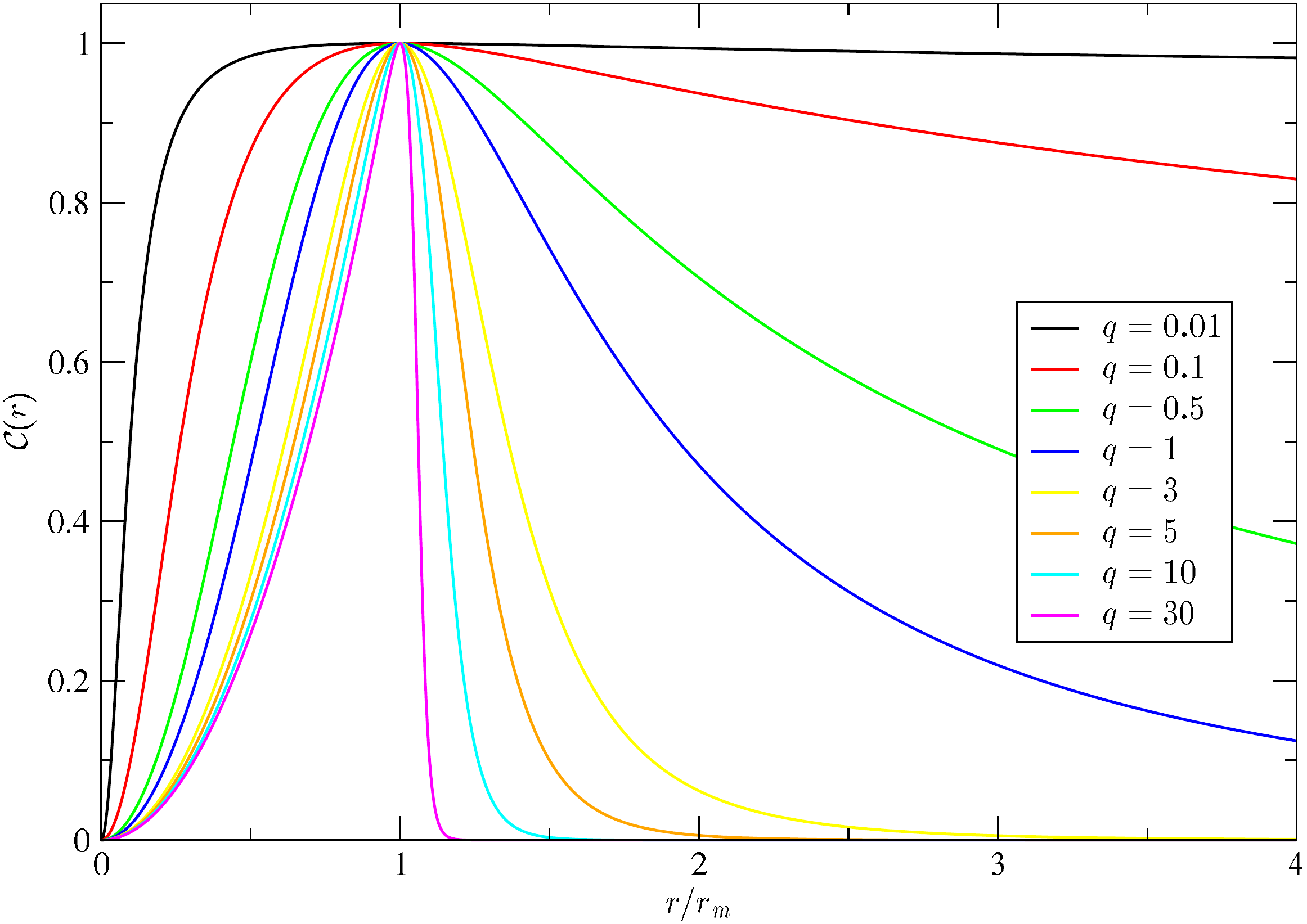} 
\includegraphics[width=0.5\linewidth]{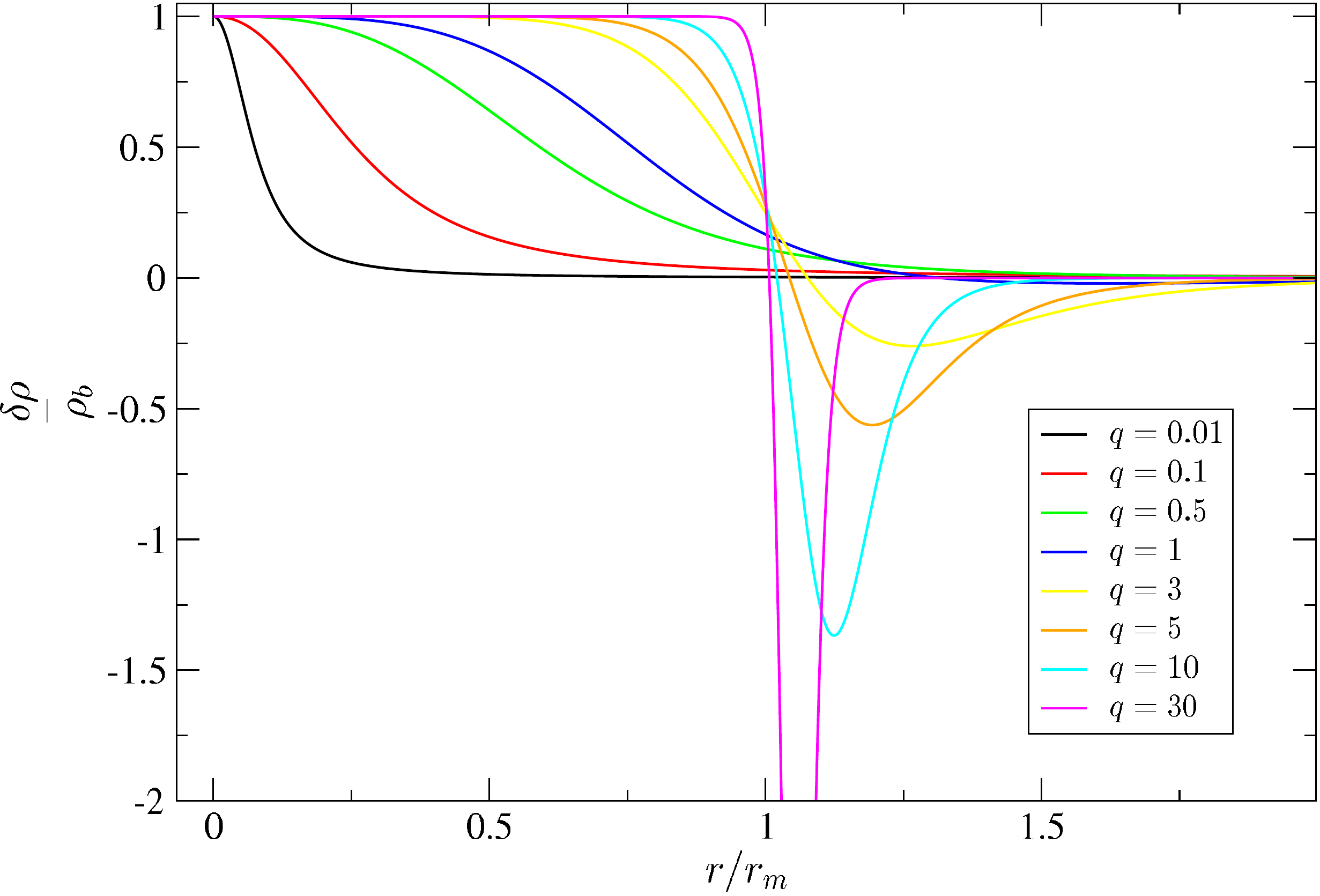} 
%\end{adjustwidth}
\caption{{\textbf{Left-panel}}: Profiles of $\com(r)$ of Equation~\eqref{4_basis_pol} for several values of $q$. The values $\delta_m$ are normalised to $\delta_m=1$. {\textbf{Right-panel}}: Density contrast $\delta \rho / \rho_{b}$ for the same profiles as the top panel. The profiles are normalised to the peak $\delta \rho (r=0) /\rho_{b}$. The negative peak value for $q=30$ around $r=r_m$ goes up to $ \sim$-$4$.}
\label{fig:C_profiles}
\end{figure}

On the other hand, the parameter $q$ is a dimensionless parameter, which relates the shape around the peak of the compaction function, which is defined as,
\begin{equation}
q = -\frac{\com''(r_m)r^{2}_m}{4 \com(r_m)}.
\end{equation}
In terms of the $\tilde{r}$ coordinate, the transformation in Equation~\eqref{transforms_tilde_K} can be used, as was done in \cite{Musco:2020jjb}, to obtain,
\begin{equation}
q = -\frac{\com''(\tilde{r}_m) \tilde{r}^{2}_m}{4 \com(\tilde{r}_m)\left[1-\frac{3}{2}\com(\tilde{r}_m)\right]}.
\label{eq:q_tilde}
\end{equation}
Interestingly, in \cite{universal1}, it was found numerically that profiles with the same $q$ have the same threshold $\delta_c$ upon a deviation of $O$(3--4\%) in the case $1/3\leq w \leq 1$ (and $O$(1--2\%) in the particular case of a radiation-dominated Universe), but we will discuss this in more detail in Section \ref{sec:analytics_on_the_thresholds}, since this result will be used to obtain analytical estimations on $\delta_{c}$. The benefit of Equation~\eqref{4_basis_pol} in comparison with Equation~\eqref{eq:4_lamda} is that at $r \rightarrow 0$, this fulfils the regularity conditions, whereas Equation~\eqref{eq:4_lamda} does not for some small values of $q$. Actually, the profile Equation~\eqref{4_basis_pol} can be considered as a basis profile, since it allows spanning all possible threshold values, as we will see in more detail in Section~\ref{sec:thresholds}. The profiles of $\com(r)$ for the polynomial profile can be found in Figure~\ref{fig:C_profiles}. When $q \gg 1$, the peak of the compaction function is sharp, while when $q \ll 1$, the peak is broad (which would be a homogeneous sphere). {The opposite behaviour can be found for the density contrast (bottom panel of Figure~\ref{fig:C_profiles}). In this case, for small $q$, the density contrast profile is sharper, whereas for large $q$, the shape becomes broad up to $r_m$ (a homogeneous sphere).}

\begin{figure}[H]
\centering
\includegraphics[width=0.45\linewidth]{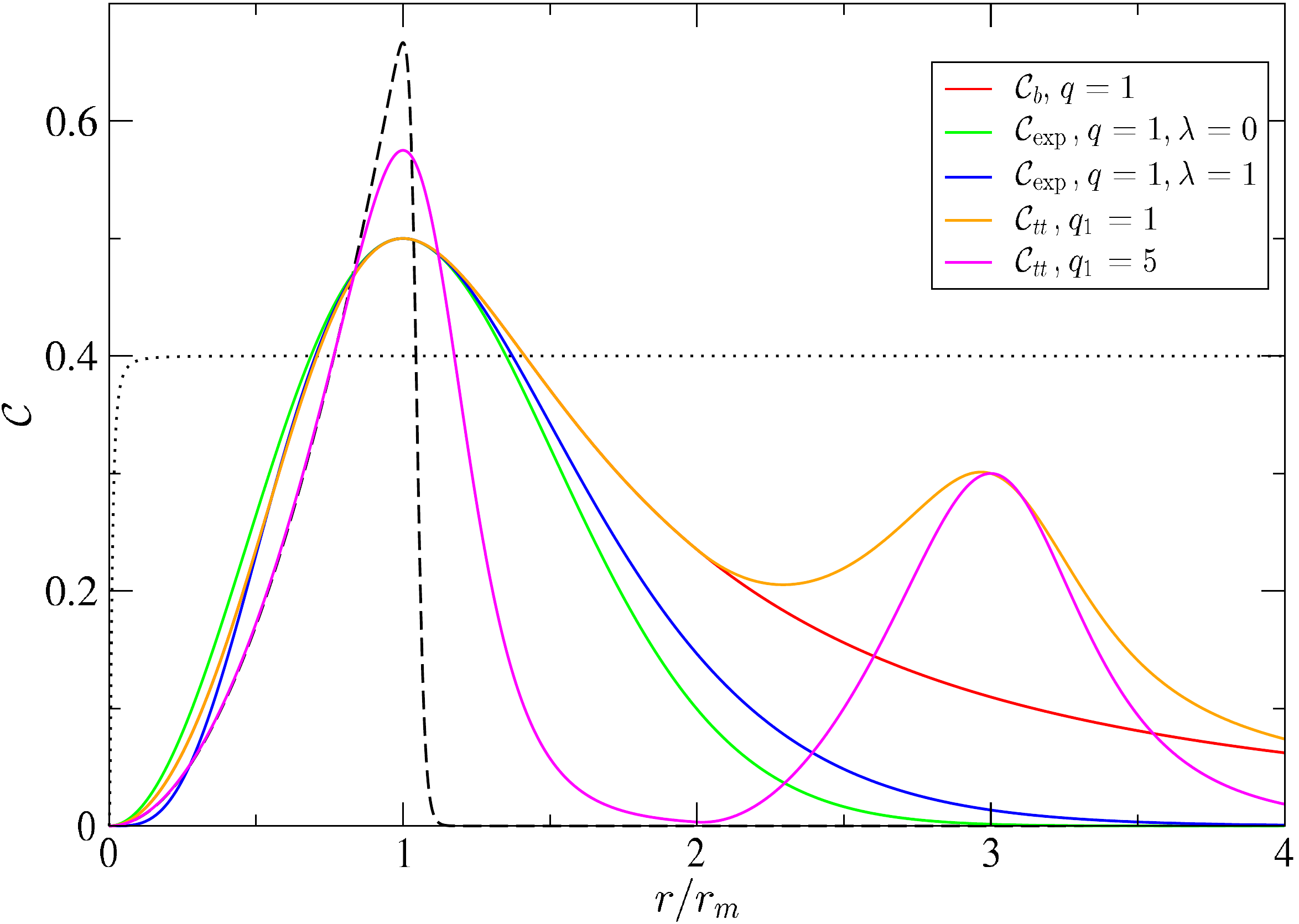} \hfill
\includegraphics[width=0.45\linewidth]{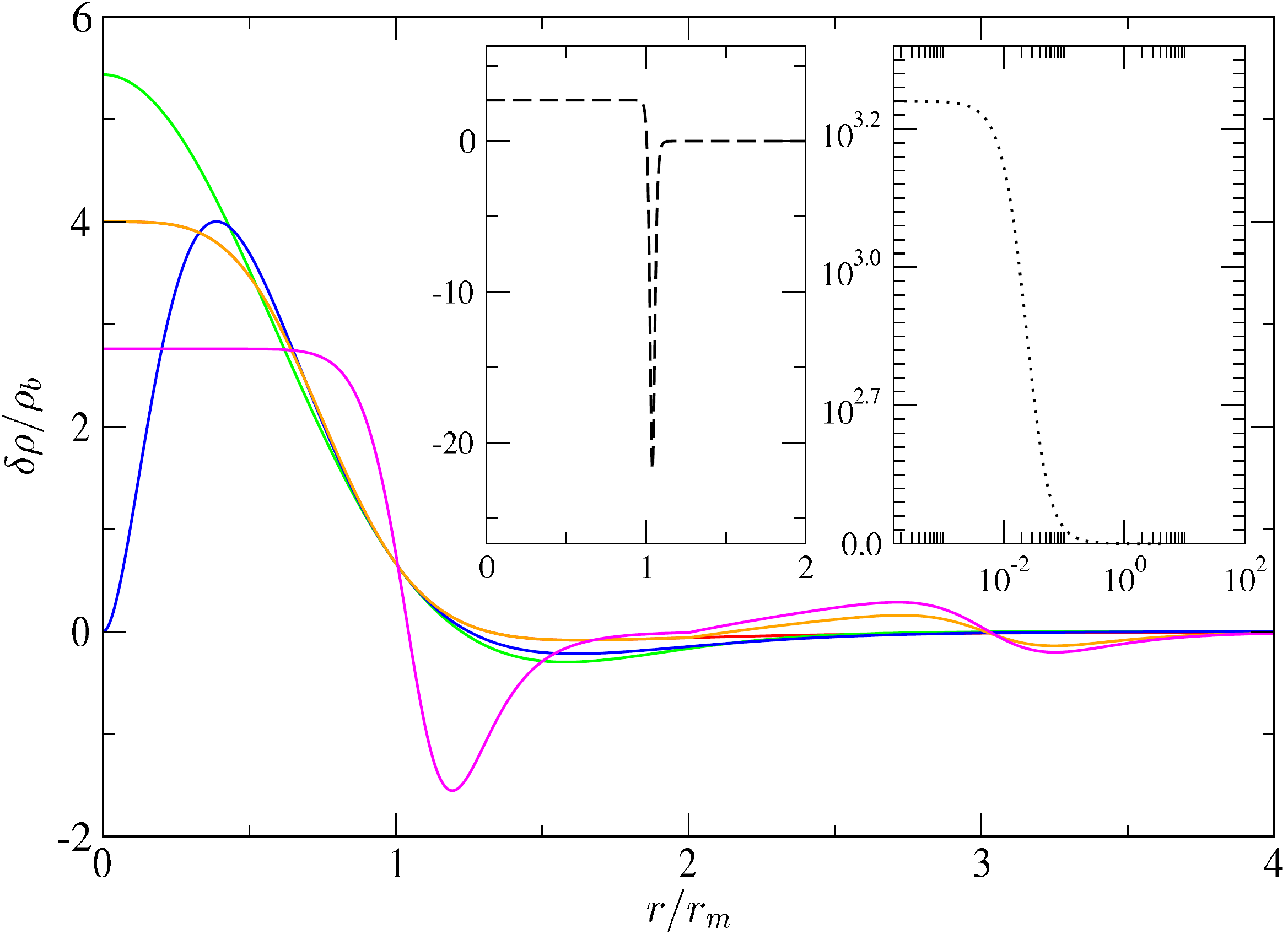}
\includegraphics[width=0.45\linewidth]{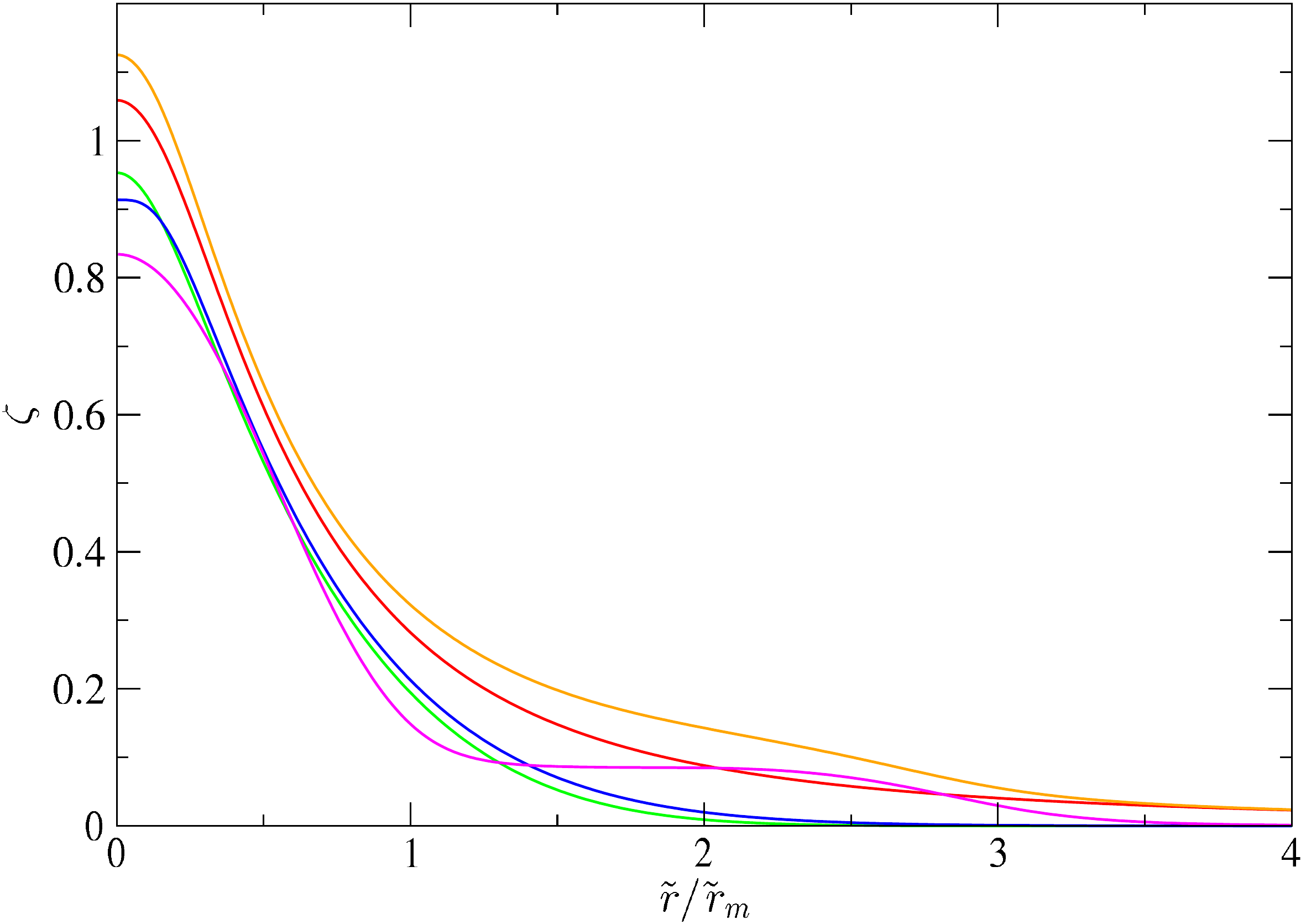}\hfill 
\includegraphics[width=0.45\linewidth]{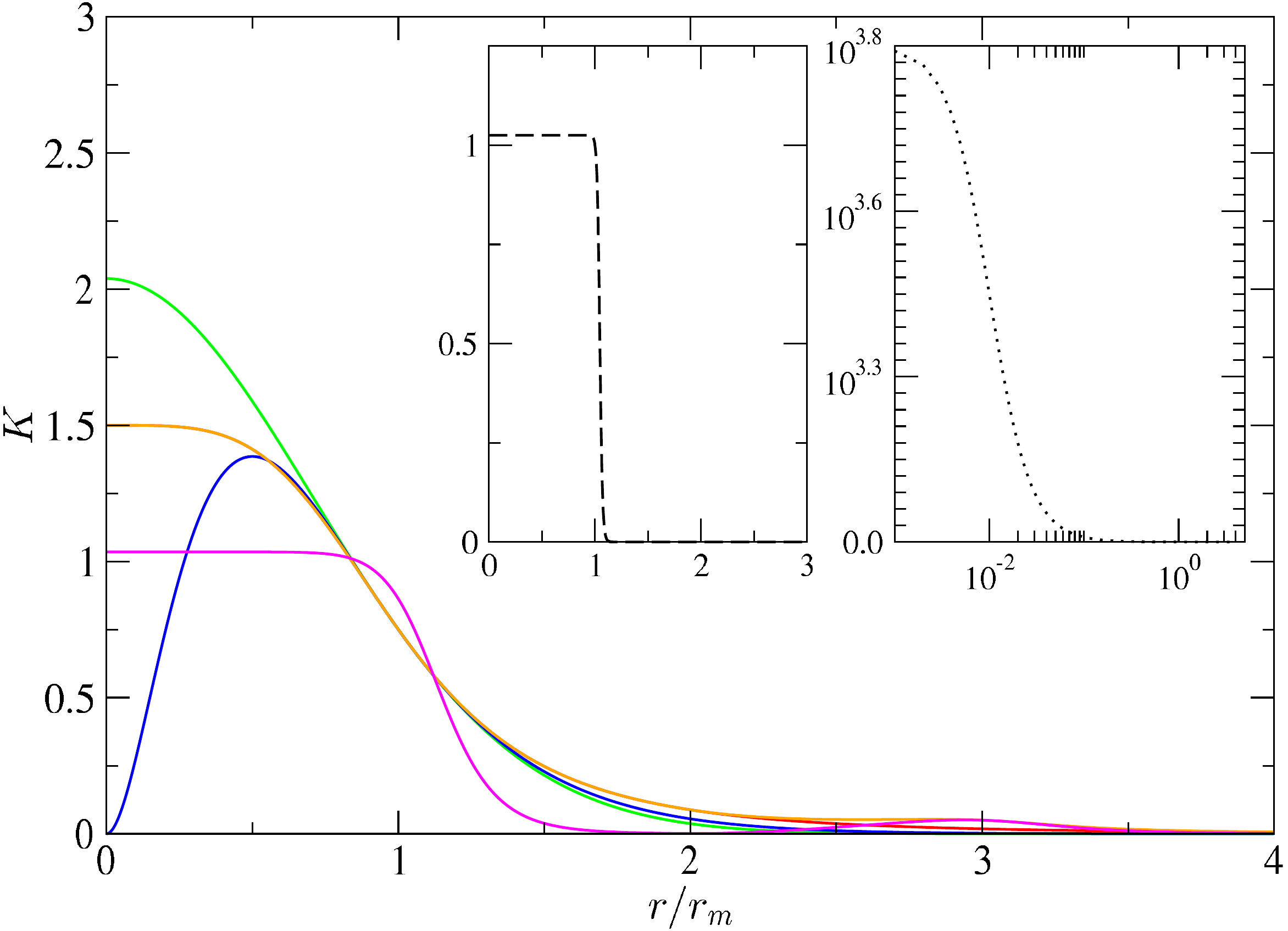} 
\caption{Plots for the profiles with Equations~\eqref{4_basis_pol},~\eqref{eq:4_lamda}, and~\eqref{eq:4_dos_torres} for $\com(r)$ (\textbf{top-left}), $\delta \rho(r)/ \rho_{b}$ (\mbox{\textbf{top-right}}), $\zeta(r)$ (\textbf{bottom-left}), and $K$ (\textbf{bottom-right}) as a function of $r/r_m$ for $\delta_m \sim \delta_{c}$. The dotted black line corresponds to $q=0$ and the dashed black line to $q \rightarrow \infty$. The values taken for $\com_{tt}(r)$ are $\delta_{1}=\delta_{c}(q_{1})$, $q_{2}=3$, $r_{m1}=r_{m2}=1$, $r_{j}=2r_{m1}$, $\com_{tt(\rm peak,2)}=0.3$, with the corresponding $\delta_{2}$ obtained from Equation~\eqref{eq:pico2} using the previous parameters, $q_{1}=5$ (violet) and $q_{1}=1$ (orange).}
\label{fig:4_profiles}
\end{figure}

A profile with two modulating peaks in $\com$ was also considered in \cite{Escriva:2021pmf} {with the aim to compare the other profiles}. It is similar to the one considered in \cite{Nakama:2014fra}, where double-PBH formation was studied. {This kind of profile would represent a perturbed region superposed on another one at much larger scales, as discussed in \cite{Nakama:2014fra}. The compaction function profile is given by,}

\begin{equation}
\mathcal{C}_{tt}(r) = \mathcal{C}_{b}(r, \delta_{1},q_{1},r_{m1})+\theta(r-r_{j}) \mathcal{C}_{b}(r-r_{j},\delta_{2},q_{2},r_{m2}),
\label{eq:4_dos_torres}
\end{equation}
where $\mathcal{C}_{b}$ is equal to:
\begin{equation}
\mathcal{C}_{b}(r, \delta_{j},q_{j},r_{mj}) = \delta_{j} \left(\frac{r}{r_{mj}}\right)^{2}\frac{1 + 1/q_{j}}{1+\frac{1}{q_{j}}\left(\frac{r}{r_{mj}}\right)^{2(q_{j}+1)}}; 
\end{equation}
the value of the second peak of $\com$ can be modulated as, 
\begin{equation}
\mathcal{C}_{tt(\rm peak,2)} = \frac{(1+q_{1})(r_{j}+r_{m2})^{2} \delta_{1}}{q_{1} r_{m1}^{2}+(r_{j}+r_{m2})^{2}(r_{j}+r_{m2}/r_{m1})^{2q_{1}}}+\delta_{2}.
\label{eq:pico2}
\end{equation}
Different profiles of the three families are plotted in Figure~\ref{fig:4_profiles}. In particular, we plot the profiles of Equations~\eqref{4_basis_pol}, \eqref{eq:4_lamda}, and \eqref{eq:4_dos_torres}, and from them, we also plot the corresponding compaction function from Equation~\eqref{C_r}, as well as the density contrast Equation~\eqref{eq:2_perturbations}. The conversion from $K(r)$ to $\zeta(\tilde{r})$ can be made using Equation~\eqref{transforms_tilde_K}.

\subsection{Non-Linear Behaviour of the Gravitational Collapse}\label{sec:non_linear_behaviour}

\begin{figure}[H]                    
\begin{center}                    
\includegraphics[width=0.45\columnwidth]{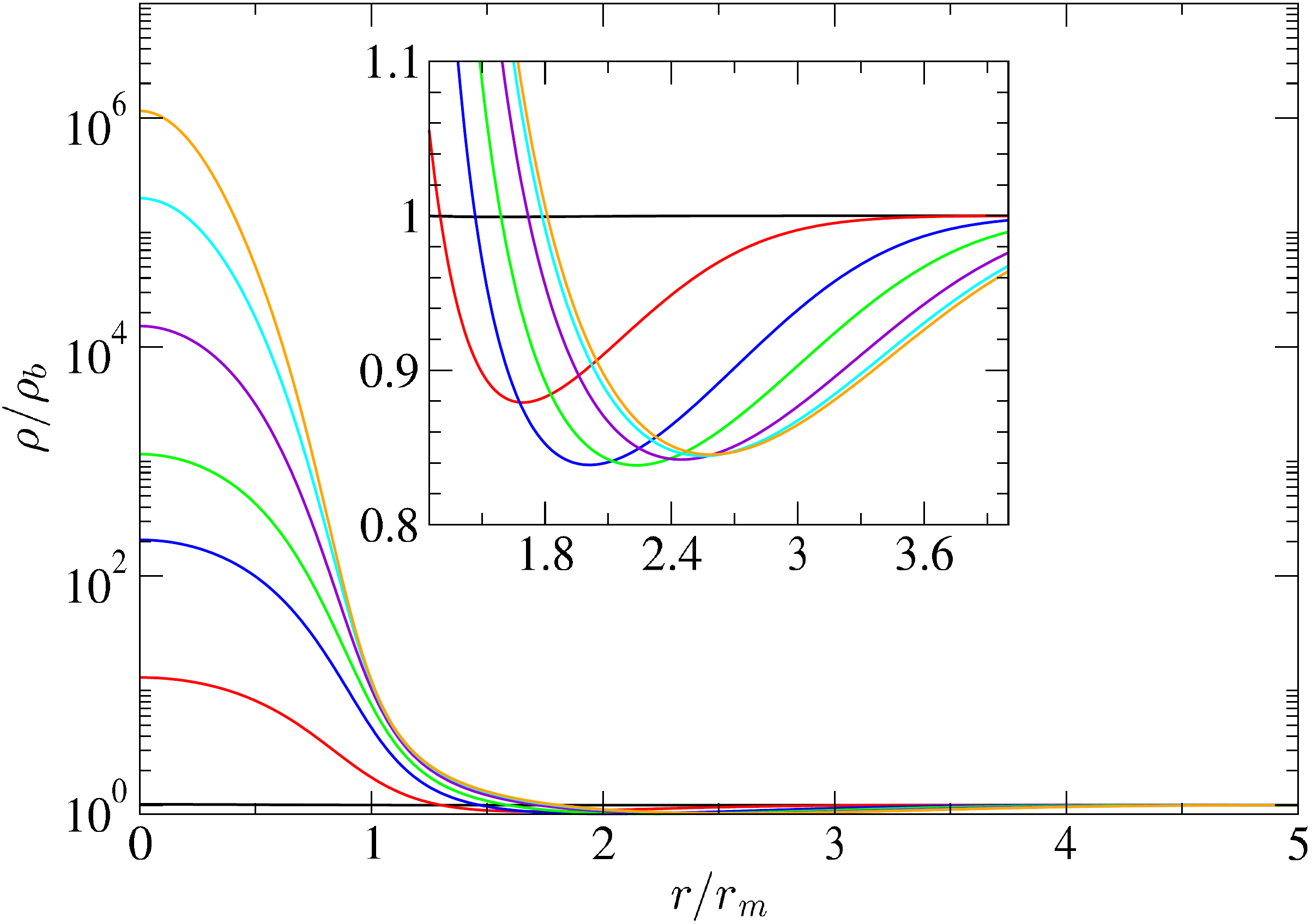}\hfill
\includegraphics[width=0.45\columnwidth]{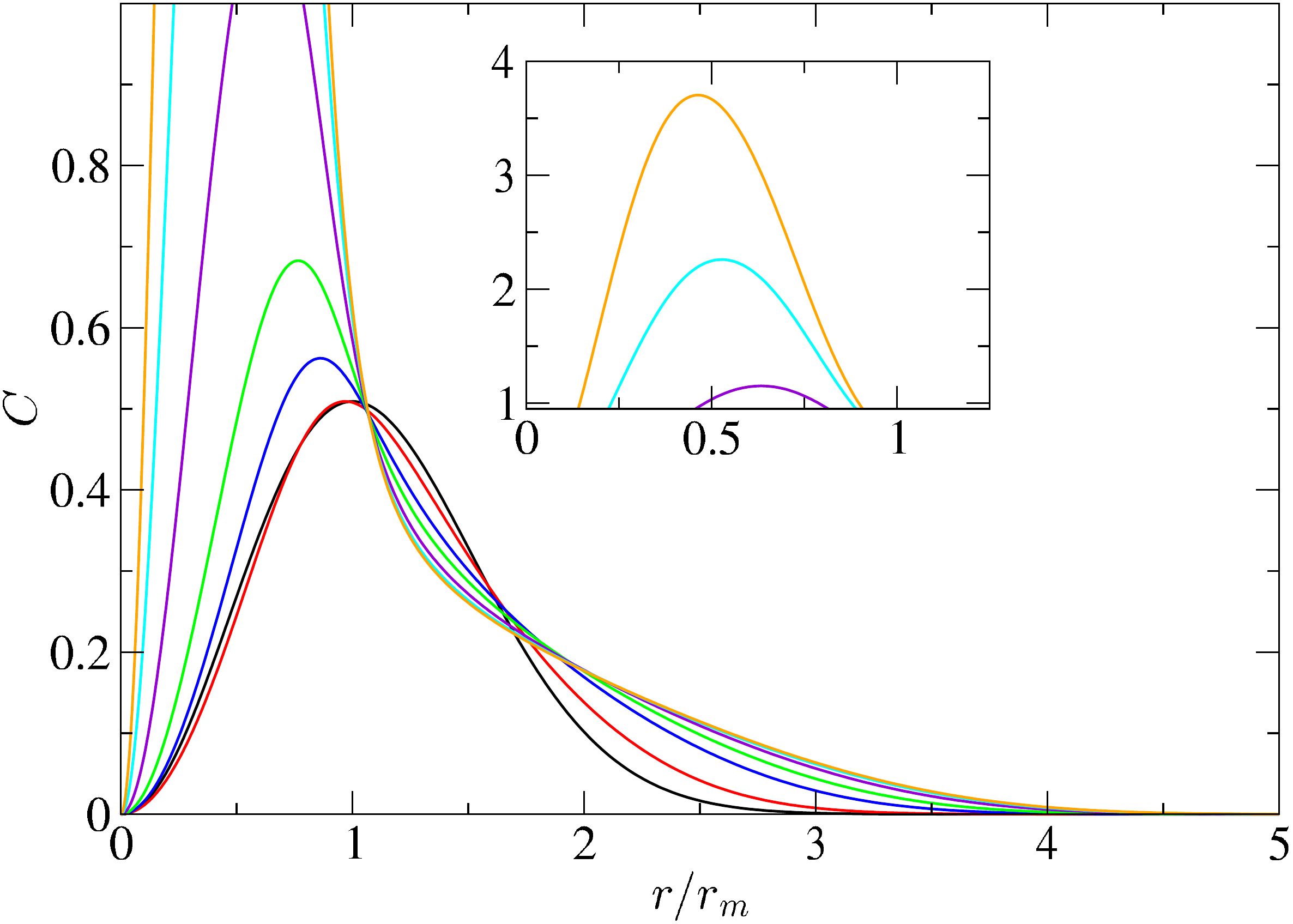}         
\includegraphics[width=0.45\columnwidth]{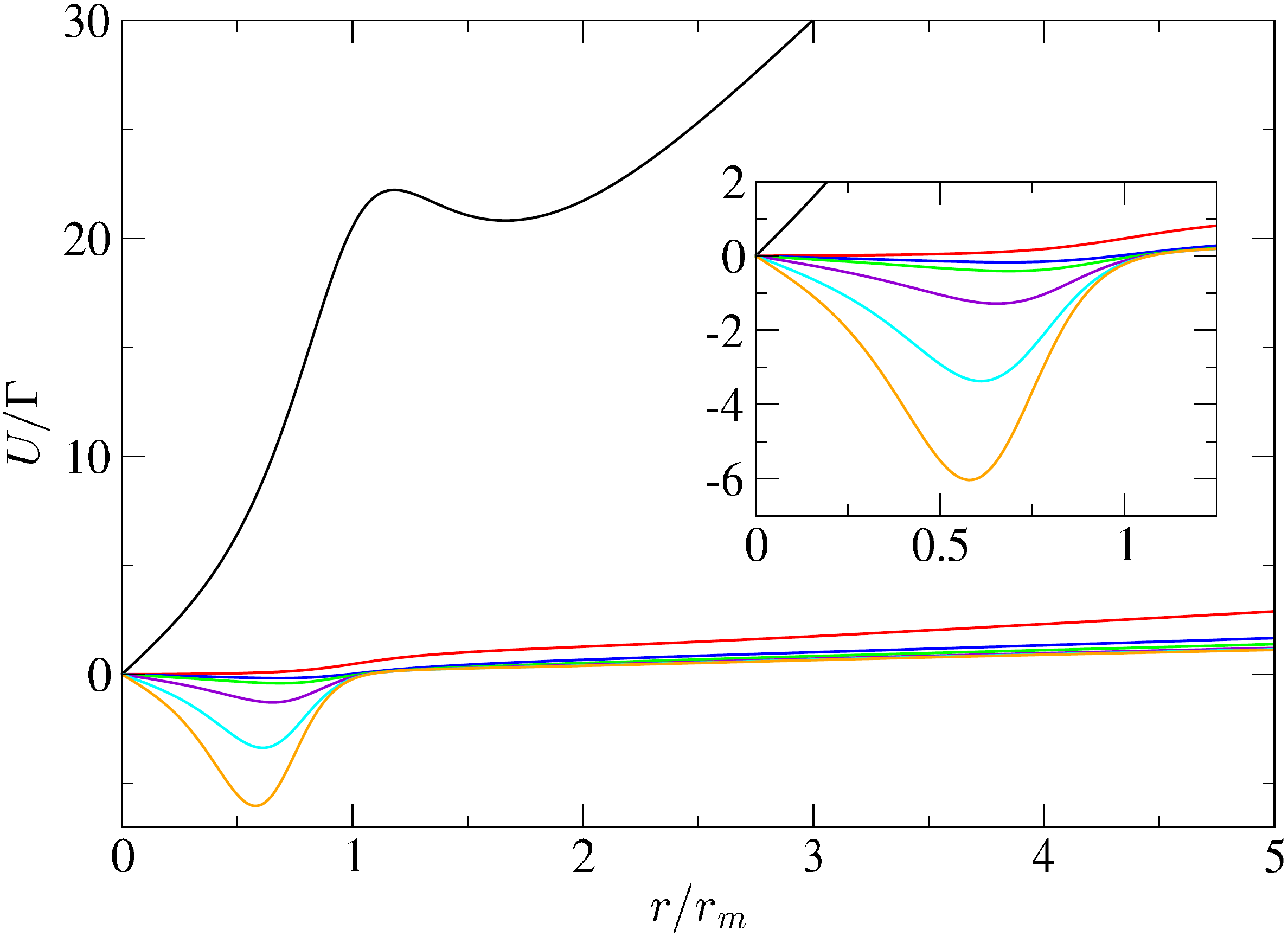}\hfill 
\includegraphics[width=0.45\columnwidth]{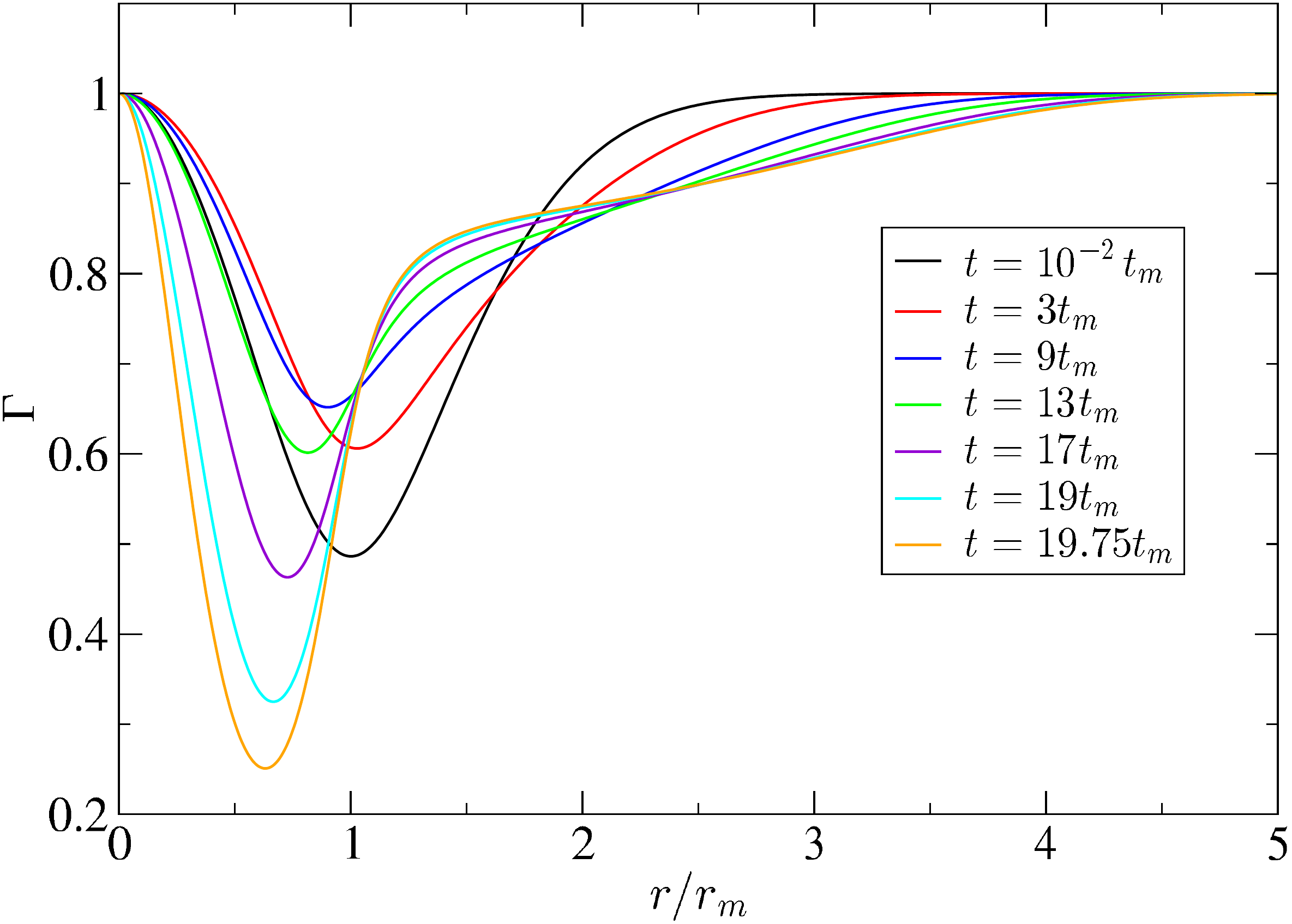}                               
\caption{{Dynamics} of different magnitudes in the case of a supercritical fluctuation (Gaussian profile) at given times $t$. In particular, $\delta_m=0.51$ and $\delta_{c} = 0.49774\pm 2\times 10^{-5}$.} 
\label{fig:2_evolutioncritical}                   
\end{center}                     
\end{figure}

\begin{figure}[H]                    
\begin{center}                    
\includegraphics[width=0.45\columnwidth]{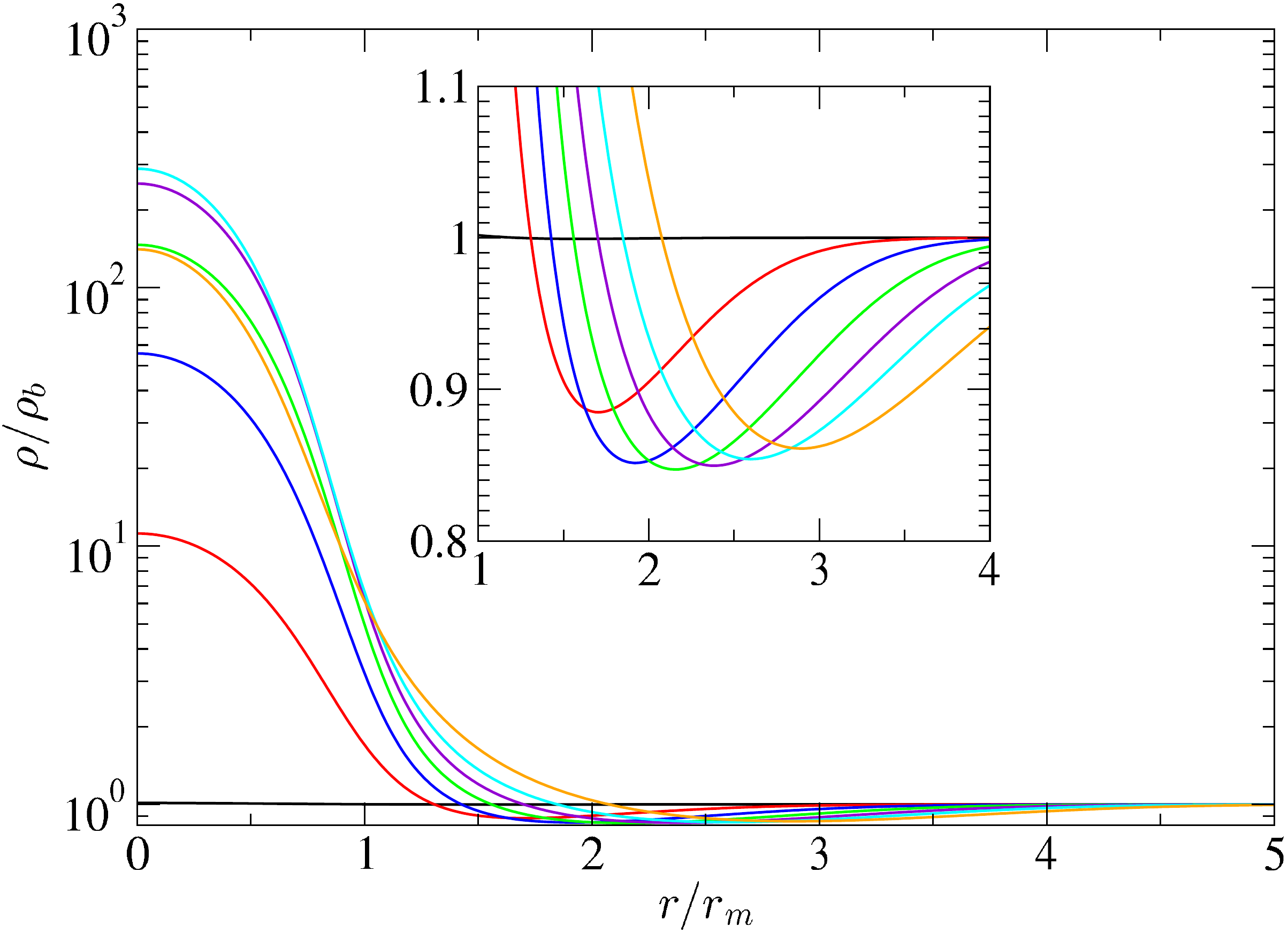}\hfill
\includegraphics[width=0.45\columnwidth]{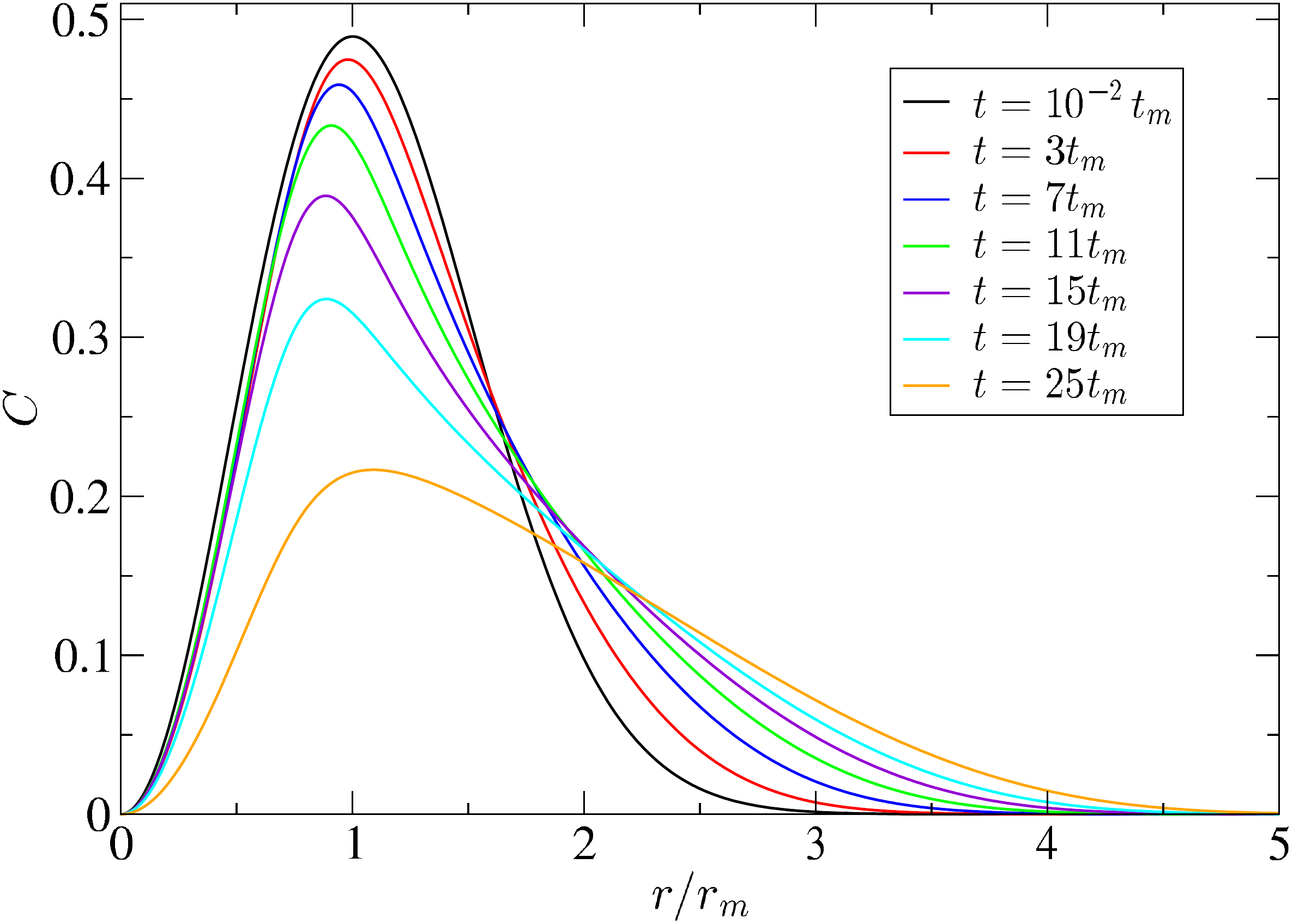}         
\includegraphics[width=0.45\columnwidth]{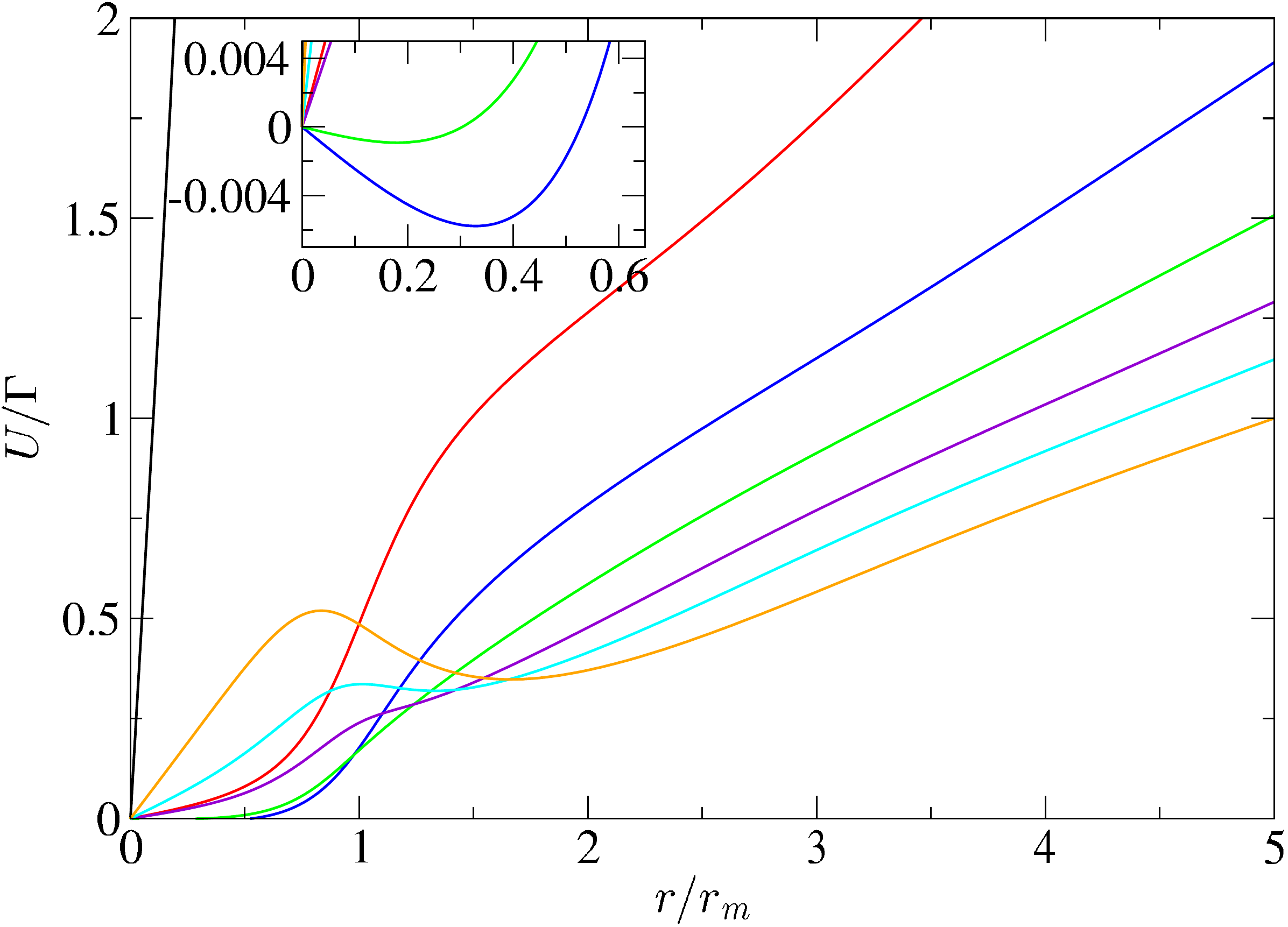} \hfill 
\includegraphics[width=0.45\columnwidth]{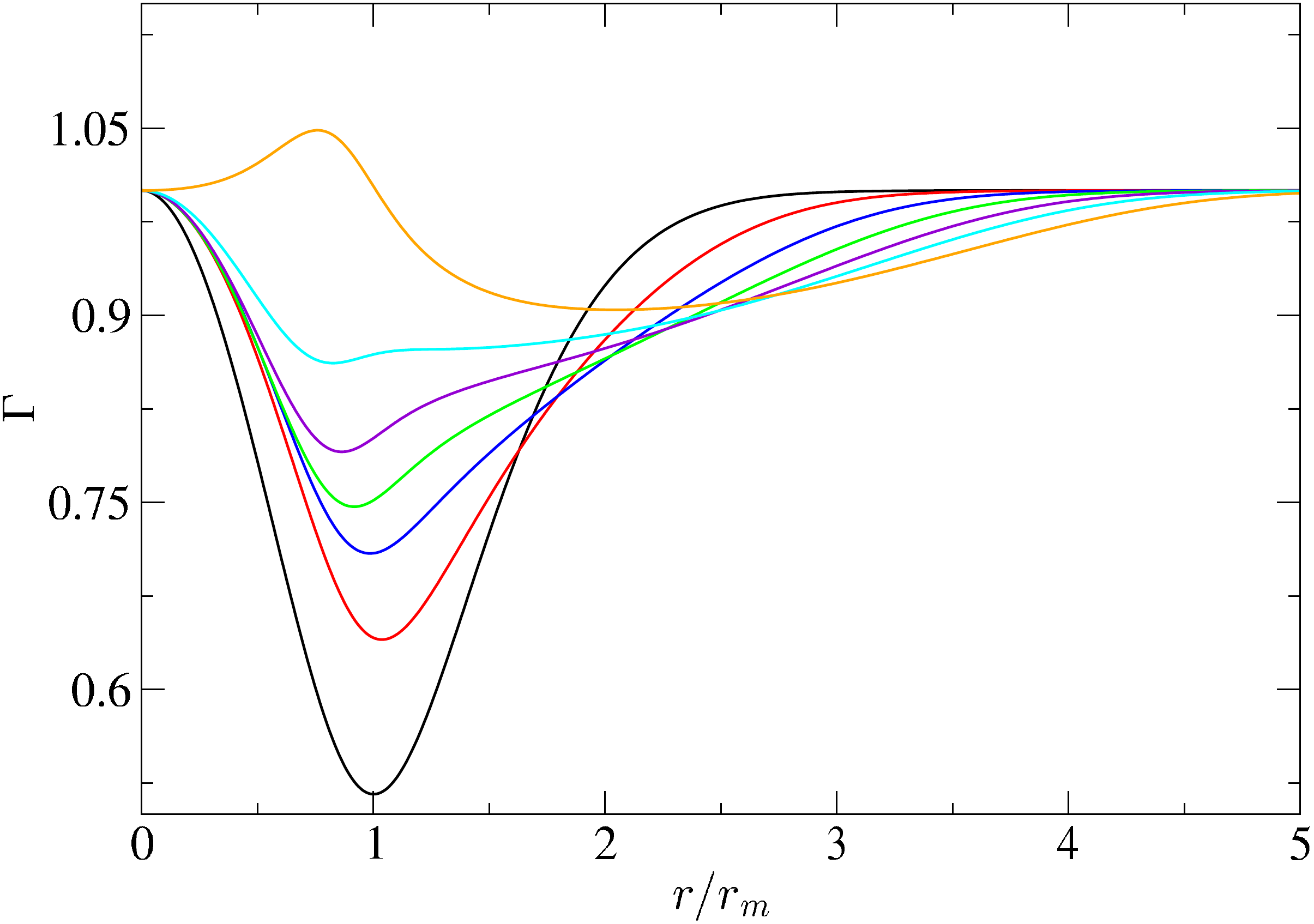}                               
\caption{
Dynamics of different magnitudes in the case of a subcritical fluctuation with $\delta_m=0.49$ (Gaussian profile) at given times $t$. In particular, $\delta_m=0.49$ and $\delta_{c} = 0.49774\pm 2\times 10^{-5}$.
} 
\label{fig:2_evolutionsubcritical}                   
\end{center}                     
\end{figure}

\begin{figure}[H]                   
\begin{center}                    
\includegraphics[width=0.45\columnwidth]{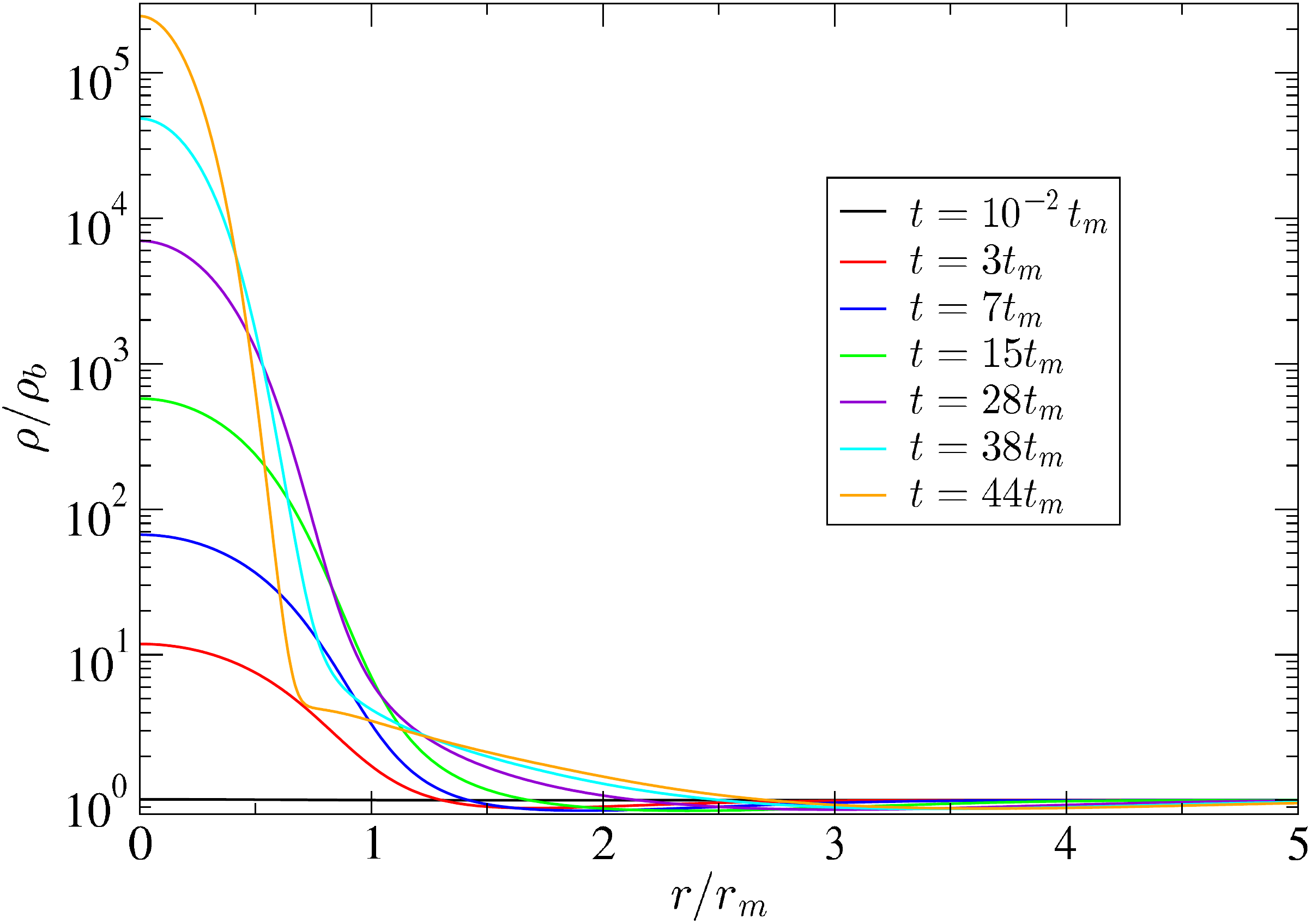}\hfill
\includegraphics[width=0.45\columnwidth]{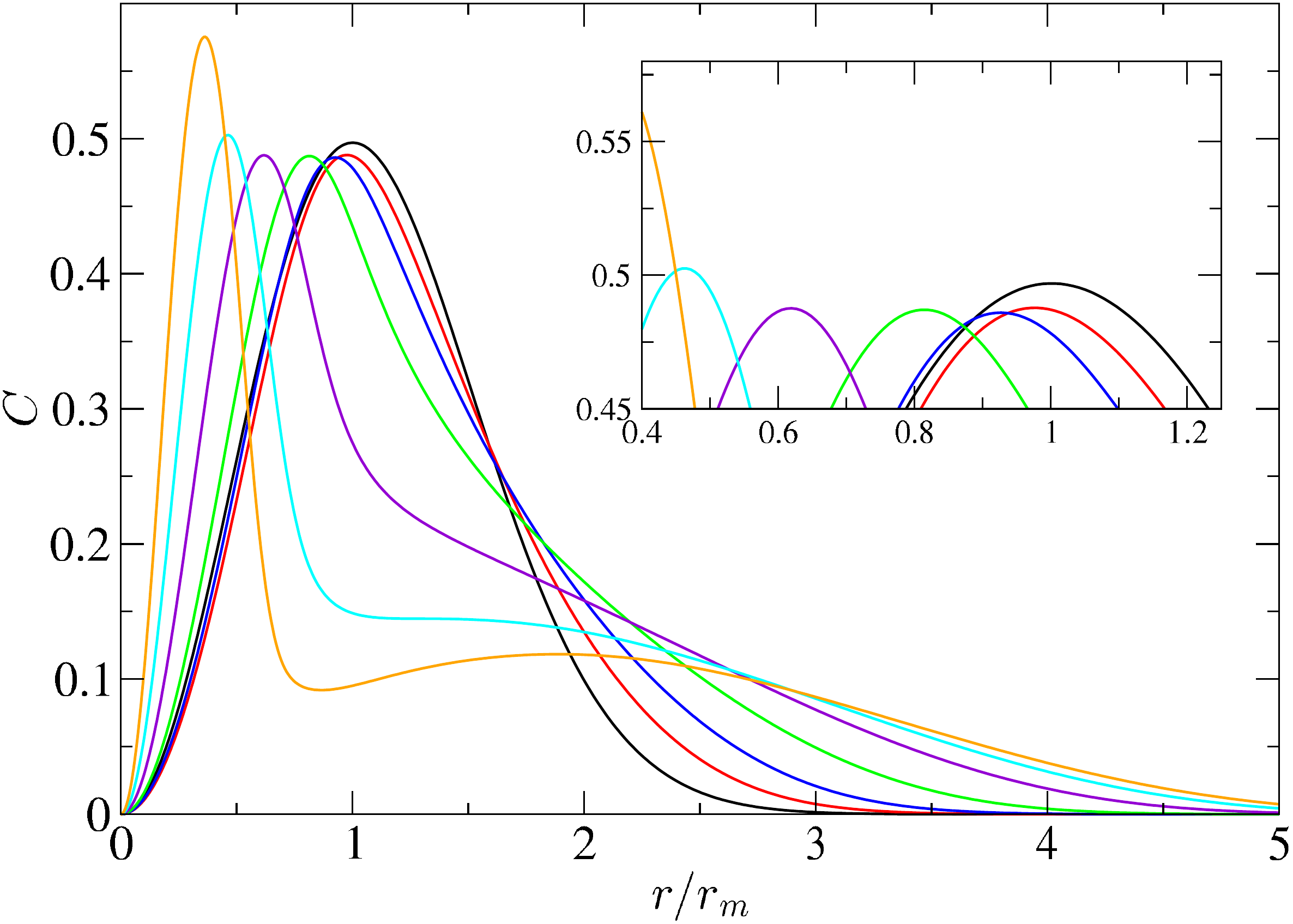}         
\includegraphics[width=0.45\columnwidth]{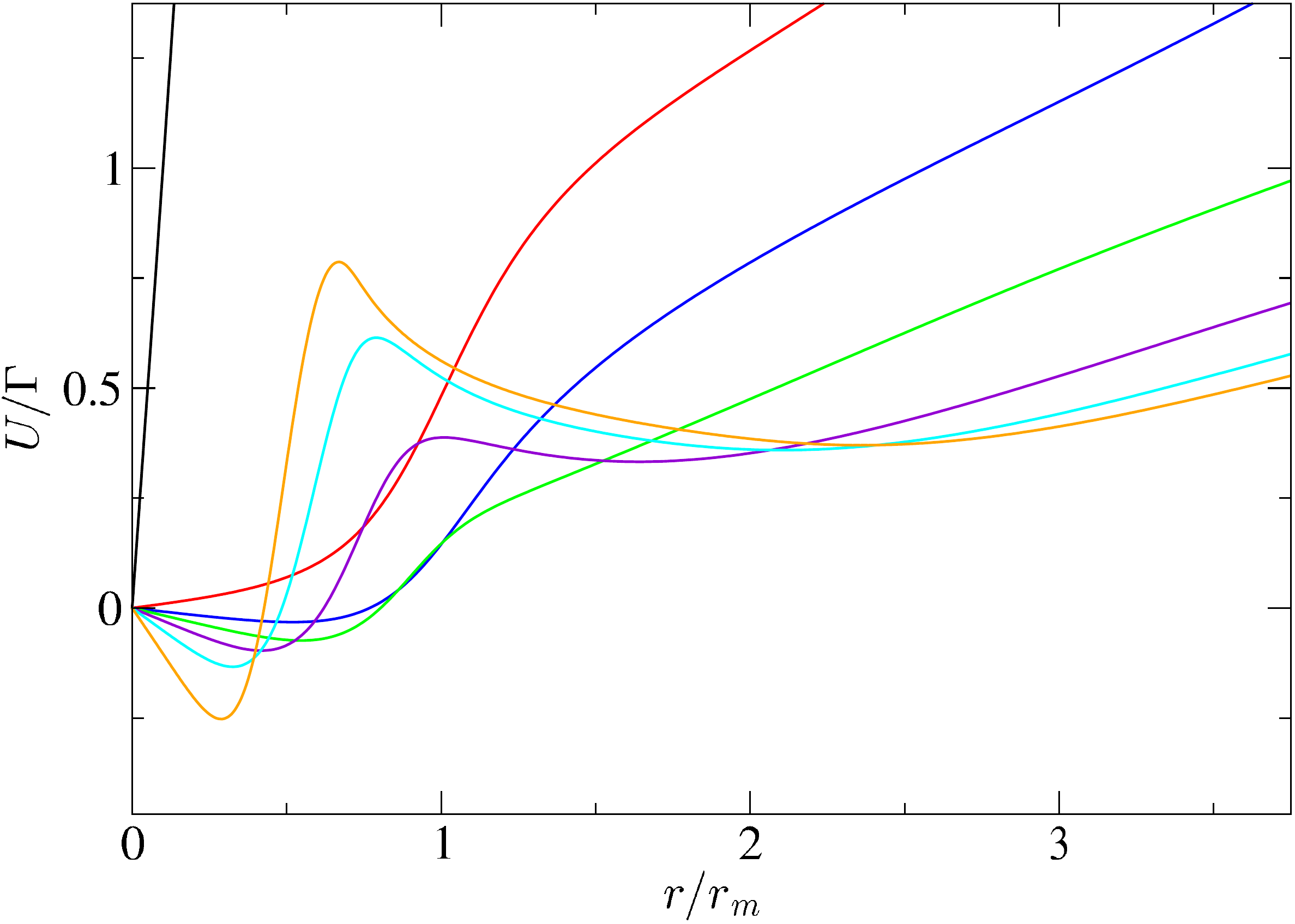}\hfill 
\includegraphics[width=0.45\columnwidth]{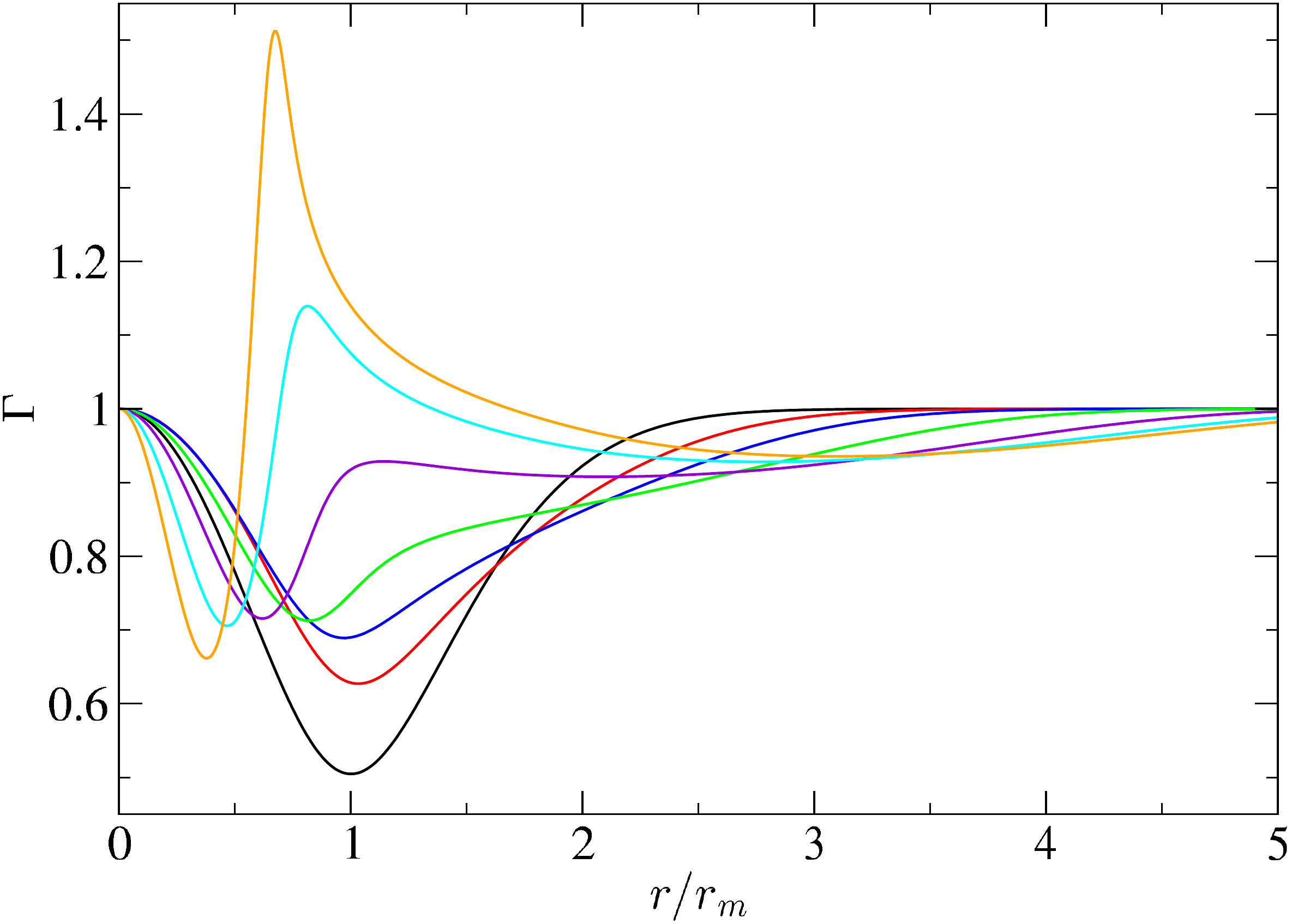}                               
\caption{Dynamics of different magnitudes in the case of a perturbation close to the critical threshold $\delta_m \approx \delta_{c}$ (Gaussian profile) at given times $t$. In particular, $\delta_m=0.49775$ and $\delta_{c} = 0.49774\pm 2\times 10^{-5}$.} 
\label{fig:2_evolutionclose}                   
\end{center}                     
\end{figure}

In this subsection, we show the behaviour of the gravitational collapse of the curvature fluctuations for three different regimes. As an example, a Gaussian profile given by Equation~\eqref{eq:4_lamda} with $q=1$ ($\lambda=0$) and for different values of $\delta_m$ and $w=1/3$ is considered. However, the same qualitative behaviour can be found for other profiles and other $w \neq 0$. In particular, three different regimes can be considered: subcritical ($\delta_m < \delta_{c}$), supercritical ($\delta_m > \delta_{c}$) with $\mid \delta_m-\delta_c \mid \gg O(10^{-3})$ in both previous cases, and critical $\mid \delta_m-\delta_c \mid \leqslant O(10^{-3})$. In Figure~\ref{fig:2_evolutioncritical} (supercritical), Figure~\ref{fig:2_evolutionsubcritical} (subcritical), and Figure~\ref{fig:2_evolutionclose} (critical), we see the evolution of the variables $\rho, \Gamma, U$, and $\com$.

In the case of a supercritical evolution, which is the case of Figure~\ref{fig:2_evolutioncritical}, we can observe that the compaction function grows during the evolution. The gravitational collapse increases the energy density on the central regions, and therefore, the mass excess grows. The formation of the two horizons is also obvious, as discussed in Section \ref{sec:horizons}. The outer horizon moves outwards, and the inner one moves inwards, faster than the outwards one. The simulation breaks down when the inner horizon approaches $r=0$. This is when the excision technique has to be applied to remove the singularity.

In Figure~\ref{fig:2_evolutionsubcritical} (the subcritical case), the compaction function decreases continuously as the perturbation is diluted away due to the superiority of the pressure gradients against gravity. In general, for late times, the peak $\com_{\rm max}$ is pushed outwards due to the pressure gradients. This behaviour can also be seen with the peak value of $\rho/\rho_{b}$. In the supercritical case, the peak value is continuously increasing. In contrast, in the subcritical case, the peak value starts to decrease at some moment (which means the perturbation starts to disperse). In addition, from Figures~\ref{fig:2_evolutioncritical}--\ref{fig:2_evolutionclose}, it can be observed that $\Gamma$ is not constant during the evolution. As expected, the gradient expansion approximation fails for sufficiently late times $ t \gg t_{0}$.

\unskip

In Figure~\ref{fig:2_evolutioncritical} (supercritical case), we observe that $U/\Gamma$ decreases quickly in time, and a negative velocity mainly dominates the profile before $r_m$ for late times, which indicates the collapse of the fluid. Instead, in Figure~\ref{fig:2_evolutionsubcritical} (subcritical case), only a small negative value $U/\Gamma$ is reached for early times, and beyond that, the values are positive, which indicates that the perturbation is dispersing, avoiding the gravitational collapse. 

The most relevant behaviour is found in the critical regime, in Figure~\ref{fig:2_evolutionclose}. Here, the fluid is divided into two parts, one going outwards (positive $U$) and one inwards (negative $U$), which generates an under-dense region. This under-dense region re-attracts the fluid with the net effect of a compression and rarefaction process, which increases its velocity in time. In this situation, the gradients become very large, and eventually, the simulation breaks down if a sufficient resolution has not been set up for the simulation. 

It is interesting to mention that in the presence of secondary peaks in $\com(r)$ (not only one isolated one), with an amplitude roughly equal to or less than the first one, it was verified numerically in \cite{garrigavicentejudithescriva} that those secondary peaks decrease and disperse on the FLRW background, even in the case of supercritical perturbations. A different situation could happen when a secondary peak is sufficiently high. In such a case, the secondary peak could also collapse even if the first peak is not greater than the threshold value to collapse itself.

\begin{figure}[H]                    
\begin{center}                    
\includegraphics[width=0.49\columnwidth]{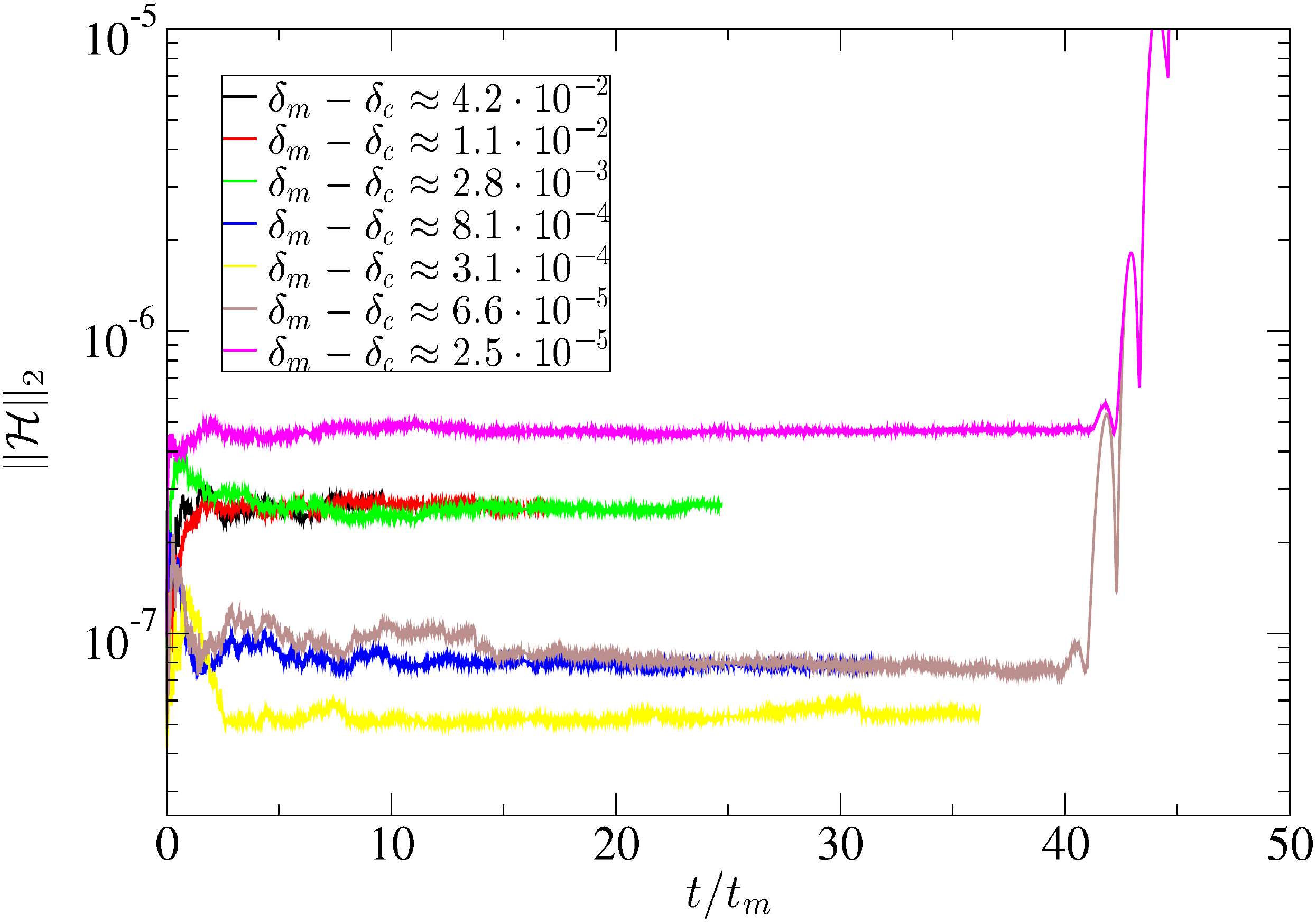}
\includegraphics[width=0.49\columnwidth]{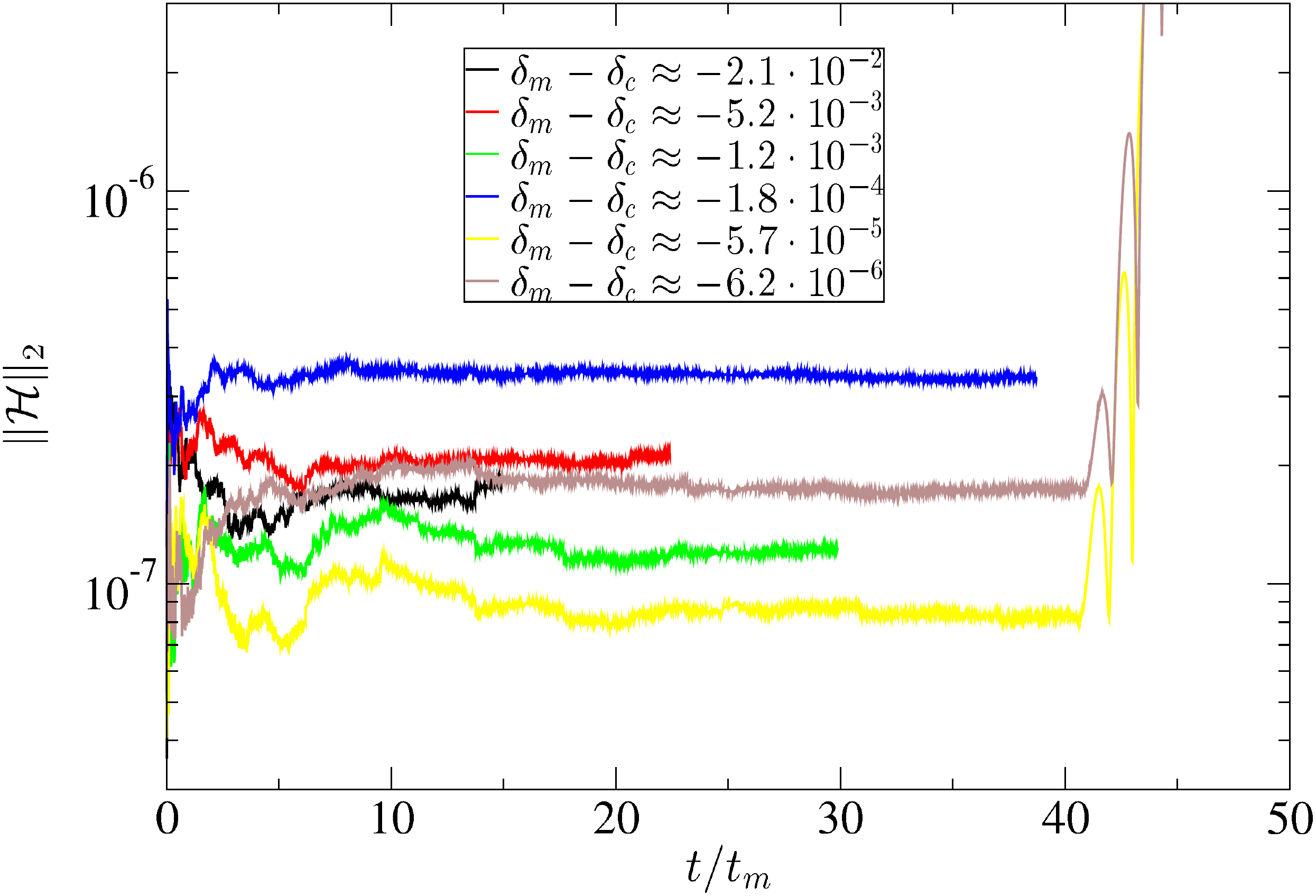}         
\caption{
\textbf{Left-panel}: Hamiltonian constraint for the numerical iterations of the bisection procedure in the case of a profile given by a Gaussian fluctuation for the supercritical case with Equation~\eqref{eq:4_lamda} and $q=1$, $\lambda=0$. \textbf{Right-panel}: The same as the left panel, but for subcritical fluctuations. The initial Hamiltonian constraint is subtracted for each evolution of $\delta_m$ in both cases.} 
\label{fig:2_constraintbisection}                   
\end{center}                     
\end{figure}

In Figure~\ref{fig:2_constraintbisection} is shown the correctness of the numerical evolution using the Hamiltonian constraint of Equation~\eqref{eq:2_constraint} at each step. It corresponds to different evolutions of the gravitational collapse, both subcritical and supercritical. In the figure, it can be seen that the constraint starts to be violated for late times for $(\delta_m-\delta_{c}) \approx O(10^{-5})$, which gives a bound on the maximal resolution for $\delta_{c}$ (in this particular example).

\begin{figure}[H]
\centering
\includegraphics[width=0.49\linewidth]{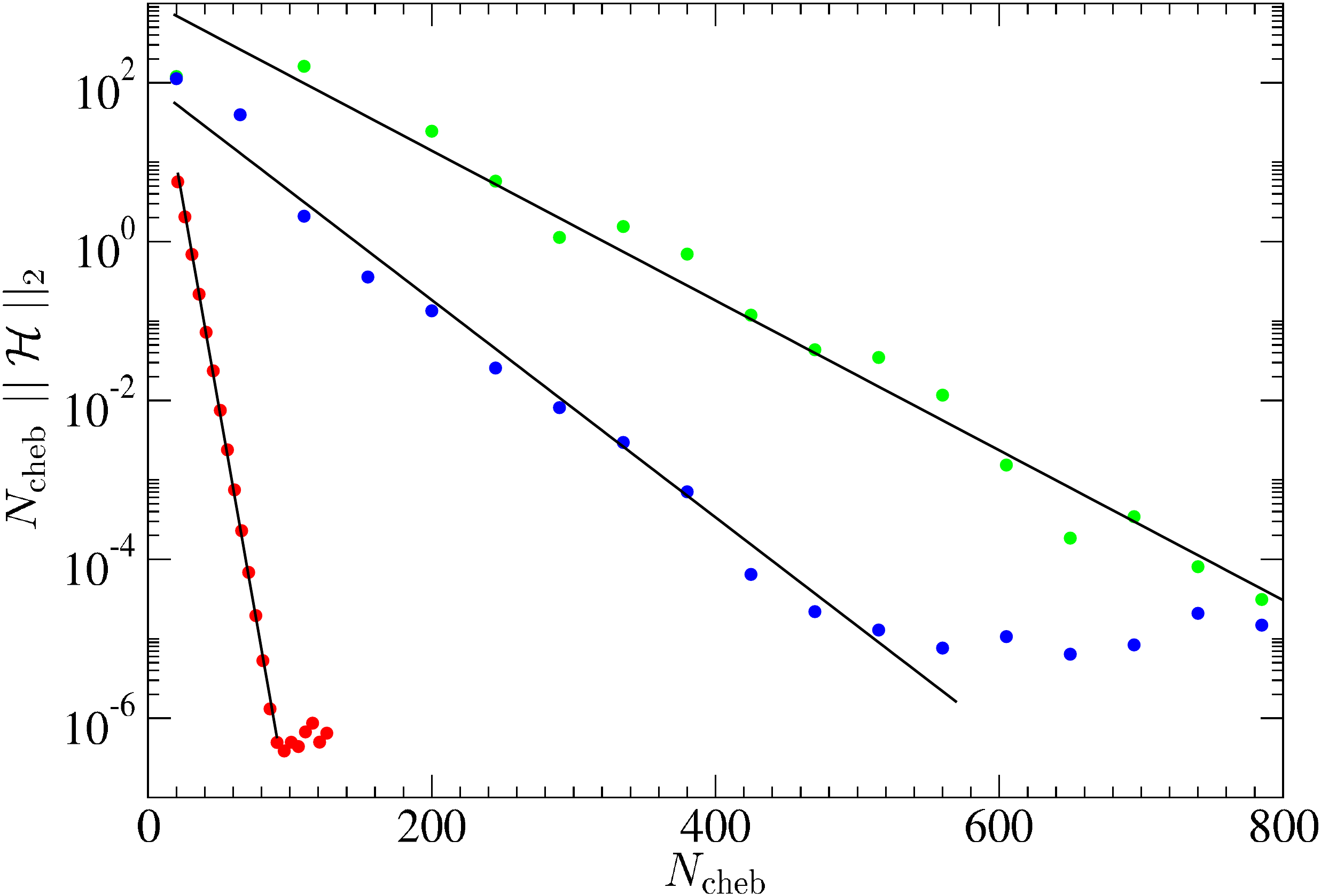} 
\includegraphics[width=0.49\linewidth]{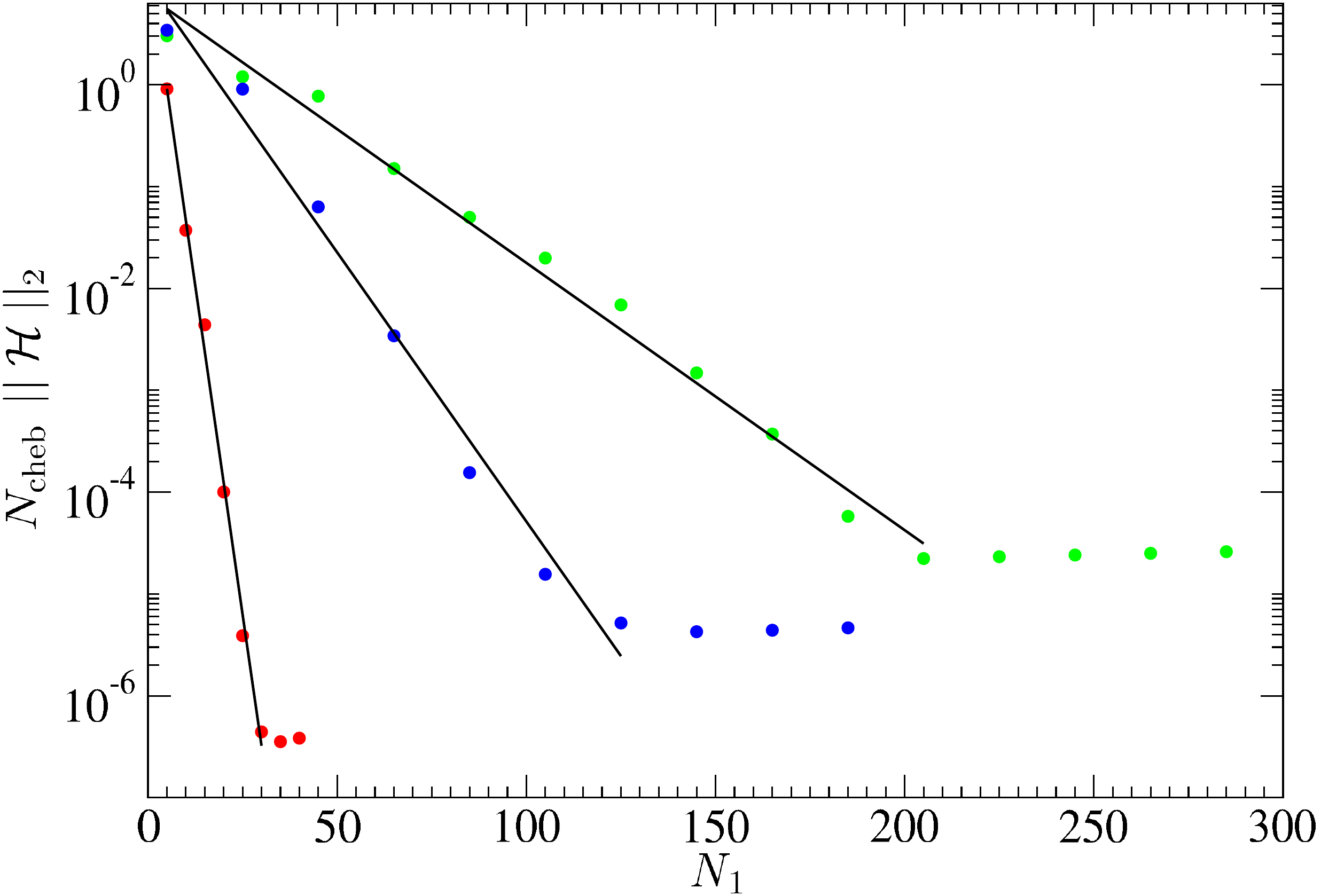} 
\caption{Spectral convergence for different profiles. Red points correspond to the profile of Equation~\eqref{eq:4_lamda} with $q=1$, blue points to $q=5$, and green points to $q=10$ ($\lambda=0$ in all cases). The black solid line is the exponential fit $\sim e^{-\alpha N_{\rm cheb}}$. \textbf{Left-panel}: Case with a single grid $N_{\rm cheb}$ and $\alpha \approx 0.23, 0.032, 0.022$ respectively for the previous cases. \textbf{Right-panel}: Case with two grids $N_1$ and $N_2=50$, with the second grid starting at $r=2 r_m$, and $\alpha \approx 0.59, 0.12, 0.06$, with the same profiles of the left panel. The simulation corresponds to $\delta_m=0.5$ and $t=t_H$ for $w=1/3$.}
\label{fig:2_spectral_acuracy}
\end{figure}

The convergence with the use of spectral methods can be found in Figure~\ref{fig:2_spectral_acuracy}. It is clear that the convergence is exponential taking into account the fits performed (see the caption of the figure). In particular, the use of a multidomain grid improves the performance and accuracy of the simulations substantially, as is shown in the right panel of Figure~\ref{fig:2_spectral_acuracy}, where the spectral accuracy is obtained with fewer points (in comparison with the left panel). Making the first grid $\Omega_{1}$ from $r=0$ up to $r=2 r_m$ allows capturing the region where large gradients are developed. Then, for the second grid, $\Omega_{2}$ goes from $r=2 r_m$ up to $r=r_f$, and only a few points are needed. 

\begin{figure}[H]                    
\begin{center}                    
\includegraphics[width=0.55\columnwidth]{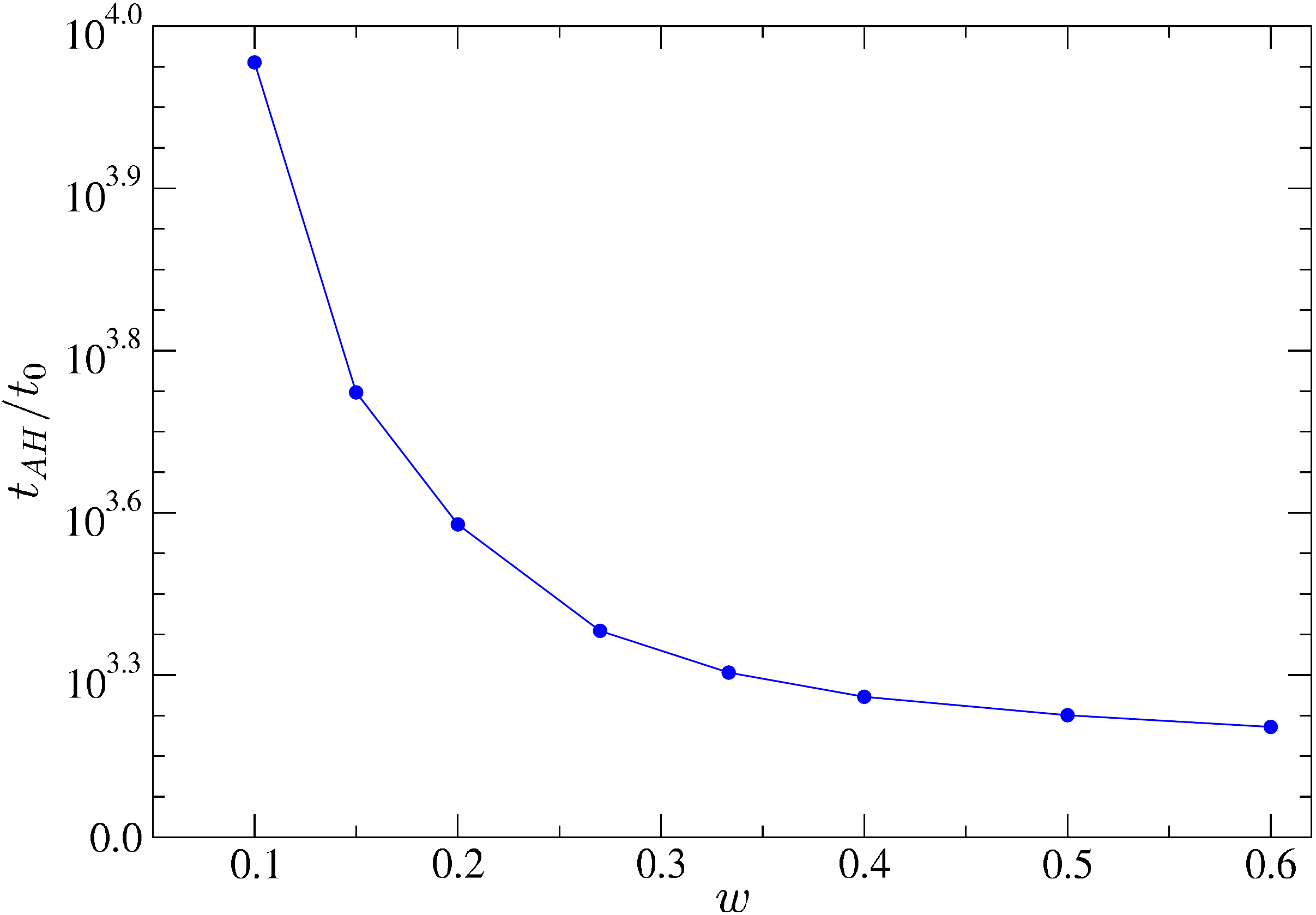}
\caption{Time that it takes the fluctuation to form the apparent horizon $t_{AH}$ in terms of $w$. The fluctuation chosen corresponds to Equation~\eqref{4_basis_pol} with $q=1$ and $\delta_{m}=\delta_{c}+10^{-2}$.} 
\label{fig:time_colapse_w}                   
\end{center}                     
\end{figure}

An interesting behaviour is found for $t_{AH}$ when $w$ is changed, as noticed in \cite{Escriva:2020tak}. Although the pressure gradients work against the gravitational collapse, since they are also a form of gravitational energy, it will mostly be favoured when the collapse is active \cite{Escriva:2020tak}. Consequently, this implies a shorter time of formation for larger $w$, as can be observed in Figure~\ref{fig:time_colapse_w}.

\unskip

\unskip

\subsection{Thresholds for PBH Formation}\label{sec:thresholds}

\begin{figure}[H]                    
\begin{center}                    
\includegraphics[width=0.49\columnwidth]{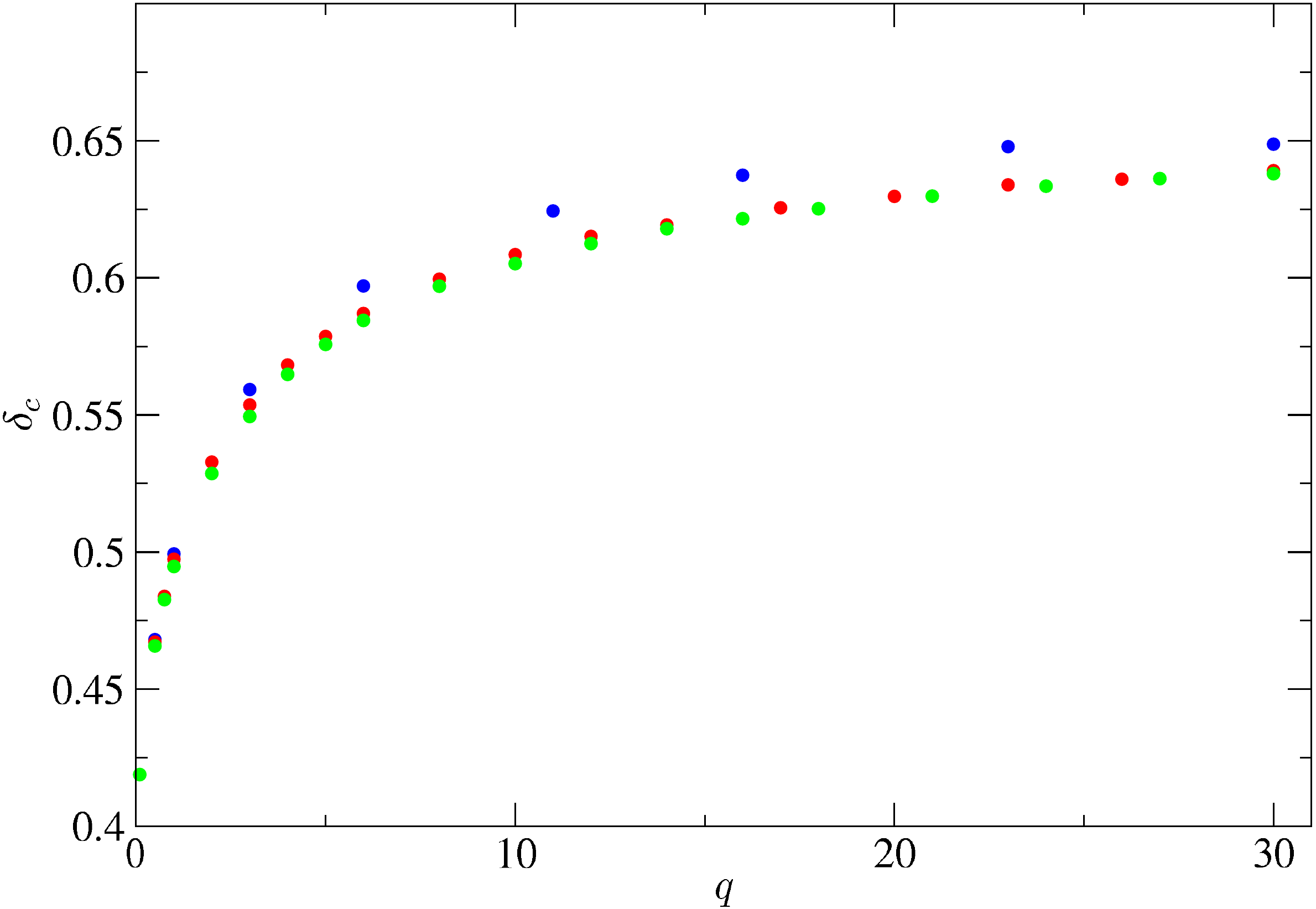}
\includegraphics[width=0.49\columnwidth]{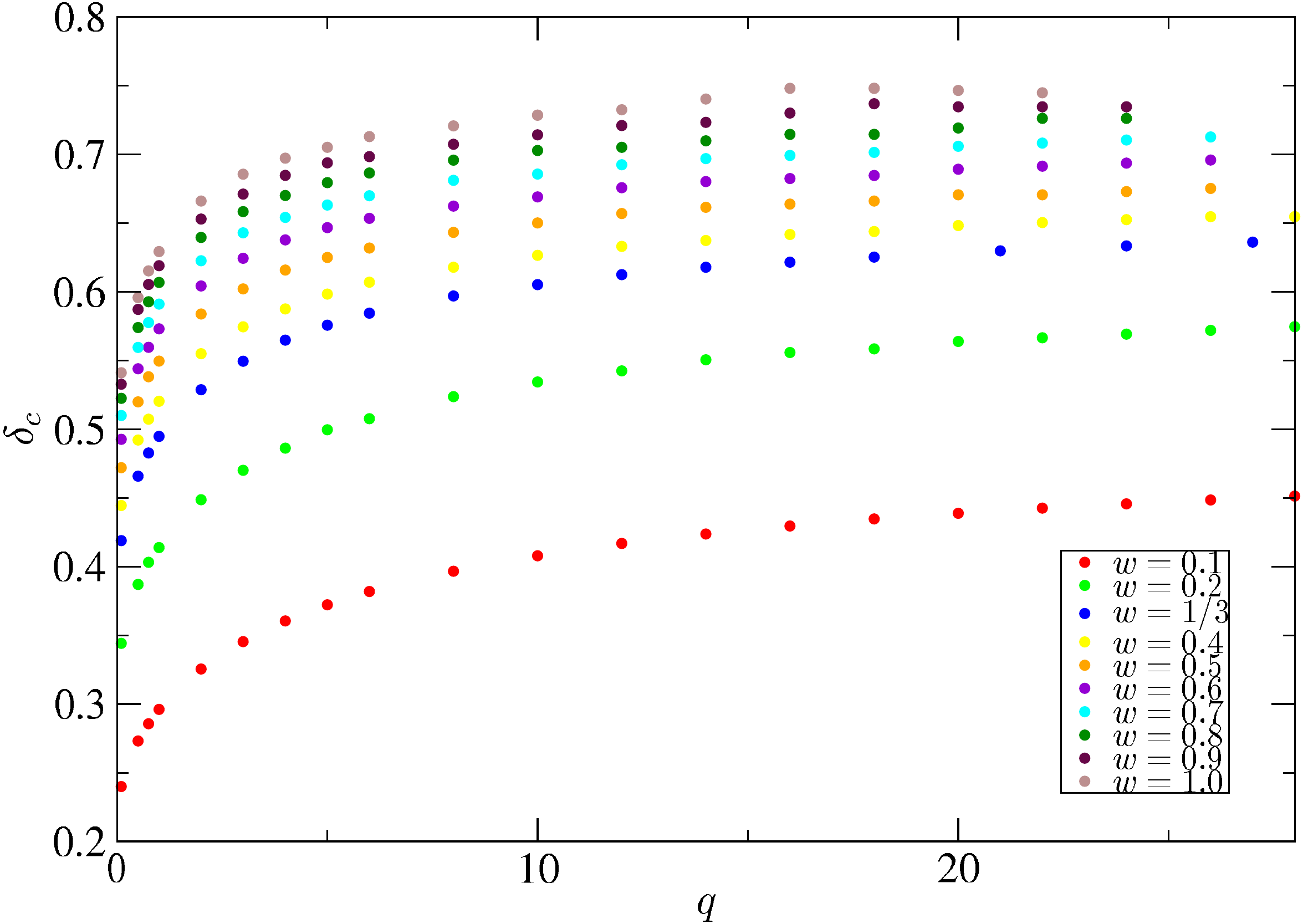}
\caption{\textbf{Left-panel}: Threshold values $\delta_{c}$ in terms of $q$ for the case of $w=1/3$. Green points correspond to Equation~\eqref{4_basis_pol}, red points to Equation~\eqref{eq:4_lamda} with $\lambda=0$, and blue to Equation~\eqref{eq:4_lamda} with $\lambda=1$. \textbf{Right-panel}: Threshold values $\delta_{c}$ in terms of $q$ for different values of $w$ corresponding to the polynomial profile of Equation~\eqref{4_basis_pol}.} 
\label{fig:thresholds}                   
\end{center}                     
\end{figure}

As we have mentioned, the threshold for PBH formation depends on the shape of the curvature fluctuation. Following the numerical procedure in Section \ref{numerical_procedure}, all the possible thresholds for PBH formation can be obtained using the basis polynomial profile of Equation~\eqref{4_basis_pol}. The results are shown in Figure~\ref{fig:thresholds} for different profiles. In the particular case of a radiation fluid, it was found in \cite{universal1,musco2018} that the range of all possible thresholds is $\delta_{c} \in [0.4,2/3]$. The minimum threshold $\delta_{c,\rm min}=0.4$ corresponds to the limit case of $q \rightarrow 0$ (broad profile in $\mathcal{C}$). Instead, the maximum threshold $\delta_{c,\rm max}=2/3$ corresponds to $q \rightarrow \infty$ (peaked profile in $\mathcal{C}$). Increasing $q$, the pressure gradients are larger, and therefore, the threshold value is higher.

Remarkably, the relative deviation between the thresholds for the different profiles in the left panel of Figure~\ref{fig:thresholds} with the same $q$ is within $2\%$. This was found for the first time in~\cite{universal1}, and it was shown that different profiles with the same $q$ (shape around the peak of the compaction function) have the same thresholds within a small deviation.

\begin{figure}[H]                    
\begin{center}                    
\includegraphics[width=0.49\columnwidth]{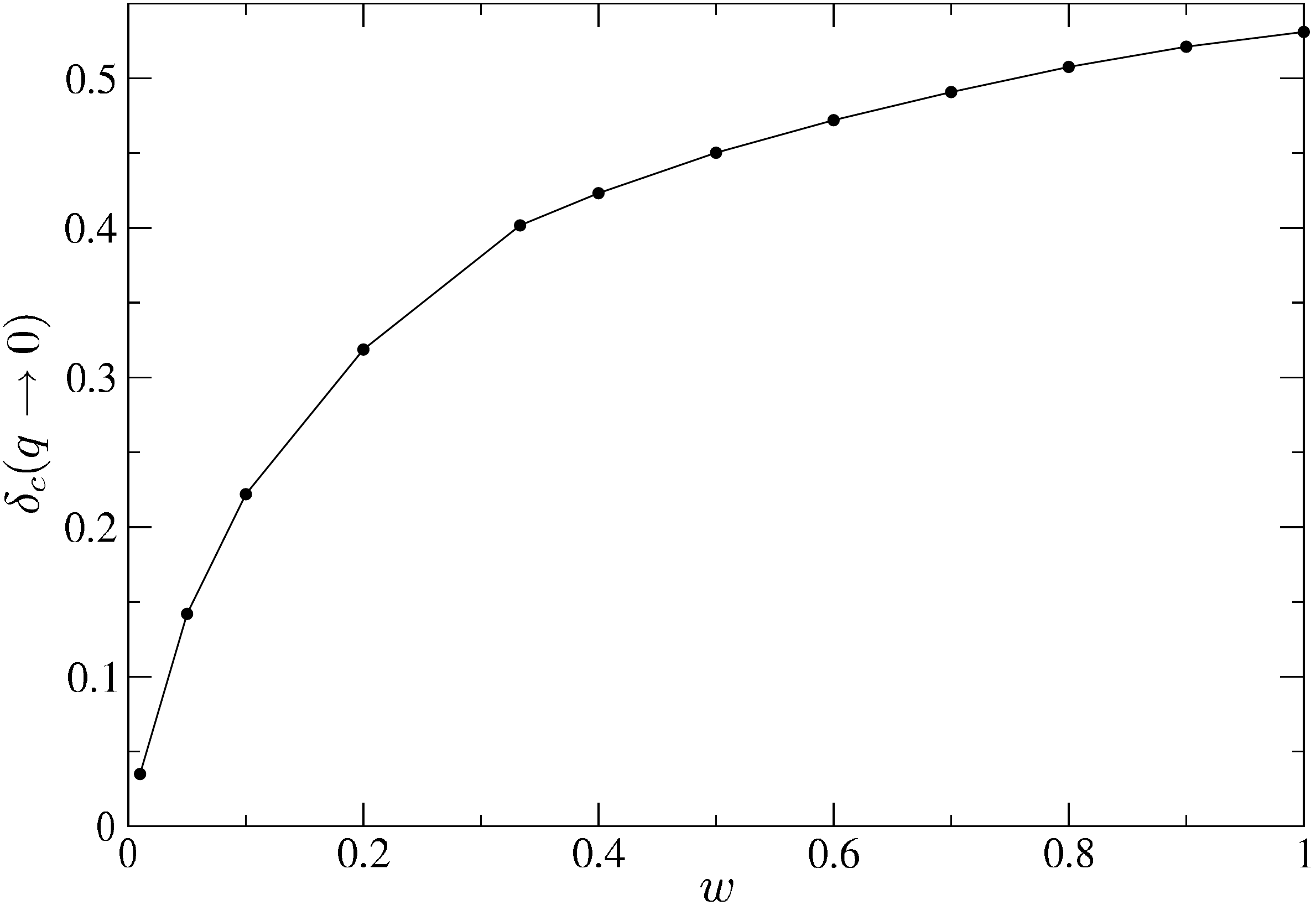}
\includegraphics[width=0.49\columnwidth]{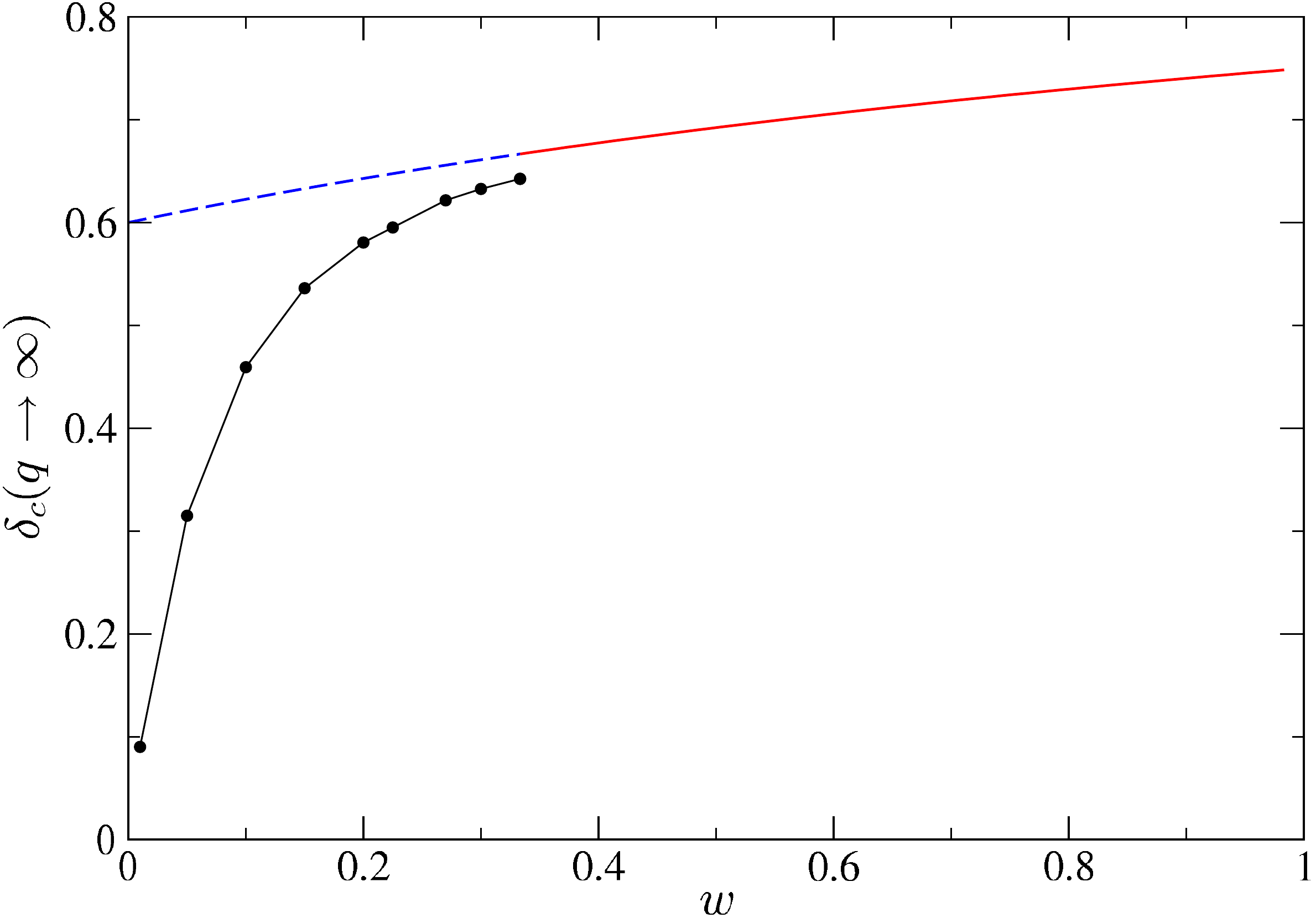}
\caption{\textbf{Left-panel}: Threshold $\delta_{c}$ in terms of $w$ for $q \rightarrow 0$. \textbf{Right-panel}: Threshold $\delta_{c}$ in terms of $w$ for $q \rightarrow \infty$; in particular, we are able to perform simulation up to $q=40$. The red line corresponds to the analytical case $f(w)$, where the black dots correspond to the numerical simulations in the region where $ \delta_c(q \rightarrow \infty) \neq f(w)$, as shown in \cite{Escriva:2020tak}. The blue dashed line corresponds to $f(w)$ on that~region.} 
\label{fig:thresholds_limits}                   
\end{center}                     
\end{figure}

In \cite{Escriva:2020tak}, simulations of PBH formation in the case of a perfect fluid with $w \in (0,1]$ were performed. The functional behaviour of $\delta_{c}$ in terms of $q$ for different $w$'s is roughly similar, as is shown in the left panel of Figure~\ref{fig:thresholds} (the case with fixed $w=1/3$), but quantitatively, the values are different. It is clear that increasing $w$, the gradients will be larger, and therefore, the threshold increases. The limits $q \rightarrow 0$ and $q \rightarrow \infty$ for the general case in terms of $w$ are both amenable to further analysis: 

\begin{itemize}

\item The case $q \rightarrow 0$ corresponds to a $\com$ that becomes approximately constant over a wide range of scales. Simulations, in this case, become difficult due to the presence of conical singularities \cite{Escriva:2020tak}. To solve this issue, in \cite{Escriva:2020tak}, a modified profile of Equation~\eqref{4_basis_pol} was used to subtract the mass excess beyond $r \gg r_m$ and fulfil that at the boundaries, the FLRW background is recovered. In Figure~\ref{fig:thresholds_limits} are shown the numerical values of $\delta_{c}(q \rightarrow 0)$ in terms of $w$, which show a strong dependence on $w$. As expected, for $w \rightarrow 0$, the threshold $\delta_{c} \rightarrow 0$;

\item The case of large $q \rightarrow \infty$ corresponds to the case of a sharply peaked profile for the compaction function. For this kind of profile, the pressure gradients acting against the gravitational collapse are maximal, and hence, the compaction function should be maximal as well. Analytically, this is shown by Equation~\eqref{C_r} that $\com(r_m) < f(w)$ (at super-horizon scales). As was proven numerically in \cite{musco2018}, $\delta_{c, \rm max} \rightarrow f(w)$ in the case of radiation $w=1/3$. The intuition is that the saturation should persist to larger values of $w$ because larger values of $w$ also imply larger pressures that fight against the collapse. Therefore, for $w \geq 1/3$, the compaction function of a peaked profile must also saturate the bound $f(w)$. This was verified numerically in \cite{escriva_solo}. This behaviour is no longer true for $w < 1/3$, something obvious in the case of dust $w=0$, since $f(0)=0.6$, but $\delta_{c}(w=0)=0$ always for any profile. This is shown in the left panel of Figure~\ref{fig:thresholds_limits}. It should be taken into account that beyond $q>40$, the simulations do not obtain the threshold with enough resolution. This is the reason why the numerical value of the threshold for $w=1/3$ does not exactly coincide with $f(1/3)$. One expects that in the asymptotic limit, {the threshold for} $ q \rightarrow \infty$ should match with $f(1/3)$, as was pointed out in \cite{musco2018} with the same argument.
\end{itemize}

\begin{figure}[H]                    
\begin{center}                    
\includegraphics[width=0.55\columnwidth]{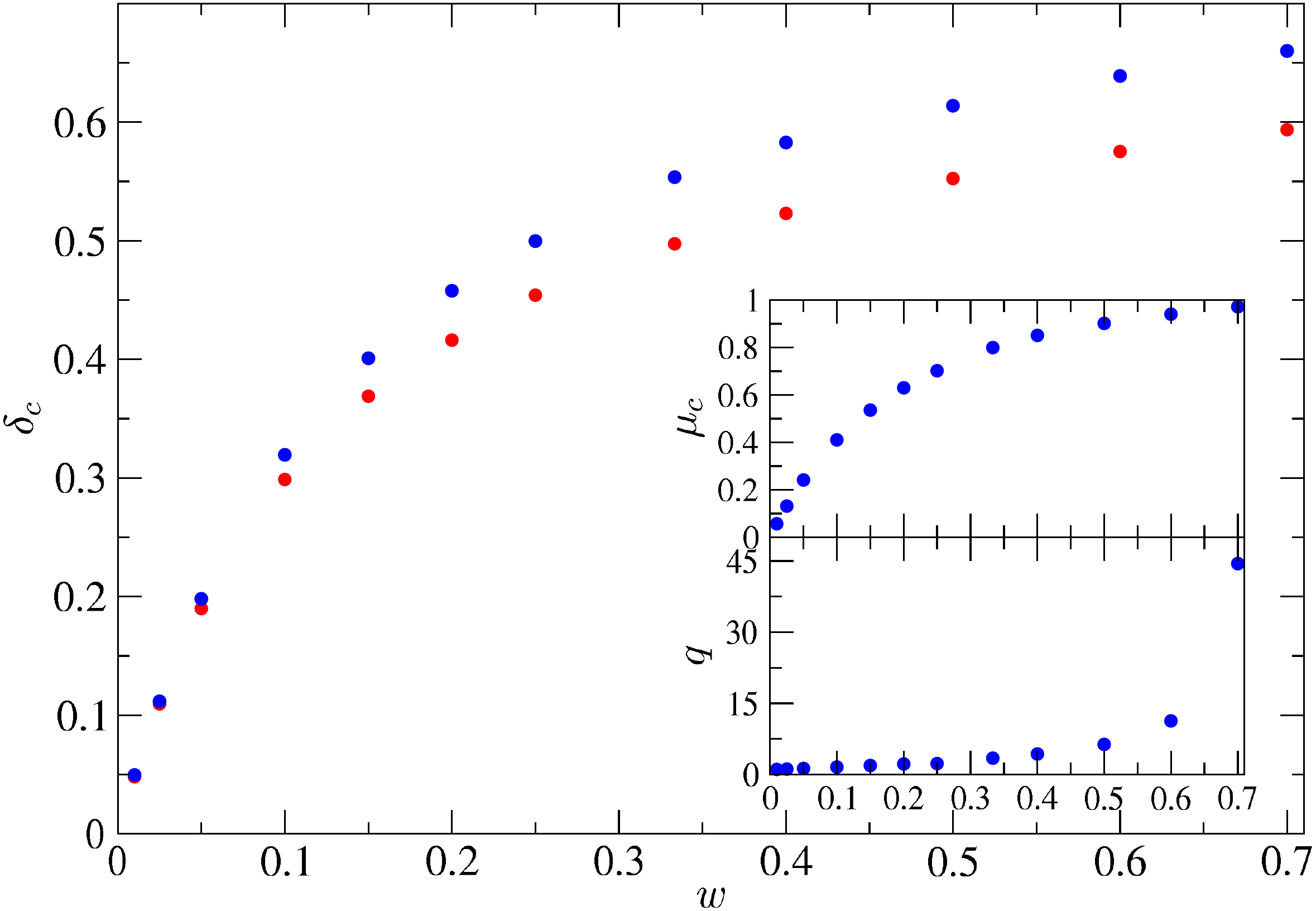}
\caption{Thresholds $\delta_c$ in terms of $w$ for two profiles. Red dots correspond to the Gaussian profile with $K(r)$ and blue dots to $\zeta(\tilde{r})$ instead. The \textbf{top-subplot} is the critical amplitude $\mu_{c}$ of the profile in $\zeta(\tilde{r})$. The \textbf{bottom-subplot} is the value $q$ of the critical profile $\zeta(\tilde{r})$ for each $w$.}
\label{fig:thresholds_zeta}                   
\end{center}                     
\end{figure}

It is interesting also to notice the difference between considering $K(r)$ and $\zeta(\tilde{r})$ {for the threshold and PBH mass}. As we have seen in Section \ref{sec:set_up}, there is a non-linear relation between $\zeta$ and $\com$, which leads to some differences between the thresholds obtained for both profiles. Consider for instance a Gaussian profile in both cases, i.e., $K(r) = \mathcal{A}\, e^{-(r/r_m)^{2}}$ and $\zeta(r) = \mu\, e^{-(r/r_m)^{2}}$. Due to the non-linear relation between $\com$ and $\zeta$ at super-horizon scales, the $\delta_{m,\zeta}$ (for this example, let us call it $\delta_{m,K}$ and $\delta_{m,\zeta}$ for both profiles) depends non-linearly on $\mu$ as,
\begin{align}
\delta_{m,K} = f(w) \mathcal{A}r^{2}_me^{-1},\nonumber \\
\delta_{m,\zeta} = 4 f(w) (e-\mu)\mu e^{-2}.
\label{deltam_zeta}
\end{align}
Due to this, the $q_{\zeta}$ value in the case of the fluctuation $\zeta$ is dependent also on its amplitude $\mu$ and $w$, 
\begin{equation}
q_{\zeta}= \frac{e^{2}(e-2 \mu)}{(e-\mu)(e^{2}-6f(w)\mu+6f(w)\mu^{2})},
\end{equation}
when for the $K(r)$ in this example, it is just $q=1$. A plot of the thresholds $\delta_{c}$ for both profiles can be found in Figure~\ref{fig:thresholds_zeta} for different $w$'s. Although both profiles are Gaussian, they have a different $\delta_c$ due to the non-linear relation with $\zeta(\tilde{r})$. The threshold for $\zeta(\tilde{r})$ is higher than $K(r)$ since the shape around the compaction function is sharper, which means a larger $q$, as can be observed in the bottom panel of the subplot in Figure~\ref{fig:thresholds_zeta}. It also has an implication for the PBH mass, as we will see in the next subsection. 

A Gaussian family of profiles with $\zeta(\tilde{r}) = \mu e^{-(r/r_m)^{2p}}$ was considered in \cite{Young:2019yug} for the estimation of PBH abundances taking into account the non-linearity of $\zeta(\tilde{r})$ for $w=1/3$. There, a range of thresholds was found, $0.442 < \delta_c < 0.656$ for $0.34 \lesssim p \lesssim 2$.

\subsection{PBH Masses}

\begin{figure}[H]
\centering
\includegraphics[width=0.6\linewidth]{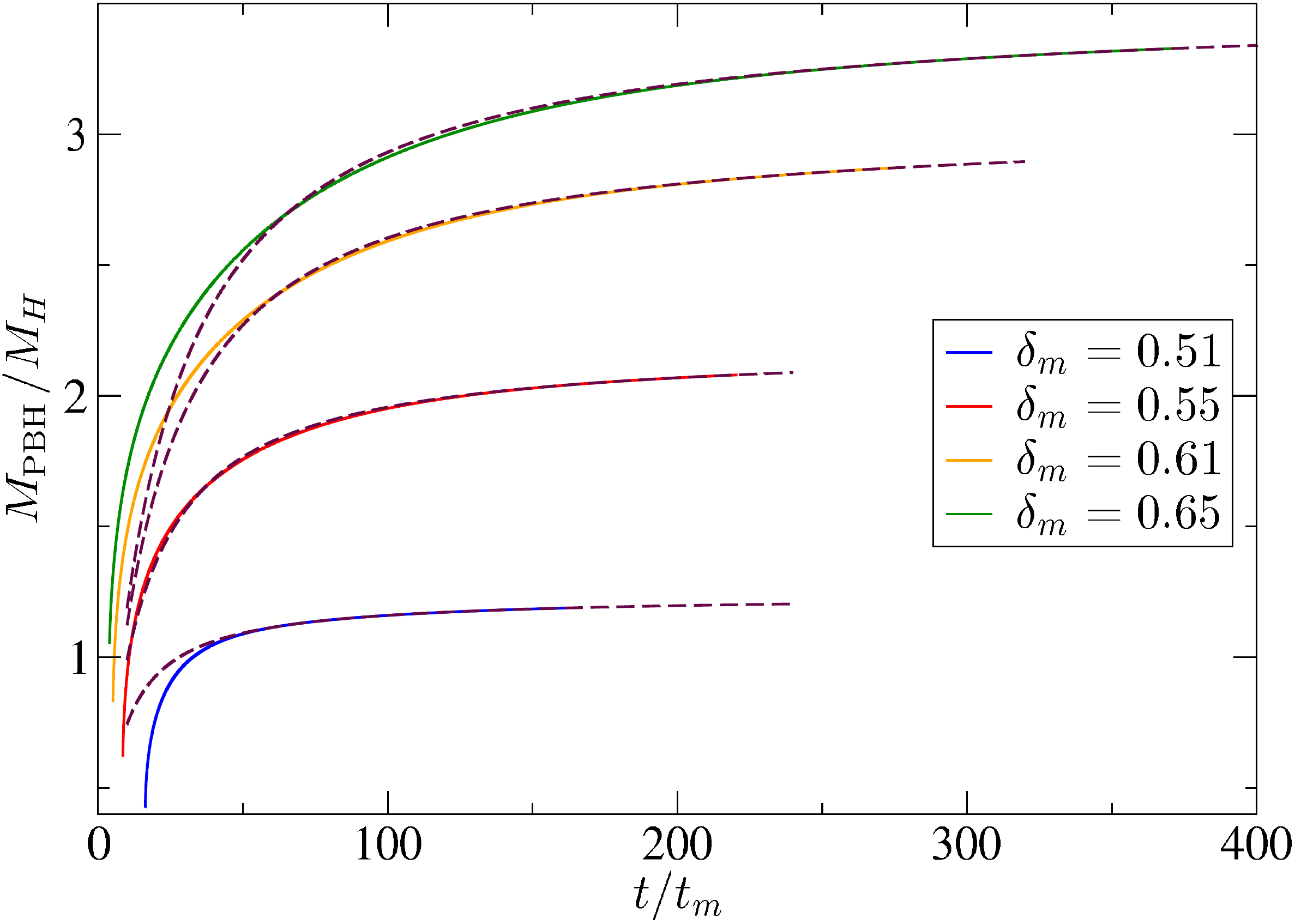} 
\caption{Evolution of the mass of the PBH for $w=1/3$ after the formation of the apparent horizon for the profile of Equation~\eqref{eq:4_lamda} with $q=1$ and $\lambda=0$. The colour lines correspond to different values of initial amplitudes $\delta_m$. The dashed line corresponds to the analytical fit using Equation~\eqref{eq:2_acretationformula}. Threshold value $\delta_{c} = 0.49774\pm 2\times 10^{-5}$.}
\label{fig:2_mass-evolution}
\end{figure}

Once the apparent horizon is formed, an excision technique following Section \ref{sec:2_excision} can be applied to remove the singularity. An example of the evolution of the PBH mass $M_{PBH}(t)$ for the profile of Equation~\eqref{eq:4_lamda} with $q=1$, $\lambda=0$, and different amplitudes $\delta_m$ can be found in Figure~\ref{fig:2_mass-evolution}. The initial time of the curves corresponds to the initial mass of the PBH $M_{PBH,i}$ at the time of formation of the apparent horizon $t_{AH}$. As can be seen in Figure~\ref{fig:2_mass-evolution}, when the amplitudes $\delta_m$ increase, the PBH mass will be higher.
The amount of time needed until reaching the stationary regime where the PBH mass remains constant could be substantial and therefore computationally expensive. As we have mentioned, to avoid that, Equation~\eqref{eq:2_acretationformula} can be used to estimate the final mass of the PBH after the accretion process accurately. 

\begin{figure}[H]
\centering
\includegraphics[width=0.55\linewidth]{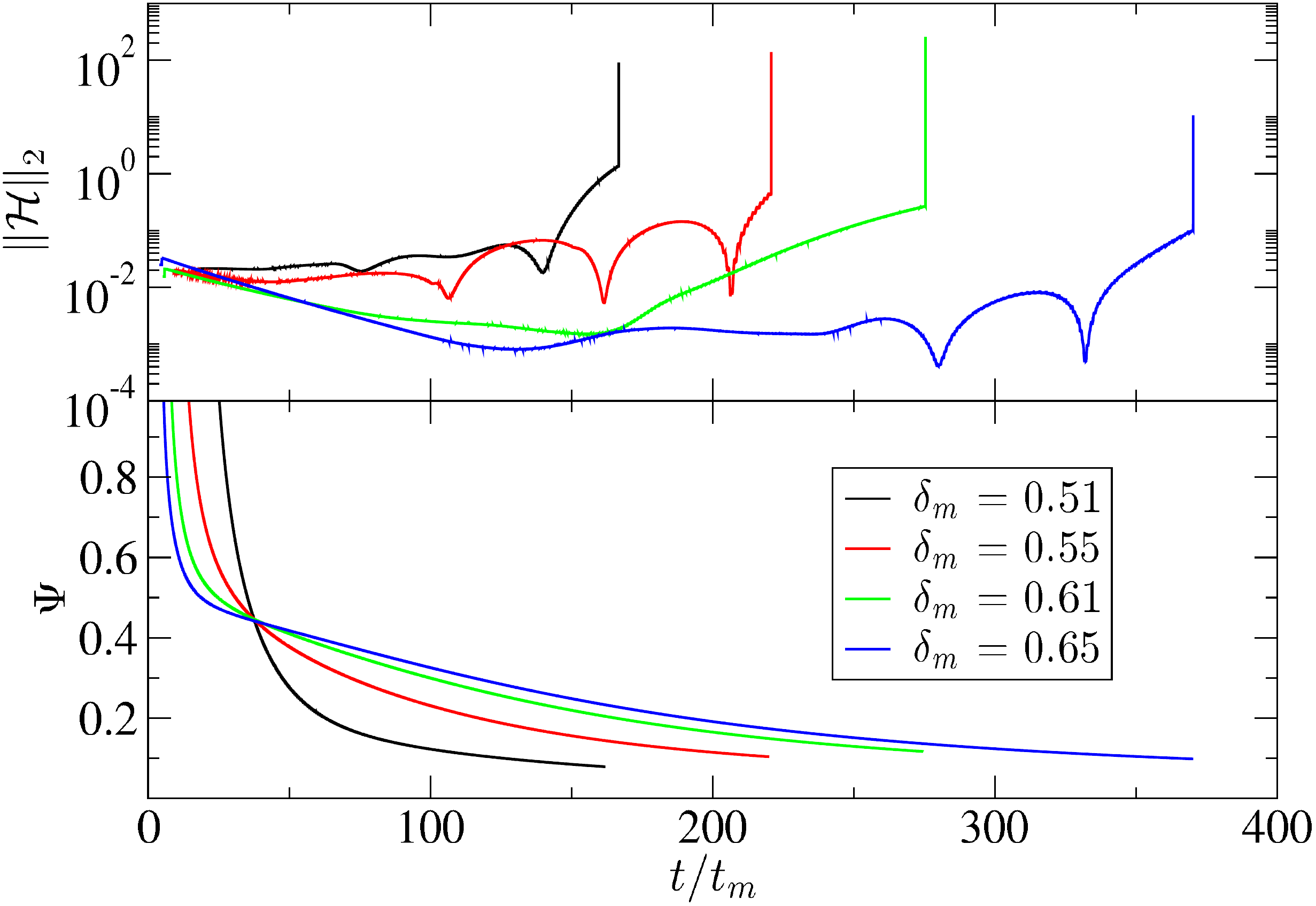} 
\caption{\textbf{Top-panel}: Time evolution of the Hamiltonian constraint for different amplitude values $\delta_m$ during the excision procedure. \textbf{Bottom-panel}: Numerical evolution of $\Psi$ in time. The crossing point is around $t/t_{m}\approx 37.5$. The profile used is the same as in Figure~\ref{fig:2_mass-evolution}. {The excision starts at very early times, at $t \approx 10 t_m$.}}
\label{fig:2_constraint-excision}
\end{figure}

\begin{figure}[H]
\centering
\includegraphics[width=0.6\linewidth]{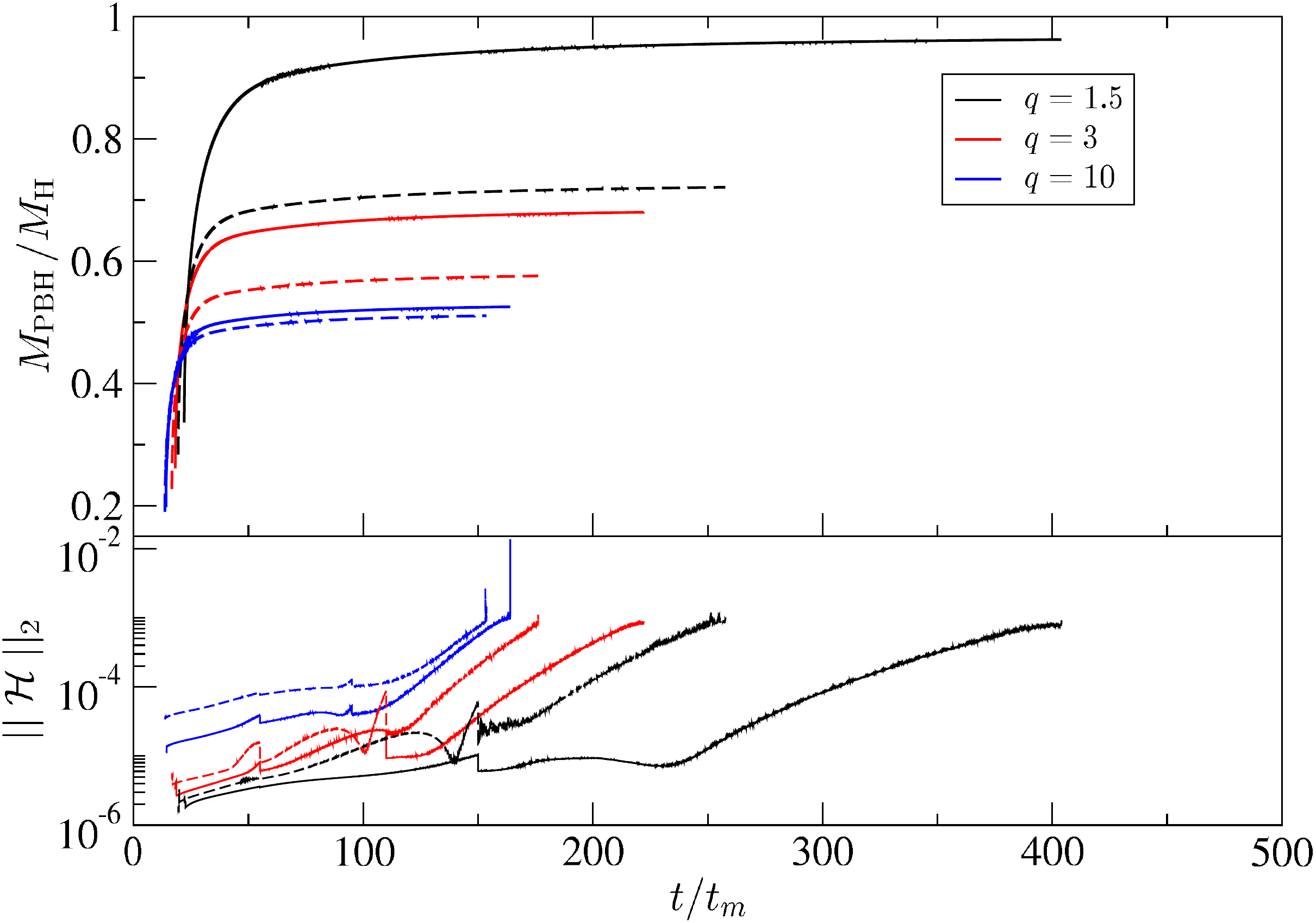} 
\caption{\textbf{Top-panel}: Evolution of $M_{PBH}(t)$ for the profiles $q = 1.5$, $q = 3$, and
$q = 10$. The dashed line corresponds to Equation~\eqref{eq:4_lamda} with $\lambda = 0$ and the solid line to Equation~\eqref{4_basis_pol}. \textbf{Bottom-panel}:
Time evolution of the Hamiltonian constraint. In all cases, $\delta_m-\delta_{c} = 5 \times 10^{-3}$.}
\label{fig:mases_extra2}
\end{figure}

In order to find the domain in time where the approximation of Equation~\eqref{eq:2_acretationformula} is valid, in \cite{escriva_solo}, the ratio of the increment in time of the black hole mass with respect to the Hubble scale $\Psi = \dot{M}/H M$ was computed, which is predicted to be $\Psi<1$ when the evolution of the PBH mass satisfies this regime. During the application of the excision technique, the Hamiltonian constraint is {computed using Equation~\eqref{eq:2_constraint} to check the correctness of the numerical evolution}. In the simulations in \cite{escriva_solo,Escriva:2021pmf}, the constraint was fulfilled until very late times, as shown in Figure~\ref{fig:2_constraint-excision}. Nevertheless, it seems that the evolution of the mass is not affected substantially by the violation of the constraint. Interestingly, in the bottom panel of Figure~\ref{fig:2_constraint-excision}, a crossing for different evolutions of $\Psi$ at a given time $t^{*}$ is observed. In Figure~\ref{fig:mases_extra2} is shown the evolution of $M_{\rm PBH}(t)$ in time for different values of $q$ and two different profiles. It is clear that the larger is $q$, the smaller is the final PBH mass $M_{\rm PBH,f}$, since the pressure gradients are stronger and prevent accretion from the FLRW background. Notice the differences of $M_{\rm PBH}(t)$ between the two profiles considered in the figure. Even with the same $q$, the values for the mass are different since the full shape of the profile will contribute to the final mass.

To obtain the parameters $t_{a}$, $M_{a}$, and $F$, a non-linear fit using the function of Equation~\eqref{eq:2_acretationformula} with the data $M_{PBH}(t)$ coming from the simulations can be performed. The range of values to make the fit are those that fulfil $\Psi \lesssim 0.1$. The values of $F$ that were obtained in the literature \cite{escriva_solo} making the non-linear fit were in the range $F \in [3.2,3.8]$ (for the case of $w=1/3$) in terms of the different profiles and amplitudes $\delta_m$. Some indicative results regarding the accuracy of the fits performed are: $\sigma_{\rm max} \approx 10^{-2}$ (variance) $s_{d}(M_{a}) = 10^{-4.2}$, $s_{d}(t_{a}) = 10^{-4.1}$, and $s_{d}(F) = 10^{-3.6}$, where $s_{d}$ is the standard deviation of the parameters. This is consistent with what was already stated in \cite{acreation1,acreation2,acreation3}: the numerical coefficient $F$ must be $O(1)$. In particular, in the case of the collapse of the domain walls, $F \approx 3.8$ was found for large black holes \cite{Deng:2016vzb}. 

\unskip

\begin{figure}[H]
\centering
\includegraphics[width=0.6\linewidth]{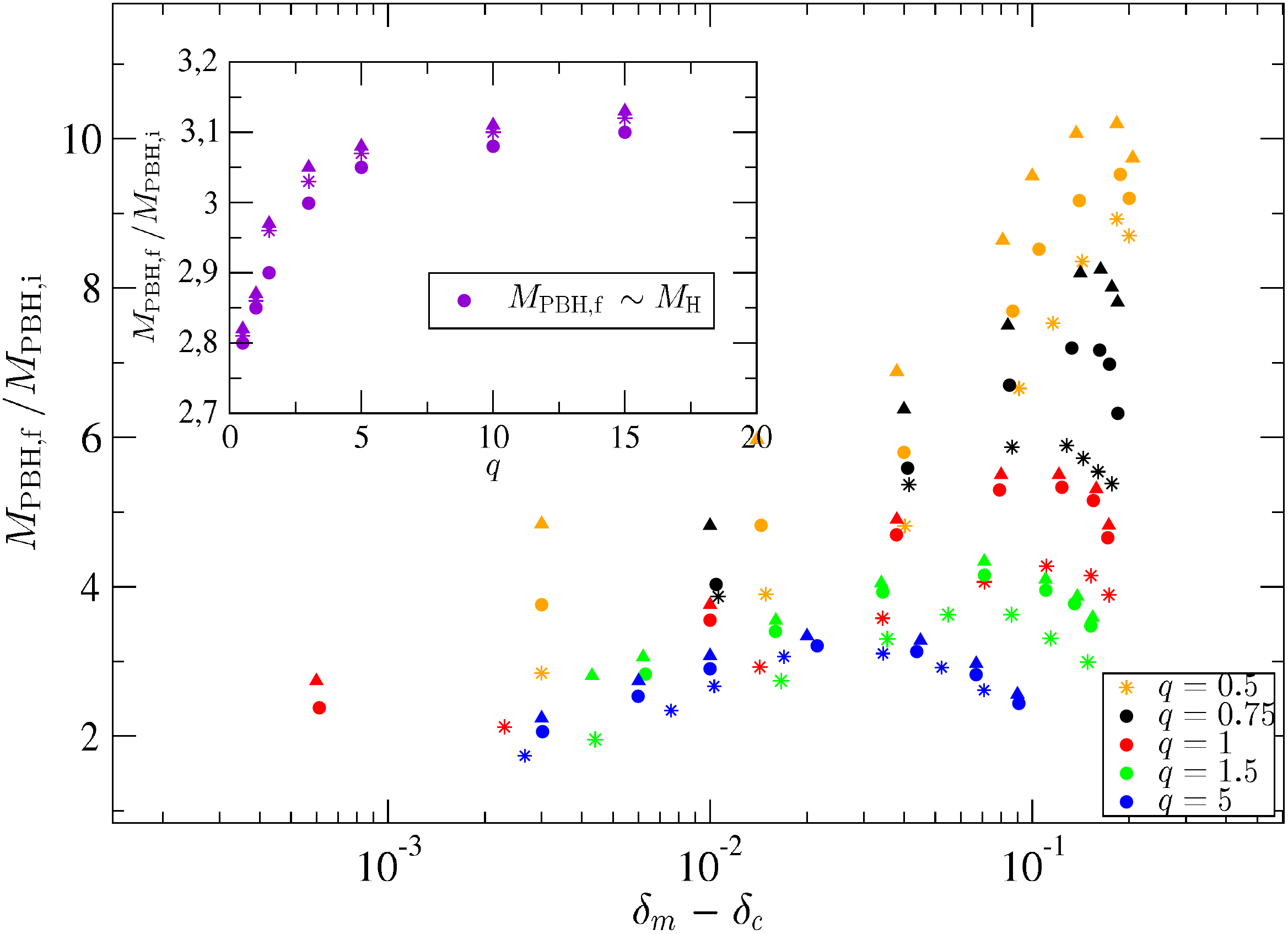} 
\caption{Accretion effect on the final PBH mass measured as the ratio $M_{\rm PBH,f}/M_{\rm PBH,i}$ in terms of $\delta_m-\delta_{c}(q)$ for different profiles with $w=1/3$. Stars correspond to Equation~\eqref{eq:4_lamda} with $\lambda=0$, circles to Equation~\eqref{4_basis_pol}, and triangles to Equation~\eqref{eq:4_dos_torres}. The ratio $M_{\rm PBH,f}/M_{\rm PBH,i}$ for PBHs with $M_{\rm PBH,f} \simeq M_{H}$ is shown in the subplot.}
\label{fig:acretion}
\end{figure}

Interestingly, in \cite{Escriva:2021pmf}, the effect of the accretion (from the FLRW background) to the growth of the PBH until reaching the stationary regime was quantified, namely $M_{\rm PBH,f}/M_{\rm PBH,i}$, for the case $w=1/3$. As was pointed out, the accretion should not be important for small PBHs, i.e., with $\delta_m$ near $\delta_{c}$. However, the situation differs for relatively large PBHs. This can be precisely seen in Figure~\ref{fig:acretion}. The accretion can be significantly important for large $\delta_m$.

Sharp profiles, corresponding to large $q$, have larger pressure gradients, and therefore, the ratio $M_{\rm PBH,f}/M_{\rm PBH,i}$ is smaller, even for large $\delta_{m}$, since the gradients prevent the accretion. For low $M_{\rm PBH,f}$, the ratio $M_{\rm PBH,f}/M_{\rm PBH,i}$ should be small, as expected \cite{hawking1,size1}. When $M_{\rm PBH,f}\simeq M_{H}$, i.e., for PBHs with a higher probability to form, $M_{\rm PBH,f} \simeq 3 M_{H}$ was obtained. In comparison, a similar result was also found in the case of the collapse of a massless scalar field \cite{scalar_harada} and in \cite{Deng:2016vzb,vacum_bubles} from BHs formed from the collapse of domain walls or vacuum bubbles, with a factor $M_{\rm PBH,f}/M_{\rm PBH,i}\leq 2$. 

\begin{figure}[H]
\centering
\includegraphics[width=0.6\linewidth]{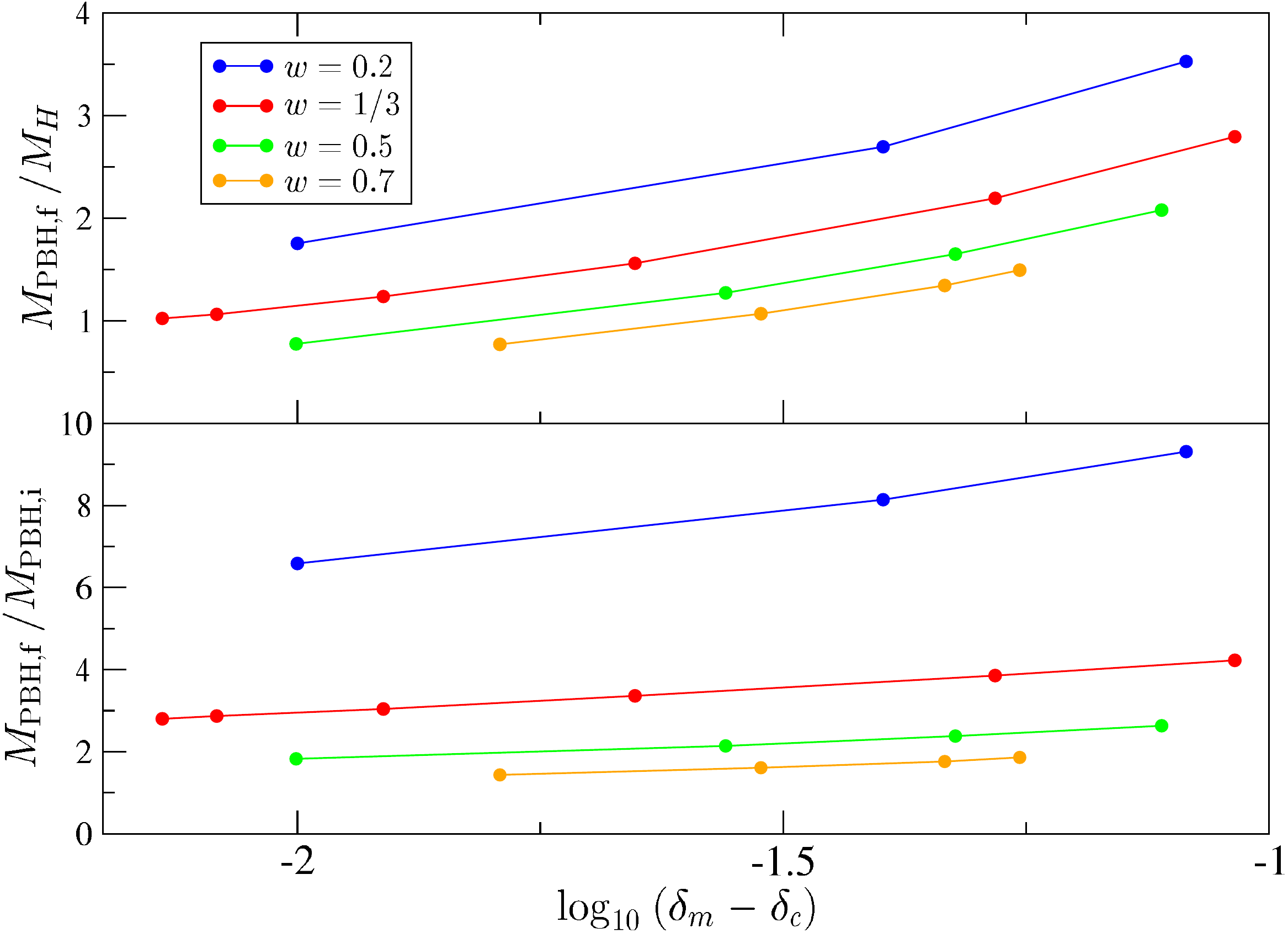} 
\caption{\textbf{Top-panel}: Values of the final PBH mass in terms of $M_H$ for some values of $w$ as a function of $\delta_m-\delta_c$. \textbf{Bottom-panel}: Accretion effect on the final PBH mass for some values of $w$ as a function of $\delta_m-\delta_c$. In both cases, the profile of Equation~\eqref{eq:4_lamda} was used with $q=1$ and $\lambda=0$.}
\label{fig:pbhmass_w}
\end{figure}

An example of the values of PBH mass for other values of $w$ can be found in Figure~\ref{fig:pbhmass_w}, where the ratio $M_{PBH,f}/M_H$ is shown using the Gaussian profile Equation~\eqref{eq:4_lamda} with $q=1$ ($\lambda=0$) for those PBHs $M_{PBH,f} \approx M_H$. The PBH mass $M_{PBH,f}$ is very sensitive to $w$, as expected. When $w$ increases, the gradients are larger, preventing the accretion and the increase of the PBH mass. The ratio $M_{PBH,f}/M_H$ is larger for smaller $w$ at the same $\delta_m-\delta_c$, since the accretion is substantially higher when the pressure gradients are decreased, as can be seen in the bottom panel of Figure~\ref{fig:pbhmass_w}. In this case, the values of $F$ differ substantially from $w=1/3$. In particular, for the set of values $w \in [0.2,1/3,0.5,0.7]$, this corresponds to $F \approx [7.2,3.8,2.0,1.37]$, respectively, which already clearly indicates the substantial accretion for small $w$. Moreover, the time needed to reach the asymptotic regime of Equation~\eqref{eq:2_acretationformula} is much higher the smaller $w$ is. For example, in the case of $\delta_m-\delta_c=10^{-2}$ for $w=0.2$ and $w=0.5$, a time of order $t/t_0 \backsim 10^{4}$ and $t/t_0 \backsim 10^{3}$ was needed, respectively, to reach the asymptotic regime.

On the other hand, in \cite{size1}, some analytical approximations were made regarding the maximum size of the PBHs formed from hydrodynamical perturbations at the time of the formation of the apparent horizon $t_{AH}$; in particular, it was found that:

\begin{equation}
\left(\frac{R_{\rm PBH,i}}{R_{H,i}}\right)_{\rm max} = \left(\frac{2}{1+3w}\right)^{3}\left[\frac{3(1+w)}{2(1+\sqrt{w}}\right]^{\frac{3(1+w)}{1+3w}}w^{3/2} .
\label{bound}
\end{equation}
This analytical result (derived for $0< w\leq 1$) was verified numerically in the case of a radiation fluid in \cite{Escriva:2021pmf}, where an extensive study on the values $M_{\rm PBH,i}$, $R_{\rm PBH,i}$, and $t_{AH}$ was also performed for different profiles and amplitudes $\delta_m$.

To compute the value of the constant $\mathcal{K}$ in front, the scaling law Equation~\eqref{eq:2_scaling} is needed to estimate the final mass of the black hole $M_{PBH,f}$ using perturbations with an amplitude near the critical regime, i.e., $\delta_m \rightarrow \delta_{c}$. A range of $10^{-3} \lesssim \delta_m-\delta_{c} \lesssim 10^{-2}$ is sufficient to estimate it, since the gravitational collapse of the fluctuations happens in the critical regime. Beyond that, the scaling law deviates. In particular, for a radiation fluid, the value $\gamma = 0.357$ \cite{gamma1,renormalizationcriticalcollapse} can be taken directly to obtain $\mathcal{K}$ from a given $\delta_m-\delta_{c}$ (for other values of $w$, check Figure~\ref{fig:gammas} and the caption).

\begin{figure}[H]
\centering
\includegraphics[width=0.43\linewidth]{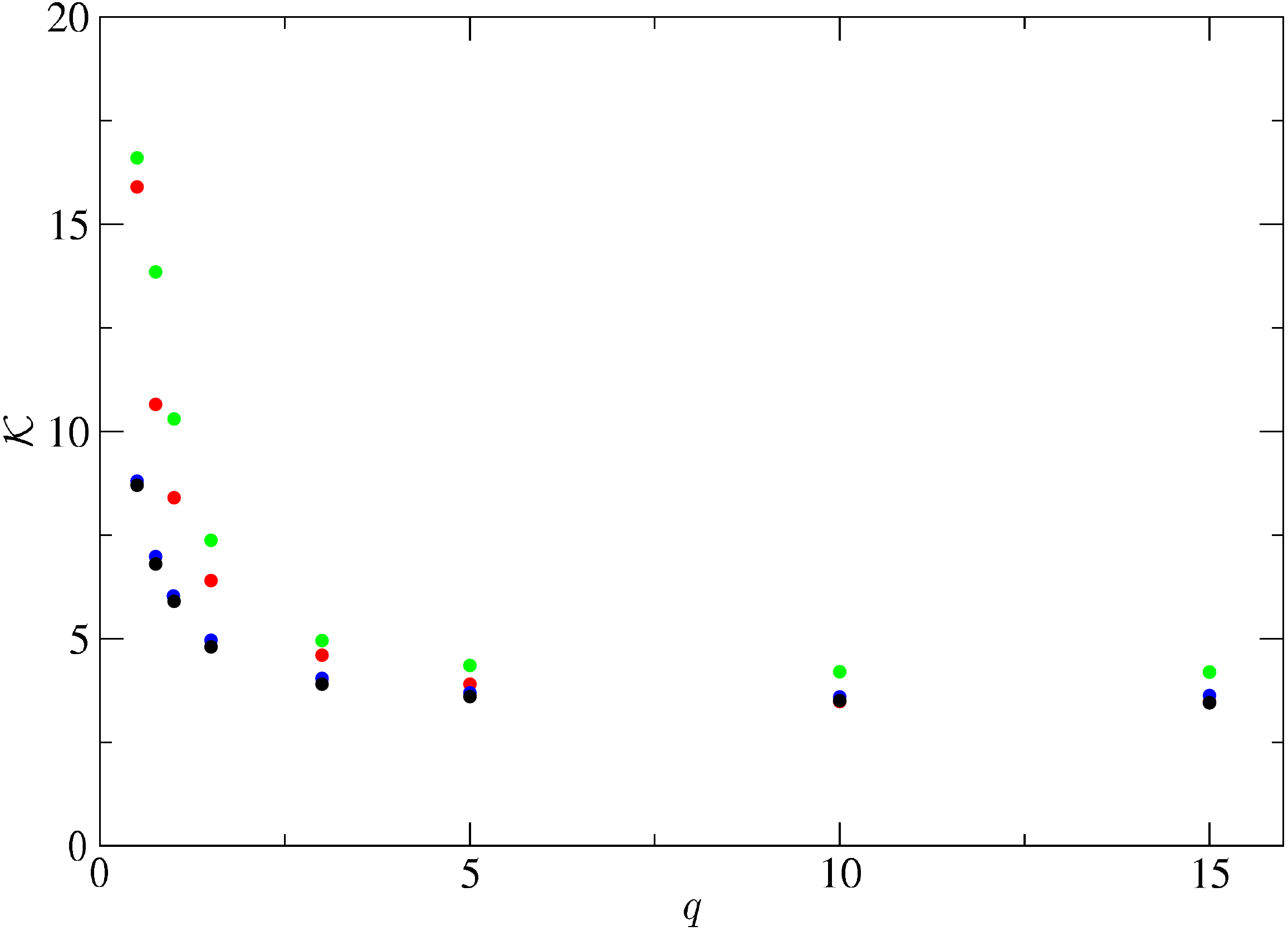} 
\caption{Dependence of $\mathcal{K}$ as a function $q$ for the profiles in Equation~\eqref{4_basis_pol} (red), Equation~\eqref{eq:4_lamda} with $\lambda=0$ (black), Equation~\eqref{eq:4_lamda} with $\lambda=1$ (blue), and Equation~\eqref{eq:4_dos_torres} (green). The parameters used for the profile $\mathcal{C}_{tt}(r)$ are $q_{2}=3$, $r_{j}=2r_{m1}$, $\mathcal{C}_{tt(\rm peak,2)}=0.3$, and $\delta_{2}$, obtained from Equation~\eqref{eq:pico2} using the previous values with and $q_{1}=q$.}
\label{fig:K}
\end{figure}

In Figure~\ref{fig:K} are shown the values of $\mathcal{K}$ for the profiles of Equations~\eqref{4_basis_pol},~\eqref{eq:4_lamda}, and~\eqref{eq:4_dos_torres} as an example of $\mathcal{K}$ in terms of the profile. The value of the PBH mass $M_{\rm PBH,f}$ (and therefore, also $\mathcal{K}$) is affected by the shape of the profile beyond the peak of $\com$ since the accretion process depends on the full shape of the fluctuation, in comparison with the threshold $\delta_{c}$, as was pointed out in Section \ref{sec:non_linear_behaviour}, which mainly depends on the shape around the compaction function peak.

In the case of the profile Equation~\eqref{4_basis_pol}, $\mathcal{K}$ tends to $\approx 3.5$ for large values of $q$. The values of $\mathcal{K}$ tend to increase as $q$ decreases, as shown Figure~\ref{fig:K}. Numerically, in \cite{Escriva:2021pmf}, it was not possible to obtain the final mass $M_{\rm PBH,f}$ for profiles $q \lesssim 0.5$, due to conic singularities, as was already found in \cite{Escriva:2020tak}.

\begin{figure}[H]
\centering
\includegraphics[width=0.6\linewidth]{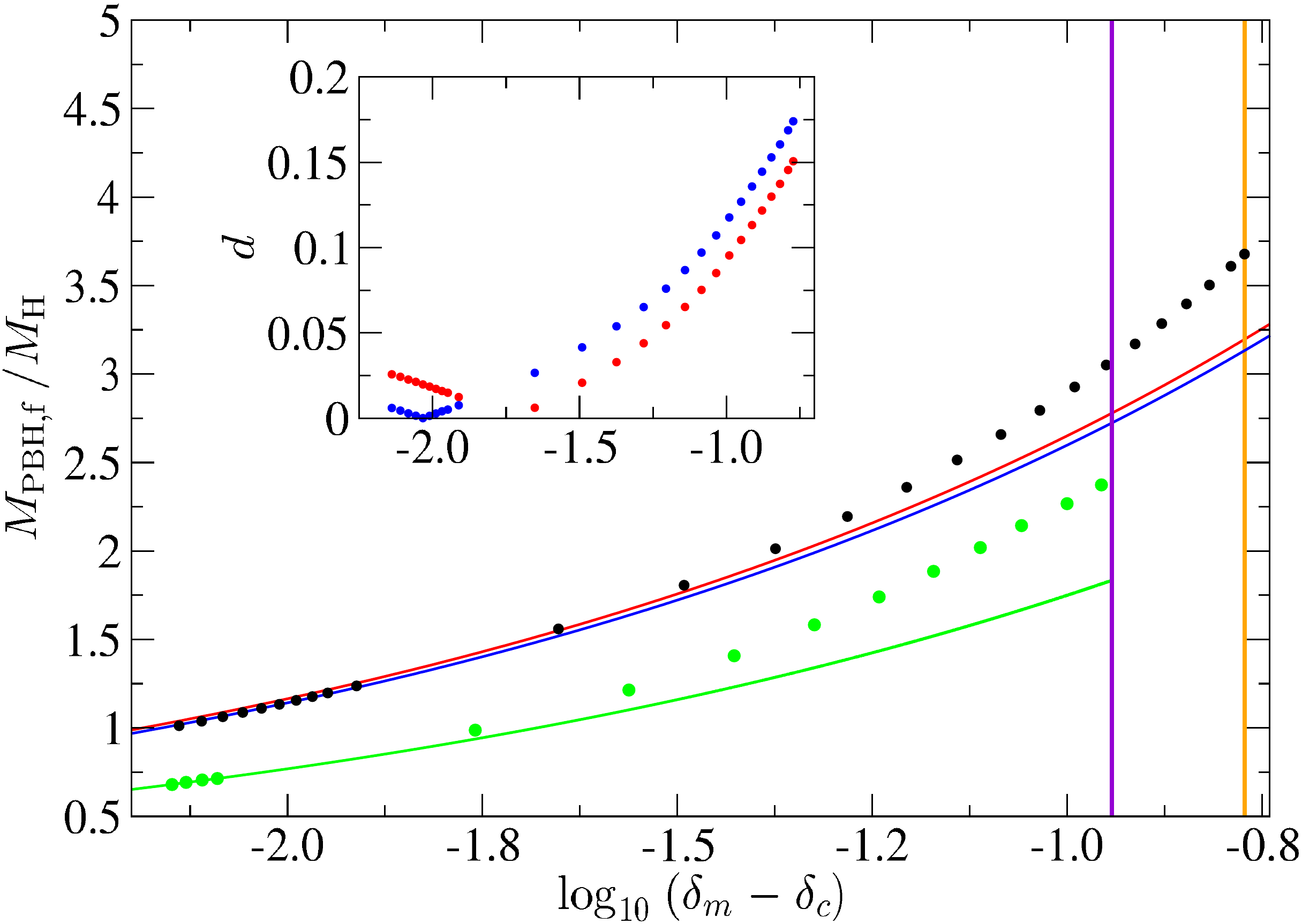} 
\caption{Values of $M_{\rm PBH_{f}}/M_{H}$ in terms of $(\delta_m-\delta_{c})$ for two Gaussian profiles in $K(r)$ and $\zeta(\tilde{r})$. Dark points correspond to the numerical values for the profile in $K(r)$. The solid red line corresponds to the scaling law behaviour with $\gamma=0.357$, $\delta_{c}=0.49774$ for ${\cal K}=6.03$ (pointed out in \cite{musco2013}) and the blue solid line with the numerical value obtained in the review with ${\cal K}=5.91$. The absolute value of the relative deviation $d$ with respect to the numerical values and the ones coming from the scaling law are shown in the subplot. The orange vertical line is the value $\delta_{\rm max}=2/3$. On the other hand, the green dots correspond to the case of the Gaussian profile in $\zeta$. The solid green line is the scaling law with $\gamma=0.357$, $\delta_{c}=0.55257$, and $\mathcal{K}=3.98$ (in \cite{Young:2019yug}, for this profile, $\mathcal{K} \simeq 4.0$ was given). The vertical magenta line corresponds to the case $\delta_{\rm max}=2/3$ for this profile. In both cases, $w=1/3$.}
\label{fit:2_scalingfit}
\end{figure}

On the other hand, in Figure~\ref{fit:2_scalingfit}, an explicit example of $M_{PBH,f}$ can be seen for a Gaussian profile (in both $K(r)$ and $\zeta(r)$) in terms of $M_{H}$ for large values of $\delta_m$ beyond the critical regime $\delta_m \gtrsim 10^{-2}$ with $w=1/3$. As can be observed in the subplot of Figure~\ref{fit:2_scalingfit}, the scaling law deviates for the highest allowed values of $\delta_m$ to $O$(15--20\%) for both cases. For these particular cases, the maximum allowed mass of the PBH formed is $M_{PBH,f} \approx 3.7 M_{\rm H}$ and $M_{PBH,f} \approx 2.3 M_{\rm H}$, respectively. It is expected in principle that this deviation from the scaling regime will not significantly modify the estimation of the PBH abundance, due to the unusualness of such perturbations since those PBHs with a higher probability to form have $M_{PBH,f} \approx M_{H}$. It is important to also notice that the masses generated with the profile $\zeta(\tilde{r})$ are lower than those with $K(r)$, since although they are both Gaussian, they have $q>1$ and $q=1$, respectively, and sharper profiles have less mass due to stronger pressure gradients, as we see in Figure~\ref{fig:mases_extra2}.

\section{Analytical Estimation for the Threshold of PBH Formation}\label{sec:analytics_on_the_thresholds}

In this section, we review in detail the current analytical estimations used in the literature for the threshold of PBH formation from curvature fluctuations with an equation of state $w$. The first analytical estimation came from \cite{carr75}, where using a Jeans length approximation, it was shown that,
\begin{equation}
 \label{eq:3_CARR}
 \delta_{\rm Carr}= w\ .
\end{equation}
Although this approximation is simple, it fulfils some properties: (i) at $w=0$ (dust), any curvature fluctuation should collapse, i.e., $\delta_{c}(w=0)=0$; (ii) {when $w$ increases}, the pressure gradients are larger, so the threshold should be higher. This approximation was obtained assuming the argument that the size of an over-density (that leads to PBH formation), at the maximum expansion, should be smaller than the particle horizon and larger than the Jean radius (Jeans criterion). In particular,

\begin{equation}
R_{\rm J} \lesssim R_{\rm max} \lesssim R_{\rm PH},
\end{equation}
where $R_{\rm PH} \backsim \sqrt{3/ 8\pi \rho_{\rm max}}$, $R_{\rm J} \backsim c_{s} R_{\rm PH}$. See also Section \ref{sec:pbh_basics} for a simple derivation.

Later on, in \cite{harada}, the previous estimation of \cite{carr75} was improved by using a sophisticated and interesting model (a three-zone model) of the collapse of a homogeneous over-dense sphere surrounded by a thin under-dense shell. We suggest the reader check \cite{harada} for more details on the model and the assumptions made. The threshold estimation that they obtained was,
\begin{equation}
 \label{eq:3_HYK}
 \delta_{\rm HYK} = \frac{3(1+w)}{5+3w}\,\sin^2\Big(\frac{\pi \sqrt{w}}{1+3w} \Big).\ 
\end{equation}
It was also provided with an upper bound $\delta_{\rm HYK}^{+}$ and a lower bound $\delta_{\rm HYK}^{-}$ on $\delta_{c}$, to take into account the different ways of applying the relativistic Jeans criterion for this model.
\begin{equation}
\delta_{\rm HYK}^{+} = \frac{3(1+w)}{5+3w}\sin\left(\frac{2 \pi \sqrt{w}}{1+2 \sqrt{w}+3w} \right).
\label{harada_plus}
\end{equation}
\begin{equation}
\label{harada_magnus}
\delta_{\rm HYK}^{-} = \frac{3(1+w)}{5+3w}\sin\left(\frac{ \pi \sqrt{w}}{1+ \sqrt{w}+3w} \right).
\end{equation}
These previous estimations (in particular, Equation~\eqref{eq:3_HYK}) have been extensively used in the literature, but as has already been noticed \cite{carr75,harada}, these estimations are not profile dependent and therefore do not take into account the specific shape of the curvature fluctuation.

\subsection{Shape-Dependent Analytical Estimation of the Threshold for Radiation Fluid}

In \cite{universal1}, it was proven that, to a very good approximation, the threshold for the \mbox{$w=1/3$} case only depends on the curvature of the compaction function at its maximum, under the assumption of a central over-dense peak in the density distribution. It was used to build an analytical formula for the threshold with a dependence on the profile considered. The formula was accurate enough and only deviated $2\%$ in comparison with the numerical~simulations. 

The two main ingredients of the procedure used in \cite{universal1} were: (i) parametrise the shape around the compaction function peak with a dimensionless parameter $q$; (ii) use the average of the critical compaction function. As was shown in \cite{universal1} and as we have already mentioned, to a very good approximation, the threshold only depends on,
\begin{equation}
\label{3_q}
 q\equiv -\frac{r_m^2\,\mathcal{C}''(r_m)}{4\,\mathcal{C}(r_m)} \ ,
\end{equation}
which is a dimensionless measure of the curvature of $\com(r)$ at its maximum. This means that different profiles, but with the same $q$ value, will have the same threshold upon a small deviation of $<$2\%.

The next point is to define a ``basis'' (or fiducial set of curvature profiles) such that, by varying $q$, this set covers the whole range of interesting thresholds and shapes with $q\in (0, \infty)$ while also being regular at $r=0$ and having $\rho'(r=0,t)=0$. In \cite{universal1}, this basis was given in terms of the exponential functions Equation~\eqref{eq:4_lamda} (with $\lambda=0$). Although it strictly does not satisfy regular conditions for $q \rightarrow 0$ at $r=0$, this limit was tested numerically with the polynomial basis of Equation~\eqref{4_basis_pol}.

\begin{figure}[H]
\centering
\includegraphics[width=0.55\linewidth]{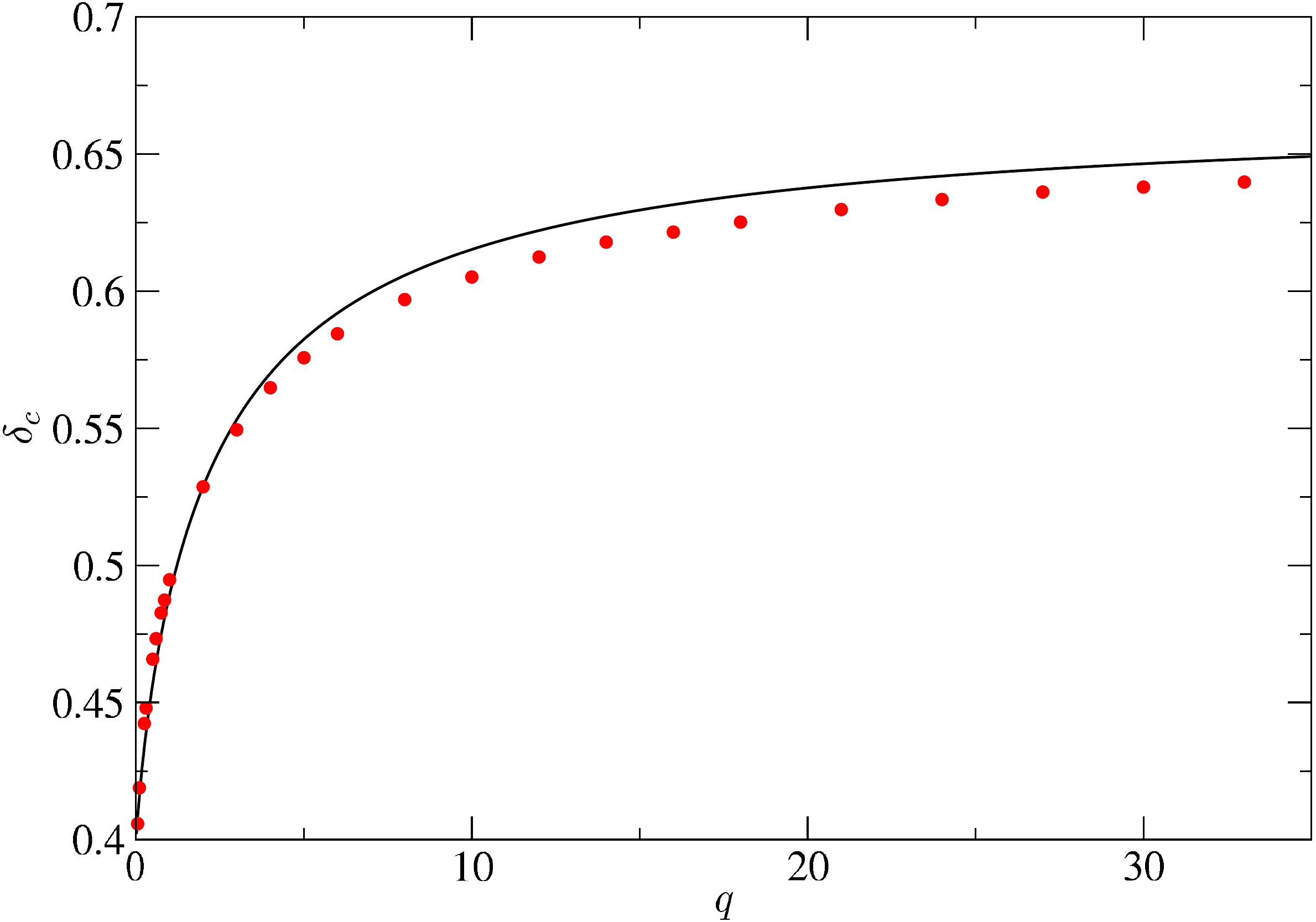} 
\caption{{Plot of $\delta_{c}(q)$ for $w=1/3$. The solid black line corresponds to the analytical estimation of Equation~\eqref{eq:threshold_analit}. Red points correspond to the numerical thresholds of Equation~\eqref{4_basis_pol} with $w=1/3$.}}
\label{fig:deltacanalitical}
\end{figure}

Taking this into account, the averaged critical compaction function integrated from $r=0$ up to the peak $r=r_m$ is defined as,
\begin{equation}
\label{3_Cbar}
 \bar\com_{\rm c}\equiv
 \frac{3}{ r_m^3 }\int_{0}^{ r_m} \com_c(r) r^2 dr\ .
\end{equation}
Introducing the profile considered into Equation~\eqref{3_Cbar}, we can solve the integral analytically to obtain:
\begin{equation}
 \bar\com_{\rm c} = \frac{3}{2} e^{\frac{1}{q}} q^{-1+\frac{5}{2q}} \left[\Gamma\left(\frac{5}{2q}\right)-\Gamma\left(\frac{5}{2q},\frac{1}{q}\right)\right] \delta_{c},
 \label{eq:averaged_int}
\end{equation}
where $\Gamma(x)$ \footnote{The gamma function $\Gamma(x)$ is defined as $\Gamma(x)=\int_{0}^{\infty}t^{z-1}e^{-t}dt$.} is the gamma function, $\Gamma(x,y)$ \footnote{The incomplete gamma function is defined as $\Gamma(x,y) =\int_{y}^{\infty}t^{x-1} e^{-t}dt$.} is the incomplete gamma function, and we have used $\mathcal{C}_{c}(r_m) \equiv \delta_{c}$. In the case of a radiation fluid, when $q \rightarrow \infty$, it was confirmed numerically that $\delta_{c}(q \rightarrow \infty)= f(w=1/3)=2/3$ \cite{musco2018}. Taking into account this analytical limit into Equation~\eqref{eq:averaged_int}, we obtain:
\begin{equation}
 \bar\com_{\rm c} ( q \rightarrow \infty) = \frac{3}{5}\delta_{c}(q \rightarrow \infty) = \frac{2}{5}.
\end{equation}
The assumption made in \cite{universal1} was to consider that $\bar\com_{\rm c}$ is a profile-independent quantity, and therefore, $\bar{\com}_{\rm c} =2/5$ for any value of $q$. The assumption was tested extensively and numerically for different profiles and was found to be consistent with numerical simulations up to a small deviation of $O(2\%)$ in $\delta_{c}$. Inverting Equation~\eqref{eq:averaged_int} in $\delta_{c}$ with $\bar\com_{\rm c} =2/5$, we finally obtain the analytical threshold formula $\delta_{c}$ in terms of the specific profile $q$, in the case of a radiation fluid,
\begin{equation}
\delta_{c}(q) = \frac{4}{15} e^{-\frac{1}{q}}\frac{q^{1-\frac{5}{2q}}}{\Gamma\left(\frac{5}{2q}\right)-\Gamma\left(\frac{5}{2q},\frac{1}{q}\right)}.
\label{eq:threshold_analit}
\end{equation}
This remarkable result in \cite{universal1} was the first analytical estimation shown in the literature for the threshold of PBH formation, taking into account the shape of the fluctuation, with enough accuracy to be used for estimations of PBH abundances \cite{nonlinear}. A plot of Equation~\eqref{eq:threshold_analit} in terms of $q$ can be found in Figure~\ref{fig:deltacanalitical}. {Interestingly, a study of the thresholds for different power spectrums (leading to perturbations in $\zeta$ with the coordinate $\tilde{r}$) was performed in \cite{Musco:2020jjb} using Equation~\eqref{eq:threshold_analit}, with the corresponding $q$ value obtained from Equation~\eqref{eq:q_tilde} and connecting with the density contrast of Equation~\eqref{eq:tilde_perturb_contrast}.}

\subsection{Generalisation of the Analytical Threshold beyond the Radiation Fluid}

Later on, the procedure used in \cite{universal1} was generalised considering PBH formation within a perfect fluid with $w \in (0,1]$ on an FLRW background, and it was found that the threshold averaged compaction function within a concentric sphere of radius $r_{m}(1-\alpha(w)) \leq r \leq r_m$ is, to a very good approximation, again universal in the case $1/3 \leq w \leq 1$, {where $\alpha(w)$ is the location where we should start to integrate the compaction function, and it should depend on $w$}. This was used to provide a new and improved analytical formula for the threshold that only depends on the normalised second derivative of the compaction function at its maximum and the equation of state $w$. This formula was proven to be accurate at order $6\%$ compared to numerical simulations and reproducing the same result for the case of radiation $w=1/3$. It was also again found that the $\delta_{c}$ for different profiles with the same $q$ are equal within a deviation less than $4\%$.

In \cite{universal1}, the basis profiles used were given in terms of the exponential functions; despite that, since the boundary conditions at $r=0$ are violated for $q<1/2$, the basis profile of Equation~\eqref{4_basis_pol} was used instead. This set of profiles satisfies the appropriate boundary and regularity conditions.

The critical compaction function, averaged within a spherical shell extending from radius $[1-\alpha(w)]\,r_m$ to $r_m$ (the peak), is defined to be: 
\begin{equation}
\label{3_Cbar33}
 \bar\com_{\rm c}(w,{\rm profile})\equiv
 \frac{3}{ r_m^3 V[\alpha(w)]}\int_{r_m[1-\alpha(w)]}^{ r_m} \com_c(r) r^2 dr\ ,
\end{equation}
where $V[\alpha(w)]=\alpha(w)\,[3+(\alpha(w)-3)\alpha(w)]$ and $\com_c(r)=\com(r)\Big|_{\com(r_m)=\delta_c}$.
Inserting Equation~\eqref{4_basis_pol} into Equation~\eqref{3_Cbar33} yields: 
\begin{equation}
 \label{3_average_formula}
 \bar\com_{\rm c}(w, {\rm basis})= \delta_c(w,q)\, g(q,w)\,
 \left[-F_1(q)+(1-\alpha)^{3-2q}F_2(q,\alpha) \right] ,
\end{equation}
with: 
\begin{equation}
 g(q,w) = \frac{3(1+q)}{\alpha (2q-3)\left[3+\alpha(\alpha-3)\right]} ,
\end{equation}
\begin{equation}
 F_1(q) = {}_2F_1\left[ 1,1-\frac{5}{2(1+q)}, 2-\frac{5}{2(1+q)},-q \right] ,
\end{equation}
and:
\begin{equation}
 F_2(q,w) = {}_2F_1\left[1,1-\frac{5}{2(1+q)},2-\frac{5}{2(1+q)},-q(1-\alpha)^{-2(1+q)}\right] ,
\end{equation}
where ${}_2F_1$ is the hypergeometric function\footnote{The hypergeometric function is defined as ${}_2F_1(a,b,c,z)= \sum_{k=0}^{\infty} \frac{(a)_{k}(b)_{k}}{(c)_{k}}\frac{z^{k}}{k!}$.}.

The key point is that as was shown numerically in \cite{Escriva:2020tak}, the dependence of the averaged critical compaction function on the profile shape is weak enough to be ignored, and therefore universal. Then, 
\begin{equation}
\label{3_assumption}
 \bar\com_{\rm c}(w,{\rm profile})\simeq \bar\com_{\rm c}(w)\ .
\end{equation}
Therefore, the analytic expression for the critical threshold value reads as: 
\begin{equation}
 \label{3_threshold-anal}
 \delta_c(w,q) = \frac{\bar\com_{\rm c}(w)}{ g(q,w)}\, \frac{1}{\left[-F_1(q)+(1-\alpha)^{3-2q}F_2(q,\alpha) \right]}\ .
\end{equation}

The functions $\alpha(w)$ and $\bar\com_{\rm c}(w)$ were found using two different procedures: (\RomanNumeralCaps{1}) a numerical double fit minimisation using the numerical results of the thresholds for different profiles and (\RomanNumeralCaps{2}) making a single numerical fit minimisation already using the analytical result that $\delta_{c}(q \rightarrow \infty$) for $w\geq 1/3$. Both procedures gave equivalent results. Therefore, the assumption of $\bar\com_{\rm c}(w)$ being universal was found good enough and verified numerically for $w\gtrsim 1/3$. This result verified the previous one obtained for a radiation fluid $w=1/3$ in~\cite{universal1}, where $\alpha=1$ and $\bar\com_{\rm c}=4$. In particular, using the procedure (\RomanNumeralCaps{1}), it was found, 
\begin{align}
 \label{eq:average_c_fit1}
 \bar\com_{\rm c}(w) &= a + b \, {\rm Arctan}(c\, w^d) \\
 \label{eq:alpha_fit1}
 \alpha(w) &= e + f \, {\rm Arctan}(g\, w^h) ,
\end{align}
with $a = -0.140381$, $b = 0.79538$, $c=1.23593$, $d= 0.357491$, $e = 2.00804$, $f = -1.10936$, $g = 10.2801$, and $h=1.113$. 
Instead, using (\RomanNumeralCaps{2}),
\begin{align}
 \label{eq:average_c_fit2}
\bar\com_{\rm c}(w) &= i + j \, {\rm Arctan}(p\, w^l) , \\
 \label{eq:alpha_fit2}
 \alpha(w) &= m + t \, {\rm Arctan}(r\, w^s) ,
\end{align}
with $i = -0.140381$, $j = 0.79538$, $p=1.23593$, $l= 0.357491$, $m = 2.00804$, $t = -1.10936$, $r = 10.2801$, and $s=1.113$.

\subsection{A Comparison between Analytical Estimates}

\begin{figure}[H]
\centering
\includegraphics[width=0.6\linewidth]{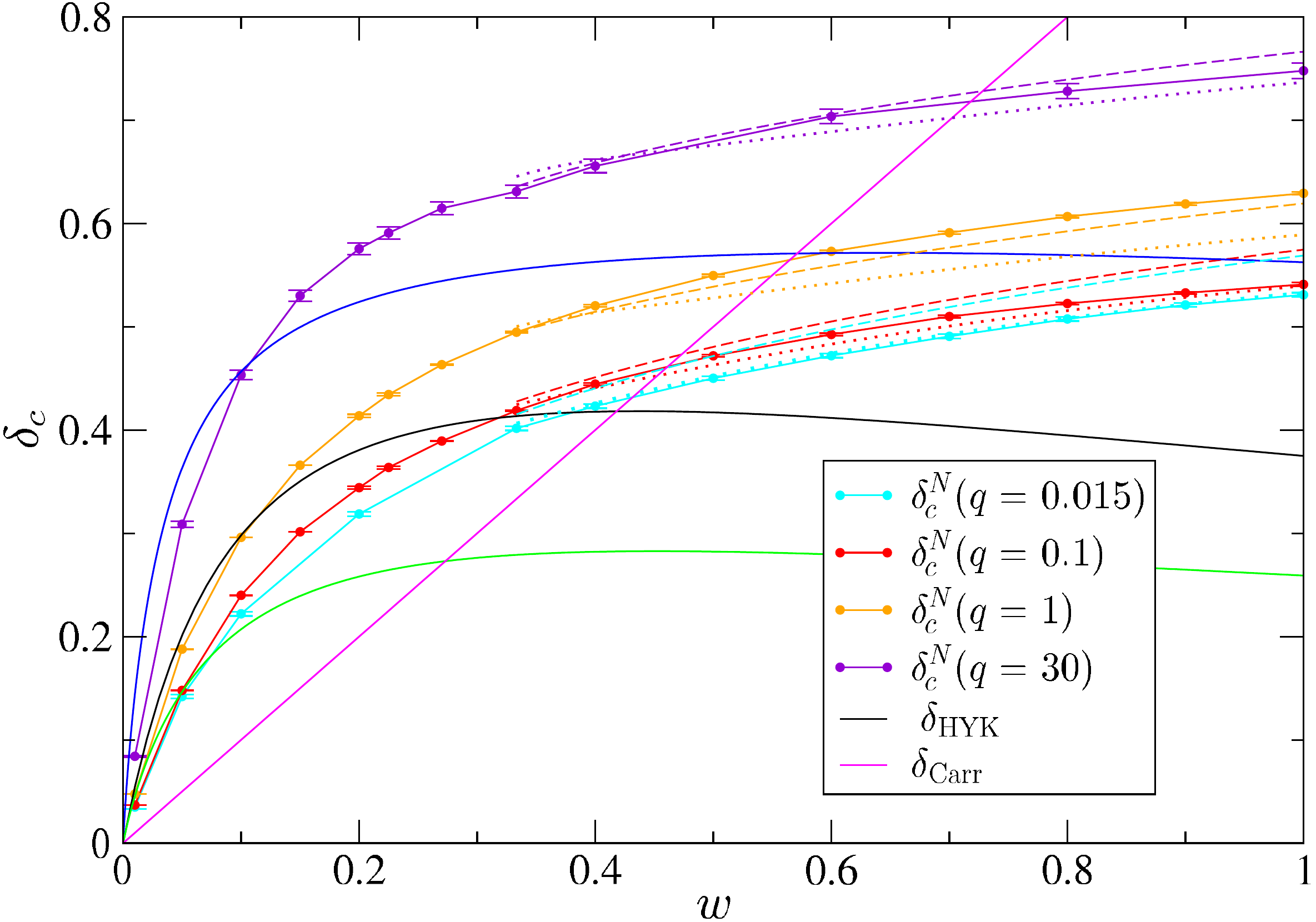} 
\caption{Dependence of threshold $\delta_c$ on $w$ when the initial profile is given by Equation~\eqref{eq:4_lamda} with $q=30$, $q=1$, $q=0.1$, $q=0.015$, and $\lambda=0$ in all the cases ($q=0$ would be a homogeneous sphere in the compaction function). The solid lines with dots and error bars show the results of the numerical simulations. The green and blue curves show the minimal and maximal bounds of $\delta_c$ from \cite{harada}, in particular Equations~\eqref{harada_plus} and~\eqref{harada_magnus}, respectively. The magenta line shows the approximation for $\delta_{\rm Carr}$ of Equation~\eqref{eq:3_CARR}. The black curve shows $\delta_{\rm HYK}$ from Equation~\eqref{eq:3_HYK}. The other curves show the approximation (Equation~\eqref{3_threshold-anal}) in which $\delta_c$ depends both on $w$ and $q$. Finally, the dotted curves use Equations~\eqref{eq:average_c_fit2} and~\eqref{eq:alpha_fit2} in Equation~\eqref{3_threshold-anal}, whereas the dashed curves use Equations~\eqref{eq:average_c_fit1} and~\eqref{eq:alpha_fit1} in Equation~\eqref{3_threshold-anal}.}
\label{fig:harada}
\end{figure}

In \cite{Escriva:2020tak}, the analytical estimation obtained with Equation~\eqref{3_threshold-anal} was compared with the previous ones in the literature, i.e., Equations~\eqref{eq:3_CARR} and \eqref{eq:3_HYK}. In Figure~\ref{fig:harada} is shown the comparison. The solid lines in Figure~\ref{fig:harada} show these approximations; the symbols with error bars show $\delta_c(w)$ from the numerical simulations of the profiles of Equation~\eqref{4_basis_pol} with $q=0.015$, $q=0.1$, $q=1$, and $q=30$. The dotted and dashed curves, which significantly better describe the result of the numerical simulations, show the result of inserting Equations~\eqref{eq:average_c_fit1} and \eqref{eq:alpha_fit1} or Equations~\eqref{eq:average_c_fit2} and \eqref{eq:alpha_fit2}, into Equation~\eqref{3_threshold-anal}. In both cases, Equation~\eqref{3_threshold-anal} (as well as the simulations) exceeds even the upper bound claimed by \cite{harada} at even lower $w$ as $q$ increases, which are Equations~\eqref{harada_plus} and \eqref{harada_magnus}.

As can be observed, the discrepancy of the estimation Equations~\eqref{eq:3_CARR} and \eqref{eq:3_HYK} for the case of small $w$ is as large as $50\%$ for the case $q=0.1$. This is the limit that was considered successful and optimal for the approximations on which Equation~\eqref{eq:3_HYK} is based. This deviation is even bigger than the one found earlier because Equation~\eqref{eq:3_HYK} was compared with a Gaussian profile (Equation~\eqref{eq:4_lamda} with $q=1$ and $\lambda=0$). Actually, in the case $w<0.15$, the solid black curve in Figure~\ref{fig:harada} give a satisfactory description of the simulations with $q=1$. However, the top-hat profile with $q \rightarrow 0$ {(top-hat for the profile in the compaction function)}, which was the one used in the analytic estimations of \cite{harada}, is quite better approximated by $q \ll 1$. For $q= 0.1$, the formula Equation~\eqref{eq:3_HYK} does not describe the simulations particularly well, and the discrepancy at $w<1/3$ is even worse when $q=0.015$ is taken. It was claimed in \cite{Escriva:2020tak} that this disagreement seems to suggest that the supposed agreement shown in Figure 3 of \cite{harada} is a result of numerical coincidences, with no physics implications.

Finally, the polynomial basis profile of Equation~\eqref{4_basis_pol} fulfils the regularity conditions for {$q \rightarrow 0 $}, as we have pointed out before, something not fulfilled with the exponential basis Equation~\eqref{eq:4_lamda} (with $\lambda=0$). Precisely, this led in previous literature \cite{musco2018} to inferring that the minimum threshold for PBH formation (in the case of a radiation fluid) must coincide with the estimation of \cite{harada}, i.e., $\delta_{\rm c,min} (w = 1/3) = \delta_{\rm HYK} (w = 1/3) \approx 0.41$. In \cite{musco2018}, {the profile leading to the minimum numerical threshold of PBH formation was given by Equation~\eqref{eq:4_lamda} with $q=0.1$ and $\lambda=0$, since as was indicated there, the profiles were too sharp to take even smaller values of $q$ and produce a numerical simulation}. This actually looks related to the fact that the exponential profile Equation~\eqref{eq:4_lamda} does not satisfy the regularity conditions at $r \rightarrow 0$ for $q<1/2$. Instead, in the case of using the polynomial profile Equation~\eqref{4_basis_pol}, it allows performing simulations even with smaller values in $q$, up to $q = 0.01$, as done in~\cite{universal1}. Therefore, this led to the conclusion in \cite{universal1} that, for a radiation fluid, the correct range of values for the threshold is $0.4 \leq \delta_c \leq 2/3$, and therefore, the minimum value was lower than $\delta_{\rm HYK}$. Indeed, the generalisation of $\delta_c$ in $w$ already indicates that the minimum threshold should not correspond necessarily to the one of $\delta_{\rm HYK}$.

\section{Other Scenarios of PBH Formation}\label{sec:other_scenarios}

In this review paper, we mainly focused on the study of PBH formation from the spherical collapse of primordial density perturbations, which is currently the case most studied in the literature and with more results and applications \cite{Carr:2020xqk}. Despite that, many other mechanisms and scenarios could have led to black holes formed in the very early Universe \cite{Carr1}. In this last section, we try to briefly enumerate other scenarios beyond the ones considered in this review.

The case $w=0$ is very special as the threshold of PBH formation, for an infinitely long matter-dominated era; it is zero {if non-spherical effects are not considered. However, as was pointed out in \cite{Harada:2017fjm}, the dust phase could induce an angular momentum on the collapsing perturbation, therefore effectively increasing the threshold value for the black hole formation, making it non-vanishing.} Thus, the study of abundances, where the Universe is matter-dominated for a finite time, differs greatly from the case of radiation. Some works have considered this scenario with scalar field types \cite{10.1093/mnras/215.4.575,Hidalgo:2017dfp,PhysRevD.96.063507,Carrion:2021yeh,Martin:2020fgl,Padilla:2021zgm}, specifically taking into account the oscillatory behaviour of the scalar field during the reheating. Recently, in \cite{deJong:2021bbo}, simulations were performed in full GR using scalar fields to modulate the perturbation and the expanding dust background of the Universe, showing that the PBH mass grows beyond the self-similar limit $M_{\rm PBH} \propto H^{-1}$, at least initially. The dust collapse has also been studied in a more oriented perfect fluid scenario \cite{KHLOPOV1980383}. The PBH formation has also been considered in the context of the Lemaître--Tolman--Bondi (LTB) model \cite{Harada:2015ewt,Harada:2001kc}. In \cite{Kokubu:2018fxy}, the effect of inhomogeneity on PBH formation in the matter-dominated era and its consequences on the PBH production were studied. On the other hand, in \cite{Harada:2016mhb,Harada:2017fjm}, the non-spherical effects and spins for PBH formation in the case of a dust-dominated Universe were studied, showing this could have a significant effect.

A non-spherical effect, in principle, could dramatically change the threshold value and PBH mass \cite{Flores:2021tmc}. Some analytical estimations taking into account an ellipsoidal collapse were performed in the past, showing that the effect could be important \cite{Kuhnel:2016exn,Harada:2015ewt}. However, in \cite{new2}, for the first time in the literature, the first numerical simulation in $3+1$ spacetime (beyond spherical symmetry) for PBH formation from the collapse of curvature fluctuations was performed, with a small (perturbative) non-sphericity initial condition (with a radiation fluid). The main conclusion was that the effect of ellipticity on the threshold for PBH formation is negligible in the standard scenario of PBH formation in the radiation-dominated Universe. This situation could dramatically change if an equation of state different from radiation is considered, as indicated in \cite{new2}. In particular, for a matter-dominated Universe, the effect of any non-sphericity could be important \cite{KHLOPOV1980383,Harada:2017fjm,Harada:2016mhb}. Another possibility is the inclusion of angular momentum to analyse the spin generation to PBHs; some analytical estimations have been performed \cite{He:2019cdb}, but still, numerical simulations to test those arguments are necessary. Recently, PBH formation with an anisotropic perfect fluid has also been considered analytically \cite{Musco:2021sva}.

The usual scenario of the collapse of cosmological density perturbations considers a constant equation of state $w$ (mainly $w=1/3$ in the literature). However, the Universe has a thermal history where the equation of state is not necessarily constant, i.e., $w(t)$. A particular (and very popular) example applied to the PBH scenario is the QCD \mbox{epoch \cite{Carr:2019hud,Widerin:1998my,Boeckel:2010bey,Jedamzik:1996mr}}. During the QCD phase transition, the equation of state $w$ could have been softened. Looking at the result of the simulations in Figure~\ref{fig:harada}, it is already clear that the threshold $\delta_{c}$ will be reduced, and therefore, this makes the formation probability of PBHs exponentially more likely. Several works have addressed this scenario with progressive refinements \cite{Carr:2019hud,Byrnes:2018clq,Gao:2021nwz,Abe:2020sqb,Clesse:2020ghq,Carr:2019kxo,Sobrinho:2016fay}, specifically with the new calculations regarding lattice QCD \cite{Borsanyi:2016ksw,Bhattacharya:2014ara}. However, numerical simulations are still needed to fully verify the effect on how much the threshold, PBH mass, and mass function modifying the equation of state in time could change.

A bit more different mechanism of PBH formation comes from the collapse of Q-balls and oscillons, which are features of supersymmetric extensions of the standard model~\cite{Kusenko:1997si}. Specific realisations of those ideas have been made through theory-motivated scalar field potentials, e.g., the axion--monodromy potential \cite{Ballesteros:2019hus}. The mechanism for PBH formation of all these cases is naively similar: small number densities of defects lead to large fluctuations relative to the background density. These fluctuations become gravitationally bound and collapse to form black holes once the relic density has come to dominate. Finally, the relics decay due to some instabilities. Precisely, it has been argued that some solitonic type solutions such as Q-balls and oscillons could produce a significant fraction of dark matter in the form of PBHs \cite{new1,Cotner:2017tir,Cotner:2018vug,Cotner:2019ykd,Flores:2021jas}, without relying on any spectrum of density perturbations. {Some recent numerical works on oscillons and Q-balls have obtained results in this direction \cite{Kou:2019bbc,Nazari:2020fmk,Kim:2021ipz}}.

On the other hand, scalar force instability can lead to the growth of structures and the formation of halos of interacting particles (even in a radiation-dominated \mbox{Universe) \cite{1992ApJ...398..407G,Gubser:2004uh,Nusser:2004qu,Amendola:2017xhl,Savastano:2019zpr}}. Actually, in \cite{Amendola:2017xhl}, it was theorised that these structures could have collapsed and formed PBHs, but later on, it was realised that they should remain as virialised dark matter \cite{Savastano:2019zpr}. Precisely, in \cite{Flores:2020drq}, it was shown that the radiative cooling due to the scalar radiation allows those halos to collapse into black holes. This was shown with a simple model with fermions interacting via the Yukawa interaction. The general relativistic formulation of the interaction between a Fermi gas and a scalar field in cosmology was carefully studied recently in detail in \cite{Domenech:2021uyx}.

All these mechanisms {could induce} small or large spins to the PBHs in terms of their formation evolution and the process of radiative cooling, as was pointed out in \cite{Flores:2021tmc}. Another proposal is the one in \cite{Dvali:2021byy}, where the dilution of heavy quarks produced during inflation, which become confined by QCD flux tubes after horizon re-entry, could have led to PBH production.

Another mechanism is through the collapse of domain walls. Domain walls are topological defects that may form when a discrete symmetry is spontaneously broken in the very early Universe \cite{PhysRevD.44.340,PhysRevD.30.712}. Several works have addressed this scenario and have shown the possibility of forming PBHs, with astrophysical consequences \cite{Liu:2019lul,Deng:2016vzb,Rubin:2000dq,Khlopov:2004sc,Tanahashi:2014sma}. Another possibility comes from the collapse of cosmic strings \cite{1976Kibble,VILENKIN1985263,1995Kibble}, which are $1+1$ topological defects that are predicted beyond the Standard Model and could have led to the production of PBHs \cite{Jenkins:2020ctp,Vilenkin:2018zol,James-Turner:2019ssu,Ozsoy:2018flq,Garriga:1992nm,Caldwell:1995fu,PhysRevD.43.1106}.

Similarly, it was also shown that vacuum bubbles could lead to a successful channel for PBH production or also be involved in related mechanisms \cite{vacum_bubles,Deng:2020mds,Dymnikova:2000dy,PhysRevD.26.2681,Jung:2021mku,DeLuca:2021mlh,Lewicki:2019gmv,Baker:2021nyl,Gross:2021qgx}. They could have been formed during the inflationary epoch through a nucleation process. After inflation and depending on their size, these bubbles could collapse or even collide, forming black holes \cite{Garriga:2015fdk,Deng:2016vzb,vicente-garriga,Kusenko:2020pcg,garrigavicentejudithescriva,Maeso:2021xvl}.

Other results have also been pointed out in the context of modified gravity, in particular in \cite{Vallejo-Pena:2019hgv}, where the PBH production in G-inflation was studied, as well as in \cite{Kawai:2021edk}, where this was studied with the inclusion of a Gauss--Bonnet term. In \cite{Chen:2019ueb}, the effects on the threshold in Eddington-inspired Born--Infeld gravity were considered. 

Finally, a new mechanism appeared in \cite{Passaglia:2021jla}, where PBHs could have been formed from the collapse of primordial CDM isocurvature fluctuations. {Related to that, a new work has numerically shown the formation of PBHs \cite{Yoo:2021fxs} from isocurvature fluctuations generated by a massless scalar field.}

\section{Conclusions}\label{sec:future_perspectives}

In this review, we focused on explaining in detail the process of PBH formation from the collapse of curvature fluctuations under spherical symmetry on an FLRW background and its numerical implementation through a particular numerical technique. Specifically, we saw that numerical simulations are essential to capture the correct and full-non-linear behaviour of the collapse and formation of PBHs. This leads to a better and accurate determination of the threshold of PBH formation and their masses compared to analytical estimations, something essential for the precise analysis of PBH abundances. In this review, we saw a simple and efficient numerical procedure based on pseudospectral methods that allows us to fully solve the non-linear gravitational collapse of those perturbations and obtain the needed values of the threshold and PBH mass in the range of practical interest.

Two aspects in the literature have been remarkably important for the development of the field: (\RomanNumeralCaps{1}) the results found from numerical simulations about the critical scaling regime for the PBH mass and (\RomanNumeralCaps{2}) the analytical estimations for the threshold of PBH formation. Both allow us to estimate the abundances of PBHs and constrain them. In particular, we saw that the threshold and PBH mass are both dependent on the specific shape of the curvature fluctuation considered, and it is crucial to take this into account for the precise analysis of PBH abundances instead of using a constant threshold or PBH mass value. Moreover, we saw that the gravitational collapse is highly dependent on the equation of state of the fluid $w$. The larger $w$ is, the smaller is the mass of the PBHs and the accretion from the FLRW background, but the higher the threshold is. 

It is interesting to mention that the compaction function, firstly used in \cite{Shibata:1999zs} years ago, has become currently essential, not only for the definition of the threshold of PBH formation, but also to obtain an analytical estimation of it, using its average.

There is still much effort to be made in other directions and different scenarios of PBH formation to consider, as the ones already pointed out in Section \ref{sec:other_scenarios}. Future theoretical developments and the implementation of numerical simulations to test and check those scenarios could lead to new possibilities for PBHs to account for dark matter and update the abundance estimations. In particular, it would be interesting to compare the results already pointed out in this review in other scenarios.

We hope that in the next few years, several scientific results in this and other directions will help to fully understand and constrain the PBH scenario and its consequences in our Universe, especially taking into account future gravitational wave observations.

\funding{AE is currently supported by "bourse de post-doctorat" from Universit\'e Libre de Bruxelles (ULB), and was supported by contract PID2019-106515GB-I00/AEI/10.13039/501100011033 (Spanish Ministry of Science and Innovation).}

\acknowledgments{I want to thank Vicente Atal, Guillem Domenech, Cristiano Germani, Claudia Gonzalez-Boquera and Marc Oncins for extremely useful comments about the manuscript. I would also like to thank discussions on the PBH topic to Vicente Atal, Sebastien Clesse, Guillem Domenech, Jaume Garriga, Cristiano Germani, Carsten Gundlach, Ilia Musco, Marc Oncins, Javier G. Subils, Ravi K. Sheth, Antonio Enea Romano, Yuichiro Tada, Shuichiro Yokoyama and Chulmoon-Yoo. Finally, I also want to thank the anonymous referees for the constructive and valuable comments that have helped to improve this work.}
%\conflictsofinterest{Declare conflicts of interest or state ``The authors declare no conflict of interest.'' Authors must identify and declare any personal circumstances or interest that may be perceived as inappropriately influencing the representation or interpretation of reported research results. Any role of the funders in the design of the study; in the collection, analyses or interpretation of data; in the writing of the manuscript, or in the decision to publish the results must be declared in this section. If there is no role, please state ``The funders had no role in the design of the study; in the collection, analyses, or interpretation of data; in the writing of the manuscript, or in the decision to publish the~results''.} 

%% Optional
%\sampleavailability{Samples of the compounds ... are available from the authors.}

%%%%%%%%%%%%%%%%%%%%%%%%%%%%%%%%%%%%%%%%%%
%% Only for journal Encyclopedia
%\entrylink{The Link to this entry published on the encyclopedia platform.}

%%%%%%%%%%%%%%%%%%%%%%%%%%%%%%%%%%%%%%%%%%
%% Optional
\abbreviations{PBHs: Primordial Black Holes\\ DM: Dark matter\\
FLRW: Friedman-Lemaître-Robertson-Walker \\
GR: General relativity\\
CMB: Cosmic Microwave Background\\
LIGO: Laser Interferometer Gravitational-Wave Observatory\\
BH: Black Hole\\
PDE: Partial differential equation\\
AMR: Adaptive mesh refinement\\
NANOGrav: North American Nanohertz Observatory for Gravitational Waves\\}

%%\noindent 
%%\begin{tabular}{@{}ll}
%%MDPI & Multidisciplinary Digital Publishing Institute\\
%%DOAJ & Directory of open access journals\\
%%TLA & Three letter acronym\\
%%LD & Linear dichroism
%%\end{tabular}}

%%%%%%%%%%%%%%%%%%%%%%%%%%%%%%%%%%%%%%%%%%
%% Optional
\appendixtitles{no} % Leave argument "no" if all appendix headings stay EMPTY (then no dot is printed after "Appendix A"). If the appendix sections contain a heading then change the argument to "yes".
%\appendixstart
%\appendix
%\section{}
%\subsection{}

%\section{}
%All appendix sections must be cited in the main text. In the appendices, Figures, Tables, etc. should be labeled, starting with ``A''---e.g., Figure A1, Figure A2, etc. 

%%%%%%%%%%%%%%%%%%%%%%%%%%%%%%%%%%%%%%%%%%
%\end{paracol}
%%%%%%%%%%%%%%%%%%%%%%%%%%%%%%%%%%%%%%%%%%
% To add notes in main text, please use \endnote{} and un-comment the codes below.
%\begin{adjustwidth}{-5.0cm}{0cm}
%\printendnotes[custom]
%\end{adjustwidth}
%%%%%%%%%%%%%%%%%%%%%%%%%%%%%%%%%%%%%%%%%%
\reftitle{References}


\externalbibliography{yes}
\bibliography{refs4.bib}

%=====================================
% References, variant B: internal bibliography
%=====================================

\end{document}